\documentclass{aa}  

\usepackage{graphicx}
\usepackage{txfonts}
\usepackage[hidelinks,colorlinks=true,citecolor=cyan]{hyperref}
\usepackage{isotope}

\newcommand{\msun}{$\rm M_{\odot}$}
\newcommand{\p}{$p$}
\newcommand{\rr}{$r$}
\newcommand{\s}{$s$}
\newcommand{\g}{$\gamma$}
\newcommand{\A}{$\alpha$}
\newcommand{\fz}{$\rm F_0$}
\newcommand{\fx[1]}{$\rm F_{#1}$}

\newcommand{\rgn}{($\gamma$,$\rm n$)}
\newcommand{\rng}{($\rm n$,$\gamma$)}
\newcommand{\rga}{($\gamma,\alpha$)}
\newcommand{\rag}{($\alpha,\gamma$)}
\newcommand{\rgp}{($\gamma,\rm p$)}
\newcommand{\rpg}{($\rm p,\gamma$)}
\newcommand{\ran}{($\alpha,\rm n$)}

\newcommand{\rgX}{($\gamma$,$\rm X$)}
\newcommand{\rXg}{($\rm X$,$\gamma$)}

\graphicspath{{./}{figures/}}

\begin{document} 

   \title{The \g--process nucleosynthesis in core-collapse supernovae}
   \subtitle{I. A novel analysis of \g--process yields in massive stars}

    \author{
        L. Roberti \inst{1,2,3,4}
        \and
        M. Pignatari \inst{1,2,5,4}
        \and
        A. Psaltis \inst{6,7,4}
        \and
        A. Sieverding \inst{8}
        \and
        P. Mohr \inst{9}
        \and 
        Zs. F\"{u}l\"{o}p \inst{9}
        \and 
        M. Lugaro \inst{1,2,10,11}
    }
    
    \institute{Konkoly Observatory, Research Centre for Astronomy and     Earth Sciences, E\"otv\"os Lor\'and Research Network (ELKH), Konkoly     Thege Mikl\'{o}s \'{u}t 15-17, H-1121 Budapest, Hungary 
    \and
                CSFK, MTA Centre of Excellence, Budapest, Konkoly Thege Miklós út 15-17, H-1121, Hungary
    \and
                INAF -- Osservatorio Astronomico di Roma Via Frascati     33, I-00040, Monteporzio Catone, Italy
    \and            
                NuGrid Collaboration, \url{http://nugridstars.org}
    \and
               E. A. Milne Centre for Astrophysics, University of Hull,     Hull HU6 7RX, UK
    \and    
            Department of Physics, North Carolina State University, Raleigh, NC, 27695, USA
    \and
            Triangle Universities Nuclear Laboratory, Duke University, Durham, NC, 27710, USA
    \and
            Max-Planck Institute for Astrophysics, Postfach 1317, 85741 Garching, Germany
    \and
            Institute for Nuclear Research (ATOMKI), H-4001 Debrecen, Hungary
    \and
               E\"otv\"os Lor\'and University, Institute of Physics,     Budapest 1117, P\'azm\'any P\'eter s\'er\'any 1/A,    Hungary
   \and
            School of Physics and Astronomy, Monash University, VIC     3800, Australia}

   \date{Received March 31, 2023; accepted June 15, 2023}

% \abstract{}{}{}{}{} 
% 5 {} token are mandatory
 
  \abstract
  % context
   {The \g--process nucleosynthesis in core-collapse supernovae is generally accepted as a feasible process for the synthesis of neutron-deficient isotopes beyond iron. However, crucial discrepancies between theory and observations still exist: the average production of \g--process yields from massive stars are too low to reproduce the solar distribution in galactic chemical evolution calculations, and the yields of the Mo and Ru isotopes are by a further factor of 10 lower than the yields of the other \g--process nuclei.}
  % aims 
   {We investigate the \g--process in 5 sets of core-collapse supernova models published in literature with initial masses 15, 20, and 25 \msun\ at solar metallicity.}
  % methods 
   {We compared the \g--process overproduction factors from the different models. To highlight the possible effect of nuclear physics input, we also considered 23 ratios of two isotopes close to each other in mass, relative to their solar values. Further, we investigated the contribution of C--O shell mergers in the supernova progenitors as an additional site of the \g--process.}
  % results
   {Our analysis shows that a large scatter among the different models exists for both the \g--process integrated yields and the isotopic ratios. We found only 10 ratios that agree with their solar values, all the others differ by at least a factor of 3 from the solar values in all the considered sets of models. The \g--process within C--O shell mergers mostly influence the isotopic ratios that involve intermediate and heavy proton-rich isotopes with $\rm A>100$.}
  % conclusions
   {We conclude that there are large discrepancies both among the different data sets and between the models predictions and the solar abundance distribution. More calculations are needed particularly updating the nuclear network, since the majority of the models considered in this work did not use the latest reaction rates for the \g--process nucleosynthesis. Moreover, the role of C--O shell mergers needs further investigation.}

   \keywords{\g--process nucleosynthesis --
                explosive nucleosynthesis --
                massive stars
               }

   \maketitle
%
%-------------------------------------------------------------------

% ==================== Introduction ==================== 
\section{Introduction} \label{sec:intro}

    %=== p-nuclei
    Neutron--capture processes \citep[see, e.g.,][and references therein]{kaeppeler:11,cowan:19} made the majority of the nuclei beyond iron in the Solar System. However, these processes are not able to make a number of rare, neutron-deficient, stable isotopes, which constitute a small fraction in mass of the heavy nuclear species. Already in the pioneering work of \cite{burbidge:57} 35 stable proton--rich nuclides\footnote{namely: \isotope[74]{Se}, 
    \isotope[78]{Kr}, \isotope[84]{Sr}, \isotope[92,94]{Mo},     \isotope[96,98]{Ru}, 
    \isotope[102]{Pd}, \isotope[106,108]{Cd}, \isotope[112,114,115]{Sn}, 
    \isotope[113]{In}, \isotope[120]{Te}, \isotope[124,126]{Xe},     \isotope[130,132]{Ba}, 
    \isotope[136,138]{Ce}, \isotope[138]{La}, \isotope[144]{Sm},     \isotope[152]{Gd}, 
    \isotope[156,158]{Dy}, \isotope[162,164]{Er}, \isotope[168]{Yb},     \isotope[174]{Hf}, 
    \isotope[180]{Ta}, \isotope[180]{W}, \isotope[184]{Os},     \isotope[190]{Pt},
    and \isotope[196]{Hg}.} (the \p--nuclei) heavier than Fe were identified and the \p--process was postulated to explain their existence \citep[see also][]{cameron:57}. 
    In the following years, \cite{arnould:76} demonstrated that \p--nuclei can be synthesised through a sequence of (\g,$n$), (\g,\p) and (\g,\A) reactions on pre-existing seed heavy material during the latest stages of the hydrostatic evolution of massive stars. \cite{woosley:78} showed that the \p--process material produced during the pre-supernova evolution is completely reprocessed by the passage of the explosive shock wave; and that the O/Ne-rich layers during core-collapse supernovae (CCSNe) explosions provide the conditions to re-make the \p--nuclei via a chain of photodisintegrations. They called this the \g--process. \cite{prantzos:90b} and then \cite{rayet:95} performed the first complete computations of the \g--process and provided the sets of yields that have been taken as the reference for the studies of the \p--nuclei for the following decades \citep[e.g.,][and references therein]{arnould:03, rauscher:13, pignatari:16}.
    
    %=== production site
   Not all the heavy proton-rich isotopes are made exclusively by the \g--process. Branching points, either activated by unstable nuclei or by stable isotopes that become unstable at stellar temperature, allow neutron captures to play a role as well. \isotope[152]{Gd} and \isotope[164]{Er} are representative of this, since they have a potentially dominant \s--process contribution from low-mass asymptotic giant branch (AGB) stars due to the branching points at \isotope[151]{Sm} (unstable in terrestrial conditions) and \isotope[163]{Dy} (stable in terrestrial conditions), respectively \citep[see, e.g.,][]{arlandini:99,bisterzo:11,bisterzo:15}. The production of \isotope[113]{In} and \isotope[115]{Sn} depends on the $\beta$--decay of the unstable isomers of the stable nuclides \isotope[113]{Cd} and \isotope[115]{In}, respectively. In this case, however, their origin is still unclear, with a possible \rr--process contribution \citep{dillmann:08}. The very long-lived \isotope[180]{Ta} isomer, with a half-life larger than $\rm 1.2\times10^{15} yr$, may be efficiently produced both by neutrino-spallation reactions on \isotope[180]{Hf} \citep{cumming:85} and the \s-process \citep[see, e.g.,][]{rauscher:02,bisterzo:11}. In the latter case, as for \isotope[163]{Dy}, \isotope[179]{Hf} becomes unstable at stellar temperatures and activates a branching point on the \s--process path. Finally, \isotope[138]{La} might receive a substantial contribution from the neutrino captures on \isotope[138]{Ba} \citep{goriely:01}. Moreover, new processes have been identified in the last decades to occur in the deepest interior of CCSNe, which may also participate in the production of the lightest \p--nuclei between Fe and Pd. These processes range from the neutrino-driven winds from the forming neutron star \citep[see, e.g.,][]{froehlich:06,arcones:11} to the \A--rich freeze-out in CCSN ejecta \citep[see, e.g.,][]{woosley:92,pignatari:16}, and the $\nu r$-process proposed very recently by \cite{xiong:23}. Modern multidimensional CCSN simulations \citep[e.g.,][]{harris:17} and calculations of the explosion in spherical symmetry with the inclusion of neutrino heating \citep[e.g.,][]{curtis:19, ghosh:22}, point toward the possibility of ejecting a significant amount of \p--nuclei yields from the innermost layers of the star, especially in the Mo - Ru region \citep[see, e.g.,][]{eichler:18}.
   
    %=== problems
    Although the \g--process nucleosynthesis in CCSNe has been explored for many decades \citep[see, e.g., the reviews][and references therein]{rauscher:13,pignatari:16a}, a complete understanding of the production of \p--nuclei in stars is still missing and several discrepancies arise when comparing theoretical model predictions with the Solar System abundances. The first fundamental problem is that the average \g--process yields from massive star underproduce by about a factor of three the Solar System abundances. Furthermore, \isotope[92,94]{Mo} and \isotope[96,98]{Ru} are systematically underproduced by one order of magnitude, compared to the other \g--process nuclei \citep{arnould:03, rauscher:13,pignatari:16}. Several works have shown that nuclear uncertainties are not the reason of the substantial disagreement between theory and observations \citep[see, e.g.,][]{rapp:06,rauscher:16,nishimura:18}.

    %=== alternative scenario
    The external layers of thermonuclear supernovae (SNIa) resulting from a white dwarf (WD) accreting material from a stellar binary companion and reaching the Chandrasekhar mass limit have also been suggested as a site of the \g--process \citep{travaglio:11,travaglio:14,battino:20}. First galactic chemical evolution (GCE) calculations made for \p--nuclei \cite{travaglio:18} confirmed that SNIa scenario would solve the issues related to the \g--process in CCSNe. However, the SNIa solution strongly depends on the number of SNIa generated from Chandrasekhar-mass progenitors within the single-degenerate scenario \citep{hillebrandt:13}. According to the observations, this number is still affected by large uncertainties and could be much lower than required to solve the missing \g--process production \citep[e.g.,][and references therein]{woods:13,battino:20}.
    
    %=== other solutions regarding massive stars: C12+C12, C-O shell mergers, rotating massive stars.    
    Given these still open, significant questions, investigations of potential solutions to boost the production of \p--nuclei in massive stars, in particular in the Mo-Ru region, are still required. For example, \cite{pignatari:13} showed that an increase by two orders of magnitude of the \isotope[12]{C}$+$\isotope[12]{C} reaction rate by \cite{caughlan:88} may lead to a Mo and Ru \g--process production up to the level of the other \p--nuclei, and to an overall increased \g--process efficiency. Also, during the latest stages of the evolution of massive stars, the convective O burning shell may ingest some C-rich material \citep{rauscher:02,meakin:06,ritter:18,andrassy:19}. This interaction may lead to the formation of an extended merged convective zone, as well as to asymmetries in the stellar structure of the progenitor that might facilitate the CCSN explosion \citep[e.g.,][]{couch:13,mueller:16a}. During this "C--O shell merger" event, the C shell material is brought down to the base of the O shell at typical O burning temperature, triggering efficient \g--process nucleosynthesis \citep{rauscher:02,ritter:18a}. The mixing of the \p--nuclei throughout the extended, mixed C--O shell preserves them from being fully reprocessed by the CCSN shock wave. Their yields may therefore increase by orders of magnitude, as compared to models without the C--O shell merger. It still needs to be investigated, however, if this scenario can solve some of the puzzles of the missing production of the \p--nuclei. Finally, because of the strong sensitivity of the \g--process yields on the seeds from the \s--process \citep{costa:00}, \cite{choplin:22} proposed that an enhanced production of \s--process seeds in fast rotating massive stars ("spinstars") at sub-solar metallicity could increase the efficiency of \g--process nucleosynthesis and that this contribution could even dominate the solar \p--nuclei abundance distribution. However, the number of such spinstars as a function of the metallicity is still uncertain and a further investigation is required.

    %=== aim of the paper
    To shed further light on the mystery of the \g--process nucleosynthesis in CCSNe, we started a research program dedicated to implementing the latest nuclear reaction rates into new calculations of CCSNe for a wide grid of explosion energies, using the most recent version of the NuGrid nucleosynthesis codes \citep[e.g.,][]{pignatari:16a,ritter:18,lawson:22}. In this first paper of the series, we present our first step of the program, corresponding to the analysis of the \g--process yields in a number of sets of CCSN models from the literature. The paper is organized as follows: in Section \ref{sec:datasets} we present the sets of CCSN models; in Section \ref{sec:f0} we describe our method; in Section \ref{sec:isoratio} we outline our results; in Section \ref{sec:nuclearphysics} we give a brief overview of the current state of the art in nuclear physics for the \g--process; and in Section \ref{sec:conclusions} we discuss the results and present the conclusions.

% ==================== Datasets ==================== 
\section{Data sets} \label{sec:datasets}

    We consider yields from five sets of CCSN models \citep{rauscher:02,pignatari:16a,ritter:18,sieverding:18,lawson:22} with progenitors of initial masses of 15, 20, and 25 \msun\ and solar metallicity. All of the models are based on one-dimensional (1D) computations using different codes and nuclear networks for the hydrostatic evolution, the explosion, and the nucleosynthesis. In the following, we briefly describe the main features of each set.

    \subsection{Rauscher+02 (RAU)}
        The data from \cite{rauscher:02} (hereafter labelled as RAU15, RAU20, RAU25 for the three masses, respectively) are a well-known and widely used set. They were produced using an adaptive nuclear network including up to a maximum number of $\sim2200$ nuclear species. The progenitors of the CCSNe were calculated by an updated version of the {\scshape Kepler} code \citep{weaver:78}, while the supernova explosions were simulated using the piston technique \citep{woosley:95}. The explosion energy of the 15 \msun\ model is $1.2\times10^{51}$ erg, according to the assumption for SN 1987A \citep[e.g., ][]{woosley:88}, while in the case of 20 and 25 \msun\ models, it is adjusted to obtain the ejection of 0.1 \msun\ of \isotope[56]{Ni}. The initial solar composition is from \cite{anders:89} (AG89). The reaction rates involving the nuclei above Fe are mostly from \cite{rauscher:00}, and from \cite{bao:00} in the case of the $\rm (n,\gamma)$ reactions. In case of missing experimental information, the Hauser-Feshbach rates obtained with the NON-SMOKER code \citep{rauscher:97} were used. In the case of the two reactions, involving \p--nuclei, \isotope[70]{Ge}$(\alpha,\gamma)$\isotope[74]{Se} and \isotope[144]{Sm}$(\alpha,\gamma)$\isotope[148]{Gd}, the experimental rates from \cite{fulop:96} and \cite{somorjai:98} were implemented, respectively. Moreover, the rates that shared the same \A--potentials of the two reactions above were re-computed using those experimental information. The 20 \msun\ model experience a C--O shell merger about 1 day before the core collapse, forming a single extended mixed convective zone. As mentioned in the Introduction (Sec. \ref{sec:intro}), the C--O shell merger triggers an efficient \g--process nucleosynthesis during the pre-supernova evolution of the stars that may dominate over the effect of the explosive nucleosynthesis during the CCSN.
        
    \subsection{Pignatari+16 (PGN)}
        The 15, 20, and 25 \msun\ progenitors presented in \cite{pignatari:16a} (hereafter PGN15, PGN20, PGN25, respectively) were computed using the {\scshape Genec} stellar evolutionary code \citep{eggenberger:08} and following the evolution up to and including the central Si burning phase. The further collapse of the Fe core was not followed and the supernova explosion was simulated using a semi-analytical approach, with prescription for the mass cut from \cite{fryer:12}. The adopted initial solar composition is from \cite{grevesse:93} (GN93). The nucleosynthesis during the evolution and the explosion was computed with the Multi-zone Post-Processing Network -- Parallel code \citep[MPPNP,][]{Pignatari:2012dw,pignatari:16a,ritter:18}, using a dynamical network with a maximum number of 5234 nuclear species. The nuclear network includes 74,313 reactions and above Fe it uses the experimental rates from KADoNIS compilation \citep{dillmann:06} wherever possible for $\rm (n,\gamma)$ reactions, and the Basel REACLIB database \citep{rauscher:00} revision 20090121 for the missing reactions. These models present some interesting features regarding the \g--process nucleosynthesis. In the PGN15 model, the mass-cut is deep enough to allow the ejection of some \A-rich freeze-out material, slightly increasing the yields of the lightest \g--process nuclei. In the PGN20 model, the shock wave accelerates at the interface between CO and He core, due to a steep density gradient. This results in an increase of the peak temperature, which leads to a re-activation of the \g--process nucleosynthesis. In the PGN25 model, the \g--process occurs in a narrow region of the star that, unfortunately, lacks in spatial resolution. This does not allow us to distinguish the abundance peaks of all the \p--nuclei. For this reason, we decided to exclude this model from the discussion in the next sections.

    \subsection{Ritter+18 (RIT)}
        The set of \cite{ritter:18} (RIT15, RIT20, RIT25) constitutes a follow-up work of \cite{pignatari:16a} except that the stellar progenitors were calculated with the \texttt{MESA} code \citep{paxton:11}. The main features of the RIT models are the same as in the PNG models (initial composition, physics inputs, explosion mechanism, and nucleosynthesis code), except for parts of the reaction network database and, as mentioned above, the code used to compute the stellar progenitors. The RIT models adopt the JINA REACLIB reaction rate library V1.1 \citep{cyburt:11} instead of the Basel REACLIB database as the PNG models. The use of a different stellar code led to several significantly different results. The RIT15 model experiences a C--O shell merger event, similarly to the RAU20 model. Also in this case, the production of the \g--process nuclei greatly increases in the pre-supernova evolution. The RIT20 model has a large amount of fallback material after the explosion, such that only the region above explosive Ne burning is ejected, and ending with a remnant mass of 2.7 \msun. Moreover, in this star, the convective C--shell is particularly extended and it is able to efficiently activate the \isotope[22]{Ne} neutron source, triggering the \s--process nucleosynthesis. The RIT25 model experience so large fallback that the explosion ends in a failed supernova, locking all the explosive nucleosynthesis material into the central remnant of 5.7\msun.

    \subsection{Sieverding+18 (SIE)}
        The set from \cite{sieverding:18} (SIE15, SIE20, SIE25) is based on stellar progenitors calculated using the {\scshape Kepler} code, like the RAU models, and with the same explosion energy equal to $1.2\times10^{51}$ erg, but with different choices for several physical and numerical parameters. The initial solar composition is from \citet{lodders:03} (L03). The nuclear network used for the stellar evolution calculations is not identical to the reaction network used for the explosion. While the former is mostly identical to the RAU models, the latter includes the JINA REACLIB reaction rate library V2.2, and an updated $\nu$--induced reaction rate set. The rate of the neutron--capture reaction \isotope[137]{La}$(\rm n,\gamma)$\isotope[138]{La} is instead from a previous version of REACLIB, i.e., \cite{rauscher:00}. The inclusion of $\nu$--induced reactions is crucial for the production of the two \p--nuclei \isotope[138]{La} and \isotope[180]{Ta}, which are directly produced by $\nu_{\rm{e}}$ capture on the corresponding isobars \citep{woosley:90b}. \cite{sieverding:18} presented a parametric study of the explosive yields using several neutrino energies. Here we include the same set as in \cite{denhartogh:22}, i.e., the models with the highest neutrino energy.

    \subsection{Lawson+22 (LAW)}
        The stellar progenitors of \cite{andrews:20} and \cite{lawson:22} (LAW15, LAW20, LAW25) were computed with a recent version of the {\scshape Kepler} code \citep{heger:10}, while for the explosion a 1D hydrodynamic code was used, mimicking a 3D convective engine \citep{fryer:99,fryer:18}. The hydrostatic nucleosynthesis was calculated via the MPPNP code (as PNG and RIT), while for the explosive nucleosynthesis the Tracer particle Post-Processing Network -- Parallel code was used \citep[TPPNP,][]{jones:19}. The initial solar composition is GN93, as in the RIT models. The stellar yields for the whole stellar set including the models used in this work are available online \citep[][]{andrews:20}. As in RIT20, also in the LAW25 model the C--shell is particularly extended and has an efficiently active neutron source. Again similarly to RIT20, also LAW20 has a large fallback and all the explosive nucleosynthesis is confined within an innermost smaller mass of 2.2 \msun, at the position of the final mass-cut. Therefore, the ejecta of this model contains only the material ejected from the C shell outward.

% ==================== F_0 ==================== 
\section{Average overproduction factors} \label{sec:f0}

    \begin{figure*}
        \centering
        \includegraphics[scale=0.5]{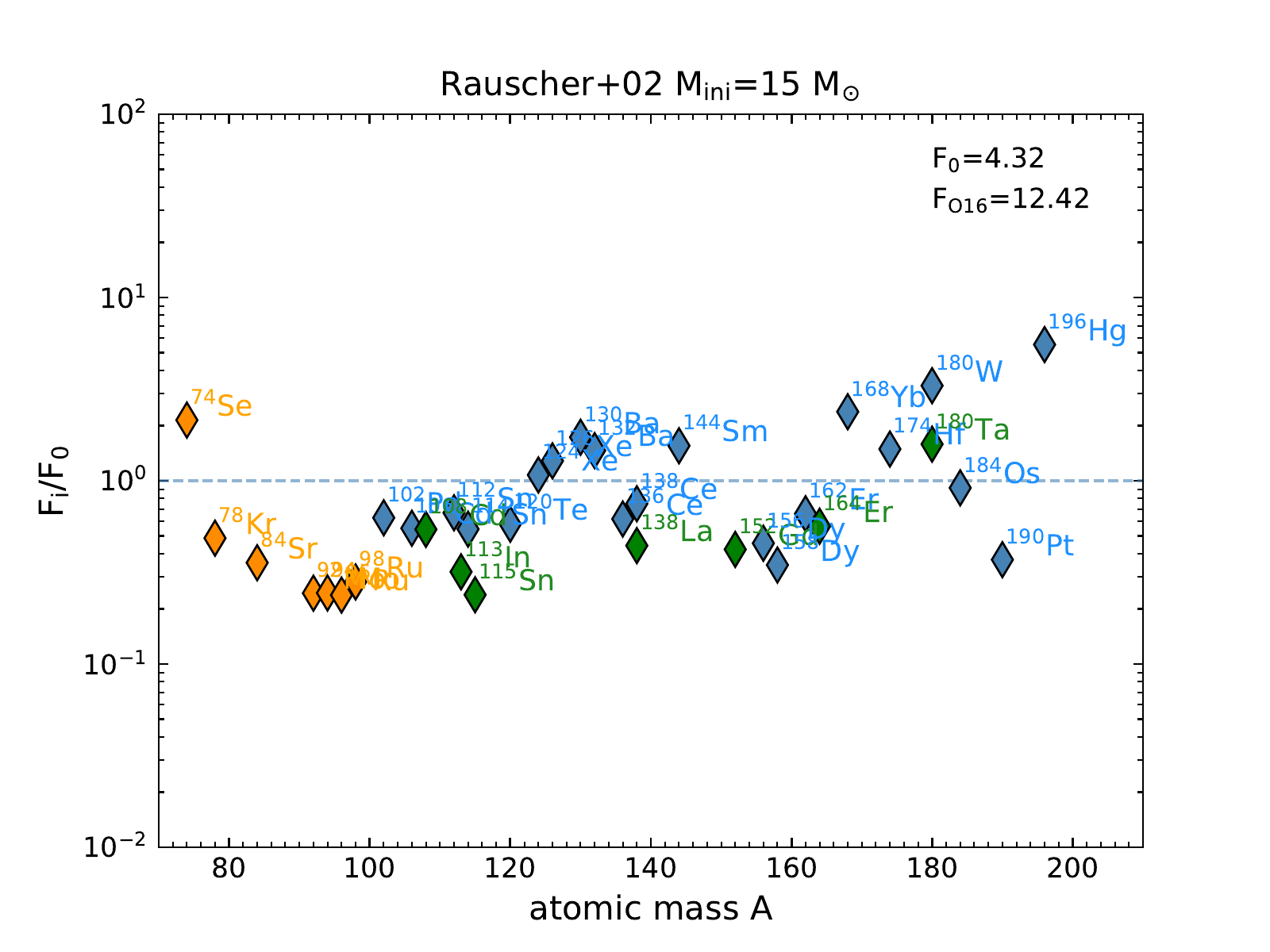}
        \includegraphics[scale=0.5]{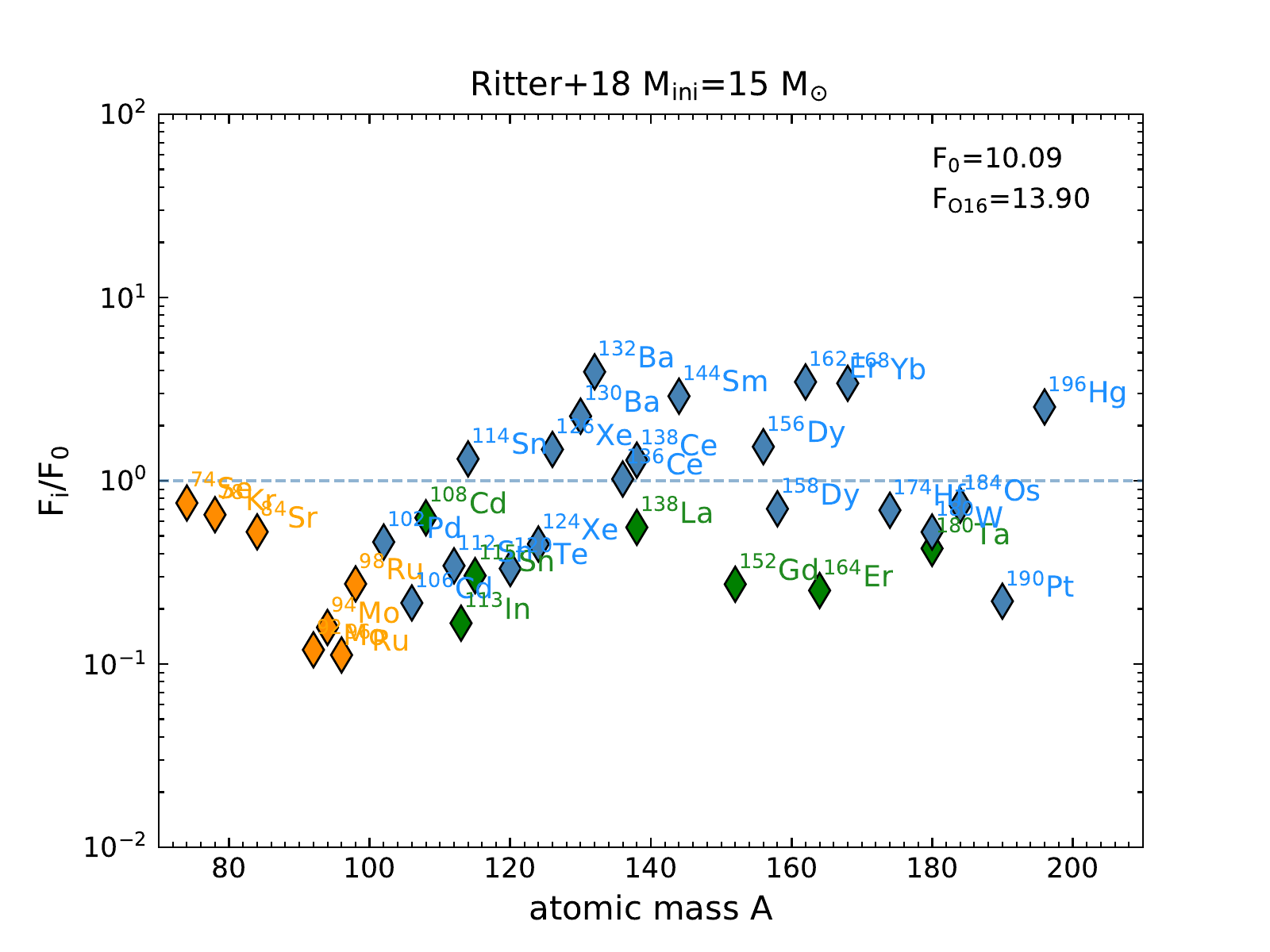}
        \includegraphics[scale=0.5]{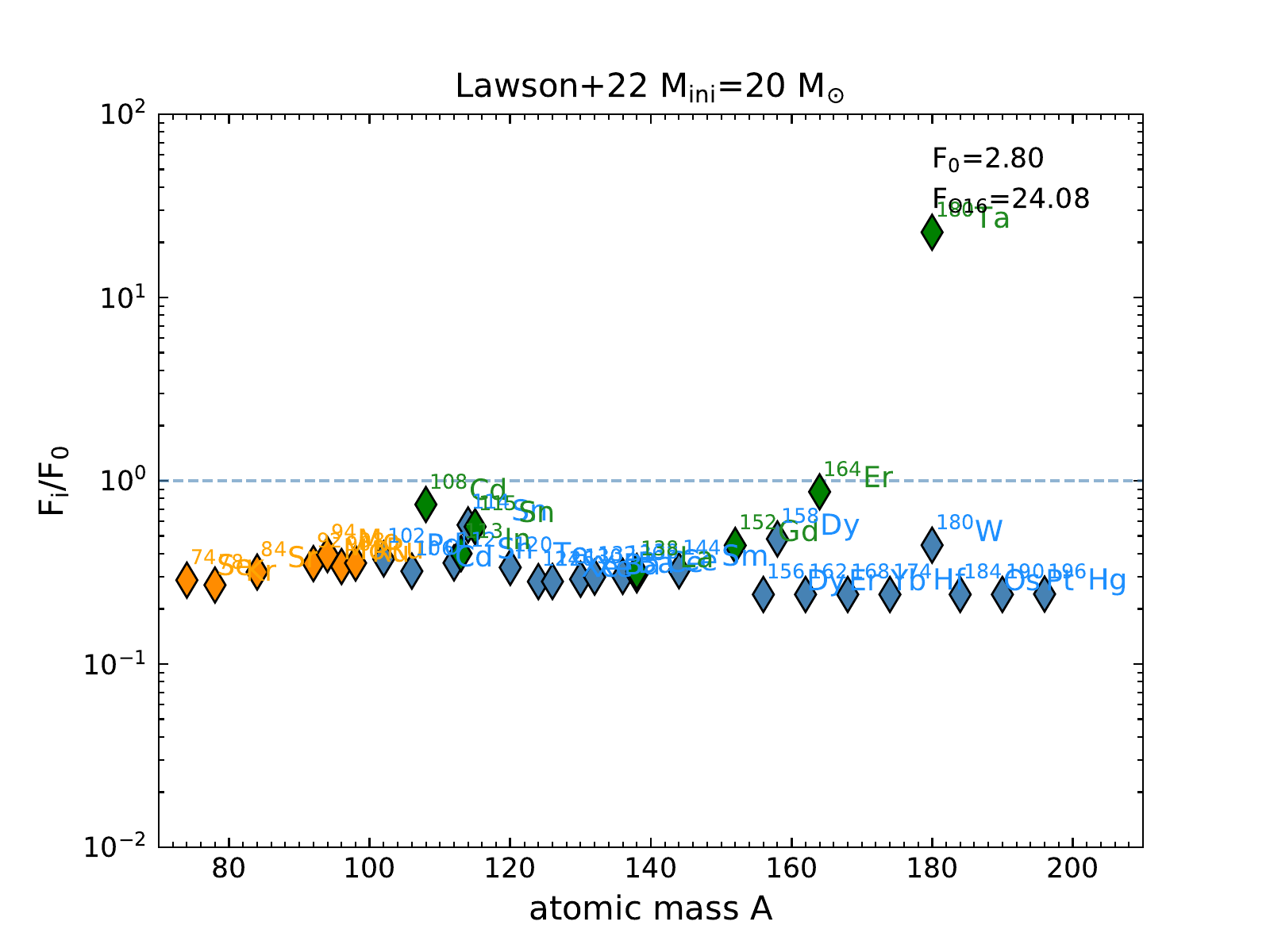}
        \includegraphics[scale=0.5]{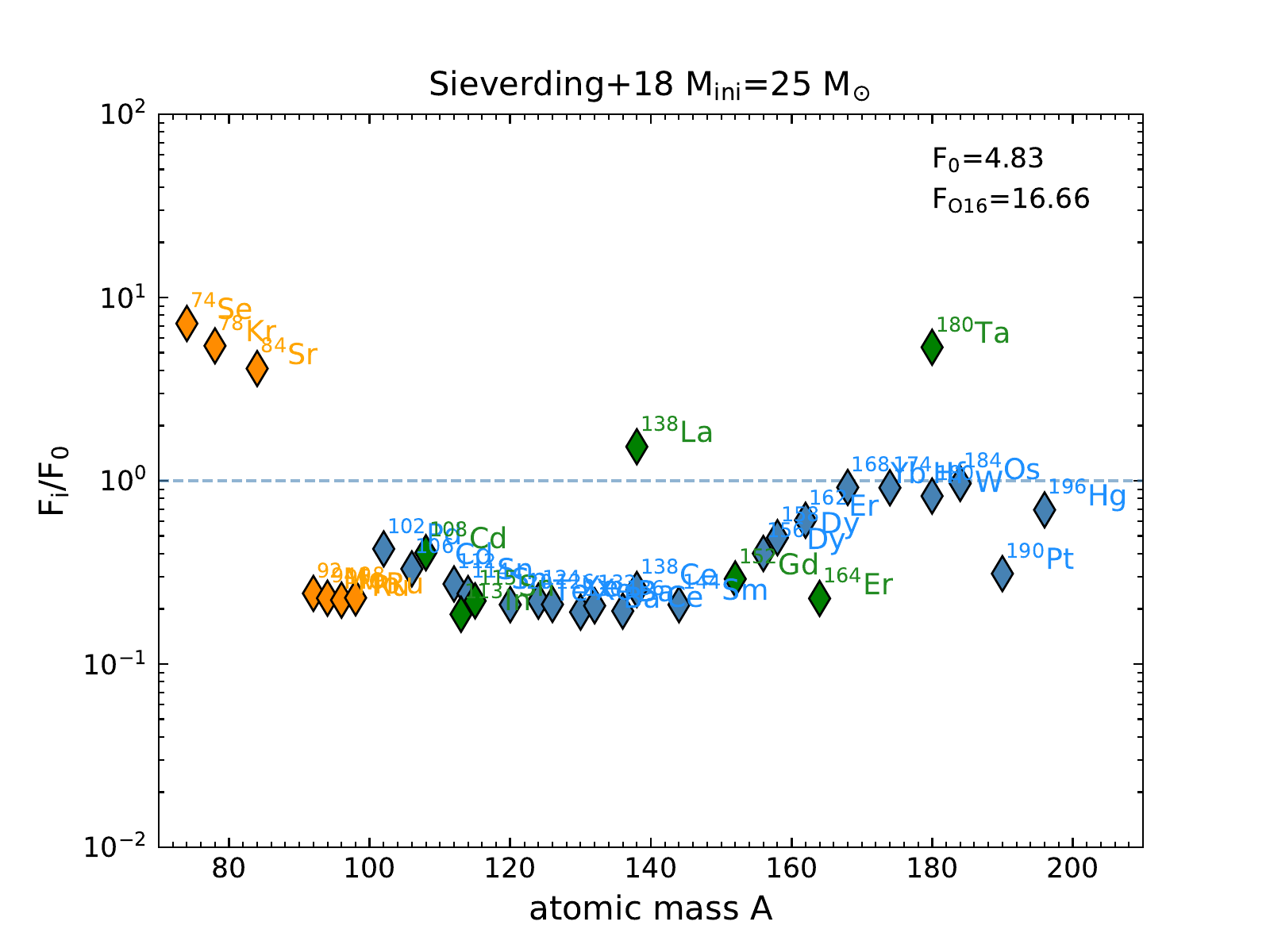}
        \caption{The \p--nuclide overproduction factors divided by their average (Eq. \ref{eq:f0}), whose value is indicated in the top right of each panel, together with the OP factor for $^{16}$O (\fx[O16]) for comparison. 
        The blue, orange, and green symbols represent, respectively: (i) nuclei produced exclusively by the \g--process, (ii) nuclei that may have an additional explosive contribution; and (iii) nuclei that have also \s--, \rr--process, or neutrino-capture  contributions. Four examples of representative models are shown: the upper left panel is RAU15, a typical model; the upper right panel is RIT15, a case where the C--O shell merger occurs; the lower left panel is LAW20, a failed supernova; and the lower right panel is SIE25, a model where the \fz\ is the same as for the typical RAU15 model shown upper left panel, but where this value is driven by two isotopes produced by neutrino captures, rather than by the \p--nuclei produced exclusively by the \g--process \citep{sieverding:18}. Note that in the two bottom panels a large number of \p--nuclei, including the Mo and Ru isotopes, share the same \fx[i]/\fz\ $\simeq 0.2-0.3$. However, as shown by the low \fz, there is no significant production of these isotopes and their pattern only reflects the solar abundance distribution used as initial in the models.}
        \label{fig:f0}
    \end{figure*}

    The overproduction factor (OP) for each isotope is defined as OP = $X/X_{\odot}$, i.e., the ratio between its mass fraction $X$ equal to the total integrated yield divided by the total mass ejected, relative to the solar mass fraction $X_{\odot}$, for which we use the values of \cite{asplund:09}\footnote{For the \p--only isotopes, these values are the same as those given in the other solar abundance compilations used in the different models as described in the previous section.}.    
    A useful parameter to evaluate the efficiency of the \g--process nucleosynthesis is \fz, defined as the average of the OP factors of all the 35 \p--nuclei, i.e.:

    \begin{equation}
        \centering
        F_0 = \frac{\sum_{\rm i=1}^{35} F_{\rm i}}{35}
        \label{eq:f0}
    \end{equation}
    
    where \fx[i] is the OP factor of the \p--nuclide $i$. The value of \fz\ has been taken over the years as the representative feature of \g--process nucleosynthesis and has been used as a normalisation factor for the single \p--nuclei OP factors, i.e., \fx[i]$/$\fz\ measures the deviation of the production of each single isotope $i$ from the average production. \figurename~\ref{fig:f0} shows the \fx[i]$/$\fz\ distribution for a selection of models. The same figure for each model can be found in Appendix \ref{app:OP}, together with a figure showing the OP factors of all the 35 \p--isotopes. 
    
    Two fundamental features are highlighted from the figure and from the distributions of all the models. First, there are large variations of both the OP factors and of \fz\ with the initial mass of the stellar progenitor. Depending on the set of models considered, \fz\ varies by up to a factor of roughly five, with even larger variation for the OP of single isotopes. Such a scatter is much larger than what was found in previous studies \citep[e.g.,][]{rayet:95}, supporting the idea that \g--process nucleosynthesis does not depend significantly on the progenitor mass. Second, large variations are present also when we compare models with the same progenitor mass but from different sets. This is because the \p--nuclei abundances do not depend only on the progenitor mass and the explosion energy. Rather, they are a complex product of the combination of conditions in the stellar progenitor layers and the interaction of those layers with the CCSN shock \citep[e.g.,][]{rauscher:02, ritter:18}. 
    
    The \fz\ has been often compared to the OP factor of the nuclide \isotope[16]{O} (hereafter \fx[O16]), one of the main products of CCSNe. This allows us to approximately quantify the contribution from massive star to the solar abundances of the \p--nuclei without using GCE calculations. If the \p--nuclei and \isotope[16]{O} share the same production site, one would expect that \fz\ $\sim$ \fx[O16], however, \cite{rayet:95} already found an underproduction of at least of a factor of 2, i.e., \fz/\fx[O16] $\simeq$ 0.5. \cite{pignatari:13} confirmed this result and pointed out that the \p--nuclei underproduction is probably more severe, since at solar metallicity secondary nucleosynthetic products such as \g--process nuclei should be synthesised twice as much as primary isotopes, such as \isotope[16]{O} \citep[see, e.g.,][]{tinsley:80}. All the models in our sets have \fz\ $\lesssim$ \fx[O16], except for PGN15, where \fz/\fx[O16] $\sim$ 1.05, still far from the factor of roughly two required to reproduce the solar abundances.

   \begin{table*}
      \caption{Percentages of \p--nuclei yields in C--O mergers.}
      \label{tab:merper}      
      \centering                          
      \begin{tabular}{lrrrcrrr}        
      \hline\hline                 
      isotope   & explosive   &    merger  &  envelope & &  explosive  & merger & envelope \\ 
                & ($\%$)      &    ($\%$)  &  ($\%$)   & &  ($\%$)     & ($\%$) & ($\%$)   \\
      \hline
      \multicolumn{1}{c}{} & \multicolumn{3}{c}{RIT15} & \multicolumn{1}{c}{} & \multicolumn{3}{c}{RAU20} \\
      \cline{2-4} \cline{6-8}
      \isotope[ 74]{Se}  &  \textbf{74.1}  &  14.9           &  10.9           &  &  47.3           & 49.0           &   3.6           \\
      \isotope[ 78]{Kr}  &  \textbf{82.1}  &   5.9           &  11.9           &  &  \textbf{62.1}  & 28.8           &   9.1           \\
      \isotope[ 84]{Sr}  &  \textbf{70.5}  &  11.5           &  17.4           &  &  42.7           & \textbf{53.2}  &   4.0           \\
      \isotope[ 92]{Mo}  &   6.4           &   8.4           &  \textbf{85.2}  &  &  10.5           & 15.0           &  \textbf{74.5}  \\
      \isotope[ 94]{Mo}  &   2.8           &  27.4           &  \textbf{69.7}  &  &   5.9           & 45.6           &  48.3           \\
      \isotope[ 96]{Ru}  &   8.2           &   4.3           &  \textbf{87.5}  &  &  12.1           &  2.4           &  \textbf{85.5}  \\
      \isotope[ 98]{Ru}  &  13.1           &  49.6           &  37.2           &  &  16.7           & 37.9           &  45.3           \\
      \isotope[102]{Pd}  &  36.4           &  40.2           &  23.3           &  &  45.1           & 25.6           &  29.2           \\
      \isotope[106]{Cd}  &  34.6           &  22.5           &  43.0           &  &  \textbf{53.5}  &  9.2           &  37.3           \\
      \isotope[108]{Cd}  &  11.4           &  \textbf{63.4}  &  25.1           &  &  19.1           & 46.2           &  34.6           \\
      \isotope[112]{Sn}  &  34.6           &  35.5           &  29.8           &  &  47.0           & 16.9           &  36.1           \\
      \isotope[114]{Sn}  &  20.7           &  \textbf{71.0}  &   8.2           &  &  18.0           & \textbf{68.7}  &  13.0           \\
      \isotope[115]{Sn}  &   0.3           &  \textbf{64.9}  &  34.8           &  &   1.3           & \textbf{61.7}  &  36.8           \\
      \isotope[113]{In}  &   5.4           &  42.5           &  \textbf{52.0}  &  &   8.5           & 46.6           &  44.7           \\
      \isotope[120]{Te}  &   8.2           &  \textbf{62.5}  &  29.3           &  &  43.0           & 29.6           &  27.3           \\
      \isotope[124]{Xe}  &  22.0           &  \textbf{59.9}  &  18.1           &  &  \textbf{70.3}  &  8.9           &  20.8           \\
      \isotope[126]{Xe}  &   7.0           &  \textbf{87.4}  &   5.5           &  &  40.1           & \textbf{54.1}  &   5.6           \\
      \isotope[130]{Ba}  &  35.0           &  \textbf{61.0}  &   3.7           &  &  \textbf{66.6}  & 24.1           &   9.3           \\
      \isotope[132]{Ba}  &   1.8           &  \textbf{95.9}  &   2.2           &  &  18.6           & \textbf{78.9}  &   2.2           \\
      \isotope[136]{Ce}  &  40.4           &  \textbf{51.0}  &   8.5           &  &  42.6           & 39.0           &  18.4           \\
      \isotope[138]{Ce}  &   8.3           &  \textbf{84.7}  &   6.8           &  &  20.7           & \textbf{74.4}  &   4.6           \\
      \isotope[138]{La}  &   0.0           &  \textbf{82.9}  &  17.0           &  &   2.0           & \textbf{91.6}  &   6.2           \\
      \isotope[144]{Sm}  &  12.6           &  \textbf{84.0}  &   3.2           &  &  16.3           & \textbf{80.6}  &   2.9           \\
      \isotope[152]{Gd}  &   0.0           &   3.9           &   \textbf{96.1}           &  &   0.5           & 13.1           &  \textbf{86.1}  \\
      \isotope[156]{Dy}  &   0.6           &  \textbf{93.7}  &   5.7           &  &   7.7           & \textbf{89.5}  &   2.3           \\
      \isotope[158]{Dy}  &   0.0           &  \textbf{84.4}  &  15.5           &  &   1.8           & \textbf{95.8}  &   1.9           \\
      \isotope[162]{Er}  &   0.3           &  \textbf{97.2}  &   2.4           &  &   3.6           & \textbf{95.0}  &   1.0           \\
      \isotope[164]{Er}  &   7.5           &  39.1           &  \textbf{53.3}  &  &  16.4           & \textbf{70.7}  &  12.7           \\
      \isotope[168]{Yb}  &   2.4           &  \textbf{94.4}  &   3.2           &  &   3.8           & \textbf{95.1}  &   0.7           \\
      \isotope[174]{Hf}  &   3.7           &  \textbf{87.5}  &   8.6           &  &   6.7           & \textbf{92.2}  &   1.4           \\
      \isotope[180]{Ta}  &   0.0           &  \textbf{98.3}  &   1.7           &  &   0.9           & \textbf{94.7}  &   4.2           \\
      \isotope[180]{ W}  &  18.1           &  \textbf{70.7}  &  10.2           &  &  35.0           & \textbf{61.8}  &   3.1           \\
      \isotope[184]{Os}  &   0.9           &  \textbf{88.9}  &  10.2           &  &   0.8           & \textbf{96.8}  &   2.3           \\
      \isotope[190]{Pt}  &   1.2           &  \textbf{58.4}  &  40.3           &  &   3.5           & \textbf{91.4}  &   4.8           \\
      \isotope[196]{Hg}  &  12.1           &  \textbf{84.4}  &   2.7           &  &  14.9           & \textbf{84.2}  &   0.6           \\
      % ending body of the table      
      \hline     
      \end{tabular}\\
      \tablefoot{Percentages of the total \p--nuclei yields of the two models with the C--O merger (RIT15 and RAU20) in three parts of the ejecta: the zone dominated by the explosive nucleosynthesis ("explosive"), the zone of the mixed C--O shell ("merger"), and the zone from the He shell outwards ("envelope"). Contributions are highlighted in bold when larger than 50$\%$.}\\
   \end{table*}

    Relative to a typical model (e.g., top left panel of Figure~\ref{fig:f0}), models with a C--O shell merger (top right panel) can significantly contribute to the production of \g--process nuclei, as evident by comparing the values of \fz\ and \fx[O16] in these two panels. Nevertheless, also in these models \fz\ does not exceed \fx[O16].
    In both the RIT15 and RAU20 models the effect of the C--O shell merger is dominant on the yields of most of the \p-isotopes with $\rm A\gtrsim110$ (see \tablename~\ref{tab:merper}). The production of the lightest isotopes is instead less clear: the RAU20 merger component is comparable to the explosive component, while the RIT15 merger component is less relevant.
    When one element has more than one isotope produced by the \g--process, the C--O shell merger favours the synthesis of the more neutron rich, i.e.: \isotope[94]{Mo}, \isotope[98]{Ru}, \isotope[108]{Cd}, \isotope[114]{Sn} and partially \isotope[115]{Sn}, \isotope[126]{Xe}, \isotope[132]{Ba}, \isotope[138]{Ce}, partially \isotope[158]{Dy}, but not \isotope[164]{Er}. The latter, in fact, has a dominant contribution from the \s--process (Sec. \ref{sub:10}). These properties indicate that the \g--process in C--O shell mergers operates at  lower temperatures (those typical of O burning at the bottom of the convective shell), as compared to the explosion. 

    \begin{table}
      \caption{Definition of average overproduction factors.}           
      \label{tab:Fval}   
      \centering                         
      \begin{tabular}{lll}        
      \hline\hline                
      F & excluded contribution & excluded isotopes \\ 
      \hline                        
      \fz              & --                                 & --                                                                              \\
                       &                                    &                                                                                 \\
      $\rm F_0^I$      & \s--, \rr--process,                & \isotope[74]{Se}, \isotope[78]{Kr}, \isotope[84]{Sr}, \isotope[92,94]{Mo},      \\
                       & $\nu$--capture, $\nu p$--process,  & \isotope[96,98]{Ru}, \isotope[108]{Cd}, \isotope[113]{In}, \isotope[115]{Sn},   \\
                       & \A--rich freeze-out                &  \isotope[138]{La}, \isotope[152]{Gd}, \isotope[164]{Er}, \isotope[180]{Ta}     \\
                       &                                    &                                                                                 \\
      $\rm F_0^{II}$   & $\nu p$--process,                  & \isotope[74]{Se}, \isotope[78]{Kr}, \isotope[84]{Sr},                           \\
                       & \A--rich freeze-out                & \isotope[92,94]{Mo}, \isotope[96,98]{Ru}                                        \\
                       &                                    &                                                                                 \\
      $\rm F_0^{III}$  & \s--, \rr--process,                & \isotope[108]{Cd}, \isotope[113]{In}, \isotope[115]{Sn}, \isotope[138]{La},     \\
                       & $\nu$--capture                     & \isotope[152]{Gd}, \isotope[164]{Er}, \isotope[180]{Ta}                         \\
                       &                                    &                                                                                 \\
      $\rm F_{\gamma}$ & \s--, \rr--process,                &                                                                                 \\
                       &  $\nu$--capture, $\nu p$--process, & all but the 3 most                                                              \\
                       & \A--rich freeze-out,               & produced \g--only nuclei                                                        \\
                       & partly the \g--process                &                                                                                 \\       
      \hline                                  
      \end{tabular}
   \end{table}

   As not all the \p--nuclei have a \g--only origin, \fz\ might not correctly reflect production from the \g--process. For example, in case of a failed supernova (such as the LAW20 shown in the bottom left panel of Figure~\ref{fig:f0}, and RIT25), \fz\ could be dominated by isotopes with additional nucleosynthetic contributions besides the \g--process, such as \isotope[180]{Ta} in LAW20. Neutrino processes producing \isotope[138]{La} and \isotope[180]{Ta} contributed the most to the \fz\ value in SIE25 (lower right panel of Figure~\ref{fig:f0}). Therefore, to better constrain the \g--process itself, we decided to define the more averages, by excluding different isotope groups, as listed and described in  \tablename~\ref{tab:Fval}. Among the various excluded isotopes, \isotope[108]{Cd} has been traditionally considered a \g--only isotope, however, in the models considered here we found that \isotope[108]{Cd} can also be produced by neutron captures starting on \isotope[107]{Ag}. This production may be not negligible since the solar abundance of \isotope[107]{Ag}, used in the models as initial, is more than 15 times larger than the solar abundance of \isotope[108]{Cd}. For this reason, we included this isotope among the nuclei with an additional contribution. Our new definitions help us to identify why some models which models do not produce \p--nuclei, and have nevertheless relatively high \fz\ values. 
   \tablename~\ref{tab:f0} lists all the different F--values in our sets of models.
   
   When comparing all these different values to \fx[O16], we found no significant improvement using $\rm F_0^I$, $\rm F_0^{II}$, or $\rm F_0^{III}$ instead of \fz. Using $\rm F_{\gamma}$, instead, we find six models with values larger than \fx[O16], of which four models (RAU20, PGN20, RIT15 and RIT20) have $\rm F_{\gamma}$/\fx[O16] $>$ 2. \isotope[196]{Hg} contributes to the $\rm F_{\gamma}$ of three out of these four models, and \isotope[130]{Ba} and \isotope[168]{Yb} to two. Overall, we conclude that F--values should be used carefully as by including many OP factors in the average can hide information. In fact, different CCSN models may efficiently produce different, single or groups of, \p--nuclei, and still result in a low value of \fz. Therefore, the use of the F--values only is not enough to investigate the properties of the \g--process nucleosynthesis and an accurate analysis of the production of the single isotopes is required instead.
   
   \begin{table*}
      \caption{The average overproduction factors in the models.}             
      \label{tab:f0}      
      \centering                          
      \begin{tabular}{lccccccc}        
      \hline\hline                 
      Model & \fz & \fx[O16] & $\rm F_0^I$ & $\rm F_0^{II}$ & $\rm F_0^{III}$ & $\rm F_{\gamma}$ & $\rm F_{\gamma}$ isotopes \\  
      \hline                        
      % inserting body of the table
      RAU15     &  4.32E+00 &  1.24E+01 &  5.53E+00 &  4.78E+00 &  4.76E+00  &  \textbf{1.61E+01}  & \isotope[196]{Hg}, \isotope[180]{W}, \isotope[168]{Yb}  \\
      RAU20$^a$ &  1.79E+01 &  2.68E+01 &  2.54E+01 &  2.06E+01 &  2.08E+01  &  \textbf{8.91E+01}  & \isotope[168]{Yb}, \isotope[196]{Hg}, \isotope[162]{Er} \\
      RAU25     &  1.56E+01 &  4.80E+01 &  1.37E+01 &  1.31E+01 &  1.68E+01  &  3.59E+01  & \isotope[196]{Hg}, \isotope[180]{W}, \isotope[130]{Ba}  \\
      PGN15     &  \textbf{5.03E+00} &  4.72E+00 &  2.93E+00 &  2.95E+00 &  \textbf{5.54E+00} &  \textbf{5.64E+00}  & \isotope[130]{Ba}, \isotope[132]{Ba}, \isotope[168]{Yb} \\
      PGN20     &  \textbf{1.86E+01} &  1.85E+01 &  \textbf{2.15E+01} &  1.79E+01 &  \textbf{2.14E+01} &  \textbf{4.93E+01}  & \isotope[196]{Hg}, \isotope[130]{Ba}, \isotope[180]{W}  \\
      RIT15$^a$ &  1.01E+01 &  1.39E+01 &  \textbf{1.43E+01} &  1.17E+01 &  1.17E+01 &  \textbf{3.63E+01}  & \isotope[132]{Ba}, \isotope[162]{Er}, \isotope[168]{Yb} \\
      RIT20     &  1.66E+01 &  1.94E+01 &  1.40E+01 &  1.34E+01 &  1.78E+01 &  \textbf{4.34E+01}  & \isotope[196]{Hg}, \isotope[130]{Ba}, \isotope[136]{Ce} \\
      (RIT25)   &  2.12E+00 &  1.94E+01 &  6.81E-01 &  2.47E+00 &  6.95E-01 &  8.70E-01  & \isotope[114]{Sn}, \isotope[158]{Dy}, \isotope[168]{Yb} \\
      SIE15     &  2.78E+00 &  6.09E+00 &  3.05E+00 &  3.02E+00 &  2.75E+00 &  5.89E+00  & \isotope[196]{Hg}, \isotope[180]{W}, \isotope[130]{Ba}  \\
      SIE20     &  3.57E+00 &  1.24E+01 &  2.53E+00 &  3.05E+00 &  3.31E+00 &  6.01E+00  & \isotope[196]{Hg}, \isotope[180]{W}, \isotope[168]{Yb}  \\
      SIE25     &  4.83E+00 &  1.67E+01 &  2.09E+00 &  2.99E+00 &  4.62E+00 &  4.51E+00  & \isotope[184]{Os}, \isotope[168]{Yb}, \isotope[174]{Hf} \\
      LAW15     &  4.13E+00 &  1.26E+01 &  4.87E+00 &  4.72E+00 &  4.10E+00 &  9.56E+00  & \isotope[190]{Pt}, \isotope[196]{Hg}, \isotope[158]{Dy} \\
      (LAW20)   &  2.80E+00 &  2.41E+01 &  8.88E-01 &  3.27E+00 &  8.98E-01 &  1.40E+00  & \isotope[102]{Pd}, \isotope[158]{Dy}, \isotope[180]{W}  \\
      LAW25     &  1.49E+01 &  5.18E+01 &  1.50E+01 &  1.62E+01 &  1.36E+01 &  3.10E+01  & \isotope[130]{Ba}, \isotope[184]{Os}, \isotope[158]{Dy} \\
      % ending body of the table
      \hline                                   
      \multicolumn{8}{l}{$^a$These two models experience the C--O shell merger}\\
      \end{tabular}\\
      \tablefoot{Values of the quantities defined in \tablename~\ref{tab:Fval} and the reference $^{16}$O overproduction factor \fx[O16] for all the models. The last column indicates the 3 most produced \g--only nuclei on which the calculation of the reported F$_{\gamma}$ is based. Values highlighted in bold are larger than \fx[O16], for each models. The models in parenthesis are not included in our analysis of the isotopic ratios presented in Sec. \ref{sec:isoratio}.}\\
   \end{table*}

% ==================== Ratios ==================== 
\section{Isotopic ratios} \label{sec:isoratio}

   \begin{table*}
      \caption{The 23 isotopic ratios.}             
      \label{tab:isoratio}      
      \centering                          
      \begin{tabular}{llccccccccc}      
      \hline\hline                 
      Isotopic ratio & Subsection & Sub-solar & Solar & Super-solar & Merger & Radiogenic & Additional components \\  
      \hline                        
      % inserting body of the table    
      \isotope[74]{Se}/\isotope[78]{Kr}   & \ref{sub:01} &  0 &  7 &  5 &     &     & \A       \\
      \isotope[84]{Sr}/\isotope[78]{Kr}   & \ref{sub:01} &  0 & 11 &  1 &     &     & \A       \\
      \isotope[92]{Mo}/\isotope[94]{Mo}   & \ref{sub:02} &  0 & 12 &  0 &     &     & \A       \\
      \isotope[96]{Ru}/\isotope[98]{Ru}   & \ref{sub:02} &  0 & 12 &  0 &     &     & \A       \\
      \isotope[102]{Pd}/\isotope[108]{Cd} & \ref{sub:03} &  0 & 12 &  0 &     &     &          \\
      \isotope[106]{Cd}/\isotope[108]{Cd} & \ref{sub:03} &  0 & 12 &  0 &     &     &          \\
      \isotope[112]{Sn}/\isotope[114]{Sn} & \ref{sub:04} &  1 & 11 &  0 &  v  &     & \s,\rr   \\
      \isotope[113]{In}/\isotope[114]{Sn} & \ref{sub:04} &  4 &  8 &  0 &     &  v  & \s,\rr   \\
      \isotope[115]{Sn}/\isotope[114]{Sn} & \ref{sub:04} &  1 & 11 &  0 &  v  &     & \s,\rr   \\
      \isotope[120]{Te}/\isotope[126]{Xe} & \ref{sub:06} &  2 & 10 &  0 &  v  &  v  &          \\
      \isotope[124]{Xe}/\isotope[126]{Xe} & \ref{sub:06} &  2 & 10 &  0 &  v  &     &          \\
      \isotope[130]{Ba}/\isotope[132]{Ba} & \ref{sub:07} &  1 &  9 &  1 &  v  &  v  &          \\
      \isotope[136]{Ce}/\isotope[138]{Ce} & \ref{sub:07} &  1 &  9 &  1 &  v  &  v  &          \\
      \isotope[138]{La}/\isotope[132]{Ba} & \ref{sub:08} &  8 &  3 &  1 &     &  v  & $\nu$    \\
      \isotope[144]{Sm}/\isotope[132]{Ba} & \ref{sub:08} &  0 & 11 &  1 &     &  v  &          \\
      \isotope[144]{Sm}/\isotope[152]{Gd} & \ref{sub:09} &  4 &  8 &  0 &     &     & \s       \\
      \isotope[156]{Dy}/\isotope[152]{Gd} & \ref{sub:09} &  6 &  6 &  0 &     &  v  & \s       \\
      \isotope[156]{Dy}/\isotope[158]{Dy} & \ref{sub:10} &  2 & 10 &  0 &     &  v  &          \\
      \isotope[162]{Er}/\isotope[164]{Er} & \ref{sub:10} &  7 &  5 &  0 &  v  &  v  & \s       \\
      \isotope[168]{Yb}/\isotope[180]{Ta} & \ref{sub:11} & 10 &  2 &  0 &  v  &  v  & \s,$\nu$ \\
      \isotope[174]{Hf}/\isotope[180]{W}  & \ref{sub:11} &  0 & 12 &  0 &  v  &  v  &          \\
      \isotope[184]{Os}/\isotope[196]{Hg} & \ref{sub:12} &  4 &  8 &  0 &     &  v  &          \\
      \isotope[190]{Pt}/\isotope[196]{Hg} & \ref{sub:12} &  9 &  3 &  0 &     &  v  &          \\
      % ending body of the table
      \hline                                   
      \end{tabular}\\
      \tablefoot{The 23 isotopic ratios are plotted in Figs~\ref{fig:00} to \ref{fig:11}. The Columns "Sub-solar", "Solar" and "Super-solar" identify the number of models that fall below, within, and above a factor of 3 from the solar \g--process component, respectively. The Column "Merger" reports "v" if at least one of the two yields of the isotopic ratio is dominated by the merger (corresponding to the bold highlighted components in \tablename~\ref{tab:merper}). The Column "Radiogenic" reports "v" if at least one of the two isotopes of the ratio has a significant radiogenic contribution. The Column "Additional components" lists the possible additional nucleosynthetic processes that may contribute to at least one of the two isotopes of the ratio; possible additional contributions are: \A\ (\A-rich freeze-out or different explosive component), \s\ (\s--process), \rr\ (\rr--process), $\nu$ (neutrino capture).}\\
   \end{table*}
   
    We analysed the correlations between 12 different couples of OP ratios (for a total of 23 ratios given that one ratio is used twice). We chose ratios of isotopes close in mass to study the local properties of the \g--process production which are also affected by the impact of the different sets of reaction rates used in the different sets of models. We avoided including in each couple of ratios more than one isotope with additional contribution(s) beyond the \g--process. 
    While the OP factor quantifies the $absolute$ production of a given isotope relative to solar and therefore indicates if this production is significant, the ratio of two OP factors quantifies the overproduction of the two isotopes $relative$ to each other. Therefore, if an OP ratio agrees with solar, then both isotopes have similar OP factors; if it does not, then the two isotopes have different OP factors. 

    Usually, comparisons with solar abundances are made using yields averaged over the initial mass function (IMF), since the solar composition is the result of galactic chemical evolution with contributions from many stars. We do not proceed in this way here because, first, a comparison among different single star yields allows us to study the \g--process in different stellar conditions in detail and to check if there exist best candidates that can reproduce the solar isotopic ratios. \cite{travaglio:18} already showed that, with the current generation of massive star yields, it is not possible to explain the \g--process solar distribution using GCE. New GCE calculations with the yields computed using the updated reaction rates for the \g--process is part of our future work. Second, our analysis is also focused on the C--O shell mergers and at present, the frequency and the relation with initial mass of such events is unknown. Third, most of the sets of models available in literature do not include a large enough number of initial masses and/or metallicities or, if so, the size of the nuclear network is too limited to produce reliable yields for the \g--process nucleosynthesis. This makes the use of IMF averaged yields not particularly meaningful in the comparison with the solar composition. 
    
    The differences in stellar modelling (e.g., different choices of convection, overshooting, etc.) and in nuclear physics (e.g., number of explicitly included nuclear species, reaction rate database, etc.), introduce a significant source of uncertainty and make it difficult to define to which level a result is in agreement with the solar distribution. To account for this, we consider models to be in relatively good agreement with solar when both following conditions are satisfied: (i) they fall within a factor of 3 from the solar ratio and (ii) they have a overproduction at least twice as solar (OP $>$ 2).

    In the following we summarise the main results of our analysis. An extended discussion for each of the 12 couples of isotopic ratios is presented in Appendix \ref{app:isoratio}. As already mentioned in Sec \ref{sec:datasets}, we did not include the two faint supernova models (LAW20 and the RIT25), because the \g--process yields remain locked in the compact remnant and only the envelope of the star is ejected in the interstellar medium. We also excluded the PGN25 model due to a lack of resolution in the \g--process abundance peaks. \tablename~\ref{tab:isoratio} collects the main results for each isotopic ratio. We note that the analysis reported below will need to be complemented by future studies where each of the isotopic ratios discussed here are carefully investigated in terms of both the nuclear physics input (see also Sec. \ref{sec:nuclearphysics}) and the different modelling approaches.

    There are 10 ratios that satisfy both of our conditions for a good agreement with the solar ratio for most of the models (\isotope[74]{Se}/\isotope[78]{Kr}, \isotope[84]{Sr}/\isotope[78]{Kr}, \isotope[102]{Pd}/\isotope[108]{Cd}, \isotope[106]{Cd}/\isotope[108]{Cd}, \isotope[120]{Te}/\isotope[126]{Xe}, \isotope[124]{Xe}/\isotope[126]{Xe}, \isotope[130]{Ba}/\isotope[132]{Ba}, \isotope[136]{Ce}/\isotope[138]{Ce}, \isotope[144]{Sm}/\isotope[132]{Ba}, and \isotope[174]{Hf}/\isotope[180]{W}). The other 13 ratios either have low OP factors or be further than a factor of 3 from solar.
    
    We find a general agreement with the solar ratios involving the lightest \p--nuclei, except in the case of Mo and Ru isotopes. No model, in fact, satisfies the condition OP $>$ 2 for \isotope[92]{Mo}, \isotope[94]{Mo}, \isotope[96]{Ru} and \isotope[98]{Ru}. The production of the isotopes between \isotope[74]{Se} and \isotope[108]{Cd} is mostly dominated by the \g--process happening during the explosion, even in the models with C--O shell mergers.

    The majority of the ratios involving heavier isotopes present a significant contribution from radiogenic species. In some cases, the radiogenic contribution can be either determinant to reach a good agreement with  the solar ratio (e.g., \isotope[120]{Te}/\isotope[126]{Xe}, \isotope[124]{Xe}/\isotope[126]{Xe}) or conversely to move away from solar (e.g., \isotope[156]{Dy}/\isotope[158]{Dy}, \isotope[184]{Os}/\isotope[196]{Hg}, \isotope[190]{Pt}/\isotope[196]{Hg}).

    As mentioned in Sec. \ref{sec:f0}, the overproduction of \p--nuclei with $\rm A\gtrsim110$ largely increases in models with C--O shell mergers (RIT15 and RAU20). Moreover, these models have generally larger overproduction of neutron richer \p--nuclei. There are 9 isotopic ratios in which the C--O shell mergers stand out from the other models. In 4 cases (\isotope[144]{Sm}/\isotope[152]{Gd}, \isotope[156]{Dy}/\isotope[152]{Gd}, \isotope[162]{Er}/\isotope[164]{Er}, and \isotope[168]{Yb}/\isotope[180]{Ta}) the isotopic ratio from the merger is closer to solar than in the standard models. The opposite occurs in the other 4 cases (\isotope[112]{Sn}/\isotope[114]{Sn}, \isotope[120]{Te}/\isotope[126]{Xe}, \isotope[124]{Xe}/\isotope[126]{Xe}, \isotope[130]{Ba}/\isotope[132]{Ba}, and \isotope[136]{Ce}/\isotope[138]{Ce}).
    
    Out of the 23 ratios, 8 of them include one isotope with an additional nucleosynthetic contribution besides the \g--process (see Sec. \ref{sec:intro} and \ref{sec:f0} and \tablename~\ref{tab:isoratio}). We used the results from the galactic chemical evolution (GCE) computation of \cite{bisterzo:14} and from \cite{dillmann:08}, \cite{nemeth:94}, and \cite{theis:98} to estimate the neutron-capture contribution to the \p--nuclei in the solar composition and derive the residual component, possibly dominated by the \g--process. There are instead no quantitative predictions for the other explosive components (e.g., \A-rich freeze-out, $\nu$\p--process). The models show a good agreement with the residual solar ratios only in 4 cases: \isotope[112]{Sn}/\isotope[114]{Sn}, \isotope[113]{In}/\isotope[114]{Sn}, \isotope[115]{Sn}/\isotope[114]{Sn}, and \isotope[144]{Sm}/\isotope[152]{Gd}. In the other 4 cases (\isotope[138]{La}/\isotope[132]{Ba}, \isotope[156]{Dy}/\isotope[152]{Gd}, \isotope[162]{Er}/\isotope[164]{Er} and \isotope[168]{Yb}/\isotope[180]{Ta}) the (residual) solar ratio is always underestimated.

% ================= Nuclear Physics ===================
\section{Nuclear Physics Considerations} \label{sec:nuclearphysics}

    The models included in our analysis (except for RAU) were computed in the last decade and the nuclear physics inputs included in these calculations cover a time span of about 25 years. The RAU models used mostly an early version of the Basel REACLIB \citep{rauscher:00} compilation and the neutron-capture rates from \cite{bao:00}. In the case of missing experimental information, they included the Hauser-Feshbach rates obtained with the NON-SMOKER code \citep{rauscher:97}. The PGN models adopted the KADoNIS compilation \citep{dillmann:06} for experimental neutron-capture rates and the Basel REACLIB (revision 20090121) compilation in all the other cases. The RIT and LAW models share the same nuclear physics for the \g--process, i.e., the KADoNIS compilation for experimental neutron-captures and the JINA REACLIB, 20120510 V1.1 \citep{cyburt:11} compilation for theoretical neutron-capture rates and charged-particle reaction rates. Instead, SIE models use a more recent version of JINA REACLIB, i.e., V2.2, but only for the explosive nucleosynthesis. The \g--process nucleosynthesis yields presented in this work were therefore computed with outdated reaction rates and new calculations are required to test the impact of the most recent recommendations discussed below.

    %\subsection{General remarks on progress on the reaction rates for the \g--process}

    Experimental data at energies relevant for \g--process nucleosynthesis are rather scarce, and for this reason there is a strong interest in the low-energy nuclear physics community to measure reaction cross sections that are related to the synthesis of \p--nuclei. Typically \A-induced reaction cross sections are smaller than their \p-induced counterparts, due to the higher Coulomb barrier, and consequently the measurements are more challenging. Since the latest \g--process nucleosynthesis review by~\cite{pignatari:16}, many new measurements on \rpg, \rag\ and \ran\ reactions that produce or destroy \p-nuclei have been performed. 

    In addition to widely used techniques, such as in-beam \g--spectroscopy~\citep[see][for some examples]{khaliel:2017, foteinou:2018, psaltis:2019, wu:2020, heim:2020}, the activation method~\citep[see][and references therein]{korkulu:2018, Gyukry:2019, scholz:2020, cheng:2021}, and the total absorption spectroscopy (TAS)/ 4$\pi$ summing method~\citep{kelmar:2020, harissopulos:2021, palmisano:2022}, other few novel methods have been successfully used recently. \cite{glorius:2019} used the ESR storage ring and silicon detectors to measure the \isotope[124]{Xe}(p,\g) reaction. \cite{fallis:2020} used the DRAGON recoil separator to measure the \isotope[76]{Se}($\alpha$,\g) reaction in inverse kinematics. Finally, \cite{lotay:2021} and \cite{williams:23} used for the first time a radioactive beam (\isotope[83]{Rb}) and the EMMA recoil spectrometer to study the \isotope[83]{Rb}(p,\g)\isotope[84]{Sr} reaction in inverse kinematics. Efforts for $\alpha$-scattering on exotic isotopes relevant for the \g-process are also underway~\citep{galaviz:21}. These advancements, along with the already established methods, will enable the low-energy nuclear physics community to measure key reaction cross sections in the near future.

    Despite this experimental progress, it remains impossible for the foreseeable future to experimentally access all the reaction rates needed in typical \g--process networks, which include thousands of nuclei and several thousands of nuclear reaction rates \citep{arnould:03}. Consequently, reaction rates derived from nuclear theory remain crucial. Furthermore, \g-induced reactions play a key role in the \g--process network, but a measurement of \rgX\ reaction cross sections in the laboratory cannot provide the \rgX\ reaction rate under stellar conditions because thermally excited states in the target nucleus play an important role under stellar conditions, but do not contribute in the laboratory experiments \citep[for a more detailed discussion see][]{rauscher:13}. Instead, the \rgX\ reaction rates are calculated from the \rXg\ capture rates using the detailed balance relation between forward and backward reaction rates \citep[see e.g.,][]{rauscher:00}. For heavy nuclei, the statistical model is appropriate and widely used to derive reaction rates. In the following paragraphs we discuss very briefly this approach, the relevant ingredients, and resulting uncertainties for the reaction rates. Note that, instead, the determination of the astrophysical reaction rates involving low Z \p--nuclei may be influenced by effects that can not be taken into account by the statistical model \citep{gyurky:14}. Therefore, in these cases, the experimental determination of the cross section is required. 

    In general, the cross section $\sigma$ of a \rXg\ capture reaction (with $\rm X = \rm n, p, \gamma$) in the statistical model is defined by the production cross section of the compound nucleus in the entrance channel and the branching ratio in the exit channel. The production cross section scales with the transmission $T_{\rm X}$ in the entrance channel. The branching $b_\gamma$ towards the \g-channel is given by $T_\gamma/\sum_{\rm i} T_{\rm i}$ where the transmissions $T_{\rm i}$ in the nominator have to take into account all open channels. This leads to the simple proportionality:
    \begin{equation}
    \centering
        \sigma(\rm X,\gamma) \sim T_{\rm X} \times \frac{T_\gamma}{\sum_i T_{\rm i}} = \frac{T_{\rm X} T_\gamma}{\sum_{\rm i} T_{\rm i}}.
        \label{eq:StM}
    \end{equation}
    This equation allows us to understand the relevance of the essential ingredients of the statistical model, which are the \g-strength function (\g SF), the level density (LD), the nucleon optical model potential (NOMP), and the \A-nucleus optical model potential ($\alpha$OMP). Note that the $T_{\rm i}$ are calculated from global NOMPs and $\alpha$OMPs for the particle channels and from the \g SF for the \g -channel. Furthermore, the $T_{\rm i}$ are composed of the sum over all bound states in the respective residual nuclei, which is approximated by the LD above a certain excitation energy; thus, all $T_{\rm i}$ depend implicitly on the LD.

    For \rng\ neutron capture reactions, typically $T_{\rm n}$ exceeds by far all $T_{\rm i}$ for the other channels. Consequently, $\sum_{\rm i} T_{\rm i} \approx T_{\rm n}$, and \rng\ from Eq.~(\ref{eq:StM}) scales mainly with the transmission $T_\gamma$ which depends on the \g SF and the LD, but is independent of the NOMP and $\alpha$OMP. Typically, an uncertainty of a factor of two is estimated for the \rng\ rates, and this uncertainty was used in recent \g--process studies, e.g.,\ \cite{rapp:06,rauscher:16}.

    For \rpg\ proton capture reactions on the neutron-deficient nuclei in the \g--process, Eq.~(\ref{eq:StM}) leads to the following. Because of the Coulomb barrier, the transmission $T_{\rm p}$ depends sensitively on the energy. At low energies, corresponding to the Gamow window for the reaction rates, $T_{\rm p}$ is smaller than $T_\gamma$, and the neutron channel is closed ($T_{\rm n} = 0$). Thus, $\sum_{\rm i} T_{\rm i} \approx T_\gamma$, and the \rpg\ cross section scales with $T_{\rm p}$, which, in turn, depends mainly on the NOMP and on the LD, but not on the \g SF. As the NOMP is well-studied, for \rpg\ rates a similar uncertainty of a factor of two as in the \rng\ case is often used, e.g.,\ \cite{rapp:06,rauscher:16}, although the main origin of the uncertainty is the NOMP for \rpg\ rates and the \g SF and LD for \rng\ rates.

    The situation for \rag\ reactions is similar to the \rpg\ reactions. Again, because of the Coulomb barrier, $T_\alpha$ is much smaller than $T_\gamma$ at astrophysically relevant energies, and thus the \rag\ cross section scales with $T_\alpha$ and becomes mainly sensitive to the $\alpha$OMP, but is practically independent of the \g SF and NOMP. Because of the higher Coulomb barrier, the sensitivity on the $\alpha$OMP is by far more pronounced than the sensitivity on the NOMP in the \rpg\ case. Much larger uncertainties of a factor of ten or even more were found from the comparison between experimental and calculated \rag\ cross sections; e.g.,\ \cite{somorjai:98}. An uncertainty of a factor of ten was used for \rag\ rates in the above mentioned \g--process studies \citep{rapp:06,rauscher:16}. It is interesting to note that also \ran\ reaction rates depend essentially only on the $\alpha$OMP, and a recent study of the weak \rr--process \citep{bliss:20} has also used a factor of ten uncertainty for the \ran\ reaction rates. Recently, the origin of this strong sensitivity was better understood \citep{mohr:2020}, and updated reaction rates with an estimated uncertainty of a factor of two using the ATOMKI-V2 $\alpha$OMP were provided in \cite{mohr:2021}. This factor uncertainty was chosen due to the success of ATOMKI-V2 $\alpha$OMP in reproducing experimentally determined cross sections of $\alpha$-induced reactions~\citep[for example][]{szegedi:21}. Using these updated \ran\ rates, significant progress in the modelling of the weak \rr--process was achieved \citep{psaltis:22}.

    Summarising the status of theoretical reaction rates from the statistical model, nowadays uncertainties of a factor of two are considered for the rates of the \rng , \rpg , and \rag\ capture reactions. Because of the relation between forward and reverse reaction rates, this uncertainty of a factor of two also holds for the \rgn , \rgp , and \rga\ photo-disintegration rates. An experimental confirmation of these claimed uncertainties should become possible at least in some well selected cases with the increasing availability of radioactive ion beams and up-to-date detection techniques. Overall, the new \rag\ and \rga\ reaction rates in \cite{mohr:2021} are significantly lower than the widely used rates by \cite{rauscher:00}, in particular for relatively heavier nuclei and lower temperatures. This clearly calls for the further investigations that we will present in our future work.

% ==================== Conclusions ====================
\section{Discussion and conclusions} \label{sec:conclusions}

    We studied the production of \p--nuclei in 5 different sets of CCSNe. Each set includes the yields from three massive star progenitors with initial masses of 15, 20, and 25 \msun. We analysed both the overproduction factor of each isotope (relative to their averages, as defined in several different ways) and the ratios of \p--nuclei close to each other in mass, relative to their solar ratio. This strategy allowed us to investigate the differences both in the production site and in the nuclear physics among the different sets of models. 

    The different sets of CCSN models present several discrepancies both in the overproduction factors and their ratios. These depend both on the assumptions on stellar physics implemented in each set, i.e., the adoption of different criteria for convection, mass-loss prescriptions, the use of overshooting, the explosion mechanism, etc., and on the different nuclear physics inputs. The different assumptions in stellar modeling mostly influence the final pre-supernova structure of the star, as well as the possibility of having merger events in the advanced phases of the pre-supernova evolution. The variations in the stellar structure also lead to different modalities of the propagation of the shock wave during the supernova explosion and hence to different explosive nucleosynthesis. The use of different nuclear physics inputs leads to local variations in the proportions of the isotopes of nearby mass, mostly influencing the isotopic ratios.

    In Sec. \ref{sec:f0}, we discussed that the \fz\ is not appropriate as a general tool to investigate the properties of the \g--process nucleosynthesis. 
    Alternative definitions of this parameter can help us to distinguish the production of \p--nuclei from different nucleosynthetic processes, but these are still not enough to study the details of the \g--process in the models. For this reason, we analysed the correlations between 12 different couples of OP ratios of isotopes close in mass.

    Ten of the considered isotopic ratios show a good agreement with the solar ratio for most of the models. These are \isotope[74]{Se}/\isotope[78]{Kr} and \isotope[84]{Sr}/\isotope[78]{Kr} (Sec \ref{sub:01}), \isotope[102]{Pd}/\isotope[108]{Cd} and \isotope[106]{Cd}/\isotope[108]{Cd} (Sec \ref{sub:03}), \isotope[120]{Te}/\isotope[126]{Xe} and \isotope[124]{Xe}/\isotope[126]{Xe} (Sec \ref{sub:06}), \isotope[130]{Ba}/\isotope[132]{Ba} and \isotope[136]{Ce}/\isotope[138]{Ce} (Sec \ref{sub:07}), \isotope[144]{Sm}/\isotope[132]{Ba} (Sec \ref{sub:08}), and \isotope[174]{Hf}/\isotope[180]{W} (Sec. \ref{sub:11}). The other 13 ratios show no agreement with solar and/or one or both isotopes have a too low overproduction factor for their origin to be attributed to the \g--process in CCSNe. In particular, the \p--only Mo and Ru isotopes are not produced in any of the considered models and their origin remain a mystery. Nuclei such as \isotope[113]{In}, \isotope[115]{Sn}, \isotope[138]{La}, \isotope[152]{Gd} and \isotope[164]{Er} are not significantly produced by the \g--process, which confirm that the bulk of their abundances in the Solar System is made by other processes. We also notice that \isotope[74]{Se}, \isotope[78]{Kr} and \isotope[84]{Sr} can be explained by the \g--process only, without requiring an \A--rich freeze-out contribution.

    The effect of the C--O shell mergers is mostly to increase the abundances of the \p-nuclides heavier than Pd.  In the case of the \p--nuclei lighter than Pd, i.e., belonging to Se, Kr, Sr, Mo, and Ru, the yield is completely dominated by the explosion and models with C--O mergers do not significantly differ from the standard models. Moreover, these events typically favour the production of the most neutron-rich isotope within \p-only pairs belonging to the same element. This is a signature of a \g--process that occurs at lower temperature, as compared to the explosive conditions.  

    In conclusion, we have shown that a large scatter among the existing CCSN model production exists and no set of models is fully able to reproduce the distribution of the \p--nuclei measured in the Solar System. Since most of the models adopts outdated nuclear reaction rates, our results point toward the necessity of an upgrade of the \g--process nucleosynthesis nuclear networks with the latest results in nuclear physics, as discussed in Sec \ref{sec:nuclearphysics}. Furthermore, it is generally assumed that \g--process yields from CCSNe depend only weakly on the energy of the explosion, however, there are no detailed studies that confirm this. Therefore, we will proceed to improve and update our \g--process nuclear network to compute new massive stars and CCSN models for a wide grid of explosion energies, to further investigate the \g--process nucleosynthesis in massive stars.
    
% ==================== Acknowledgements ====================
\begin{acknowledgements}
    We thank the support from the NKFI via K-project 138031 and the ERC Consolidator Grant (Hungary) programme (RADIOSTAR, G.A. n. 724560). LR and MP acknowledge the support to NuGrid from JINA-CEE (NSF Grant PHY-1430152) and STFC (through the University of Hull’s Consolidated Grant ST/R000840/1), and ongoing access to {\tt viper}, the University of Hull High Performance Computing Facility. LR acknowledges the support from the ChETEC-INFRA -- Transnational Access Project 22102724-ST. MP and ML acknowledges the support from the "Lend{\"u}let-2014" Programme of the Hungarian Academy of Sciences (Hungary). AP acknowledges support from U.S. Department of Energy, Office of Science, Office of Nuclear Physics, under Award Number DE-SC0017799 and Contract Nos. DE-    FG02-97ER41033 and DE-FG02-97ER41042. AS acknowledges funding by the European Union’s Framework Programme for Research and Innovation Horizon Europe under Marie Sklodowska-Curie grant agreement No. 101065891. PM, Gy. Gy., and Zs. F. acknowledges support from National Research Development and Innovation Office (NKFIH), Budapest, Hungary (K134197). This work was supported by the European Union’s Horizon 2020 research and innovation programme (ChETEC-INFRA -- Project no. 101008324), and the IReNA network supported by US NSF AccelNet (Grant No. OISE-1927130).
\end{acknowledgements}

% ==================== Bibliography ====================
%\bibliographystyle{aa}
%\bibliography{astro_tot}{}

% ==================== Appendix ====================
\begin{appendix}

    \section{OP factors}\label{app:OP} %First appendix

\begin{figure*}
            \centering
            \includegraphics[scale=0.56]{f0/Rau_F0_15.pdf}
            \includegraphics[scale=0.56]{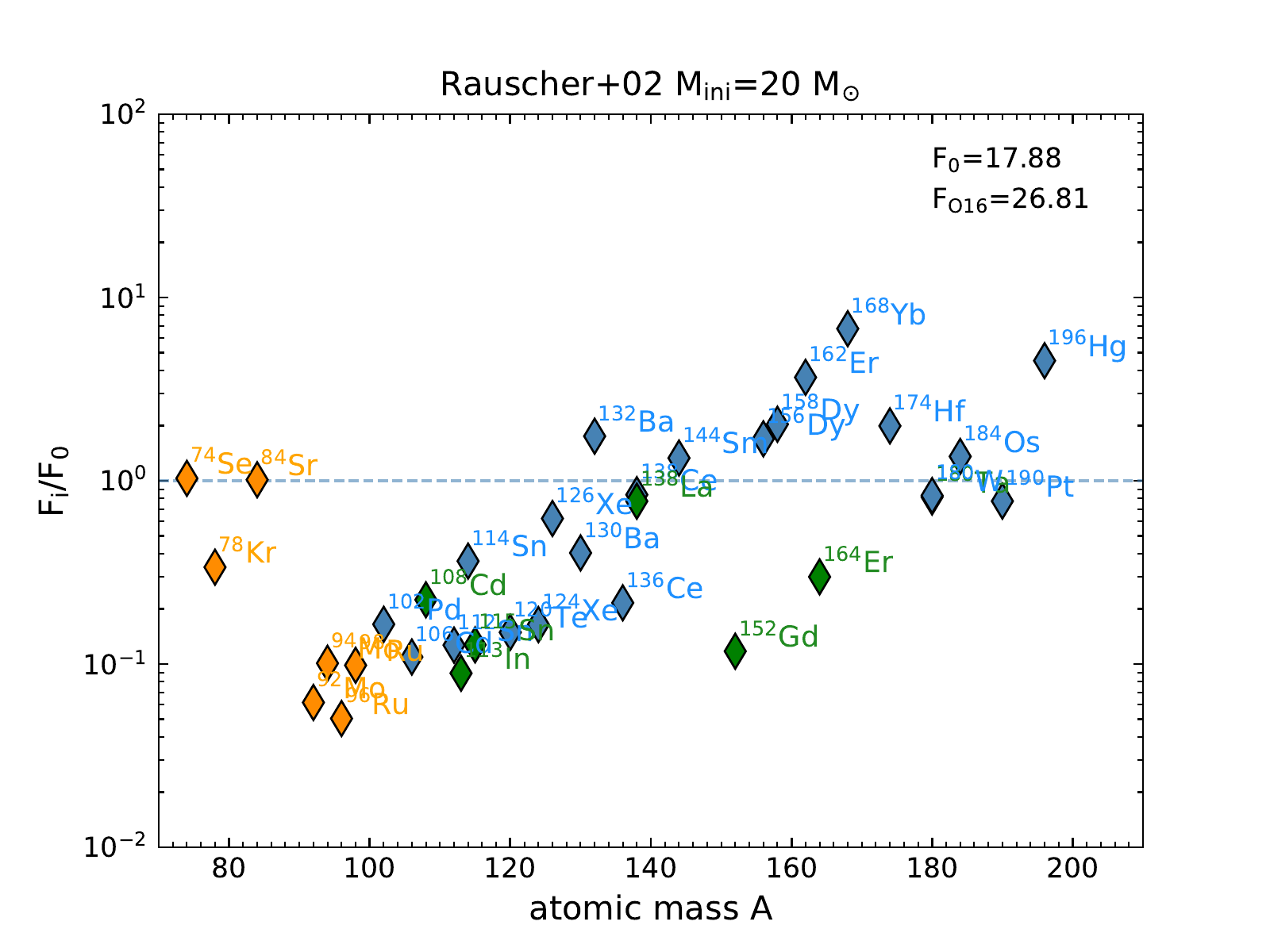}
            \includegraphics[scale=0.56]{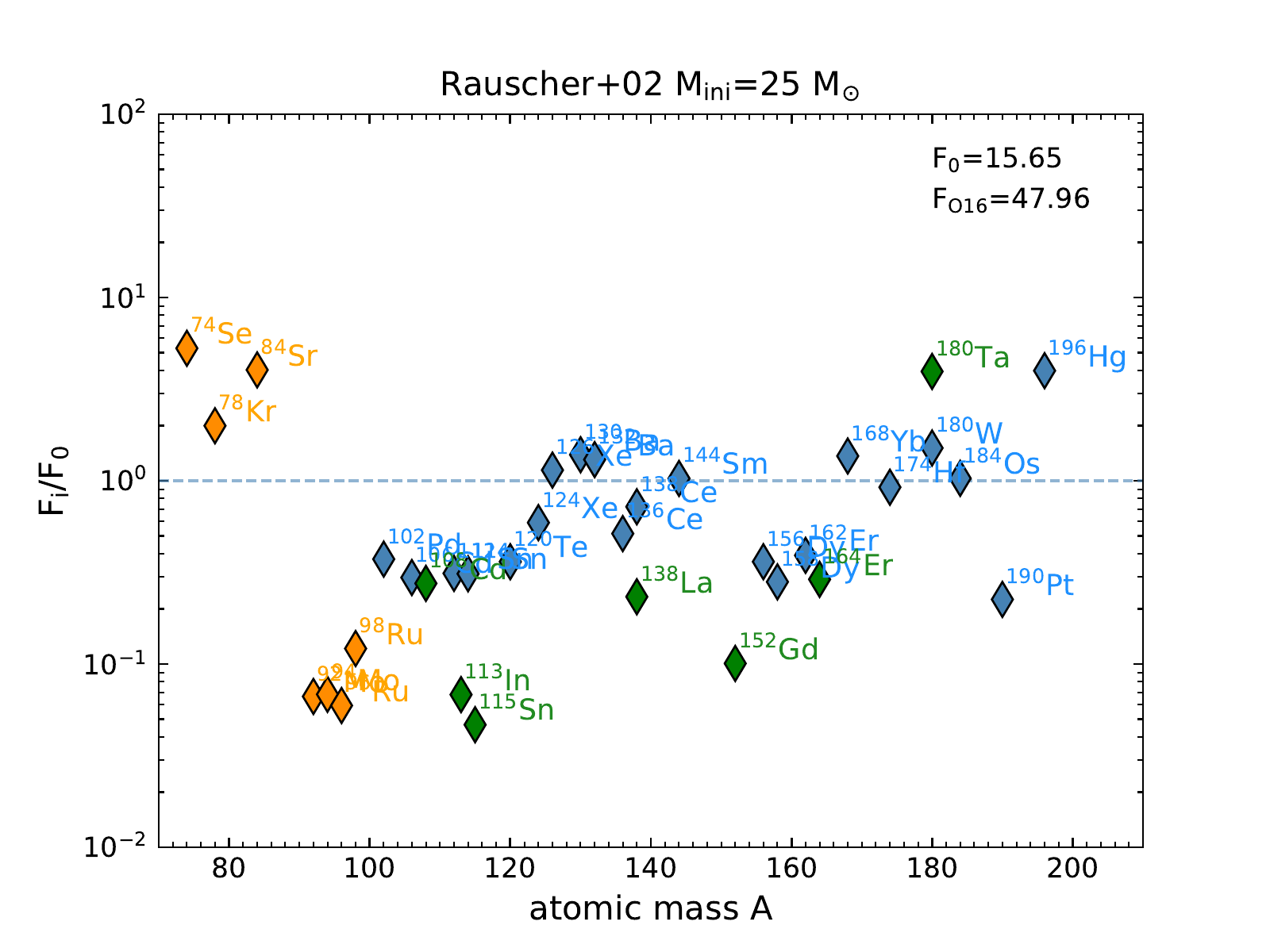}
            \includegraphics[scale=0.56]{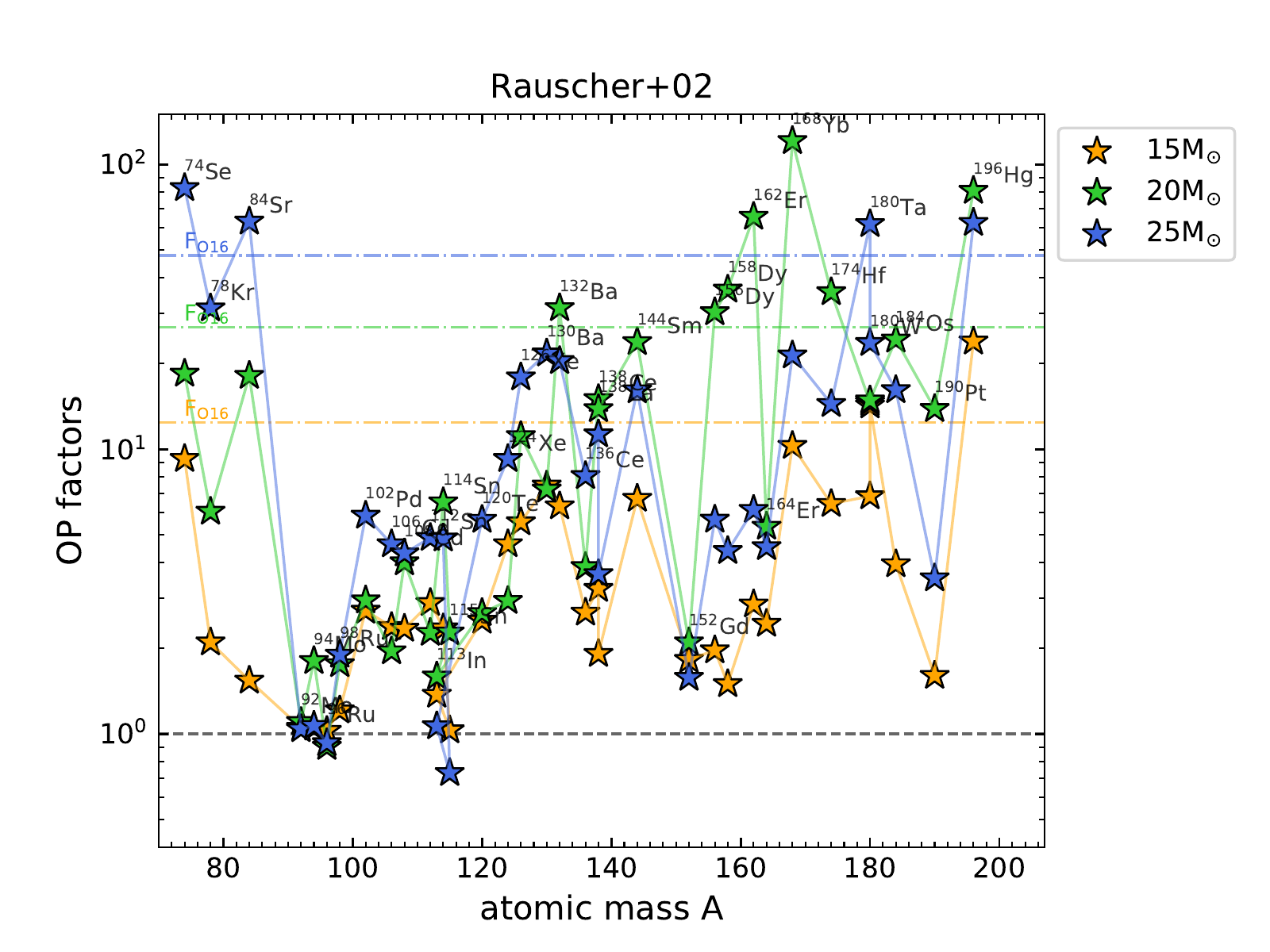}
            \caption{The F$_{\rm i}$/\fz\ of the 15 \msun\ (upper left), 20 \msun\ (upper right), 25 \msun\ (lower left) and the isotopic OP factor distributions (lower right) for \cite{rauscher:02} models. The horizontal dotted-dashed lines in the lower right panel represent the \fx[O16] in the 15 (yellow), 20 (green), and 25 \msun\ models (blue).}
            \label{fig:app_op_rau}
        \end{figure*}

\begin{figure*}
            \centering            
            \includegraphics[scale=0.56]{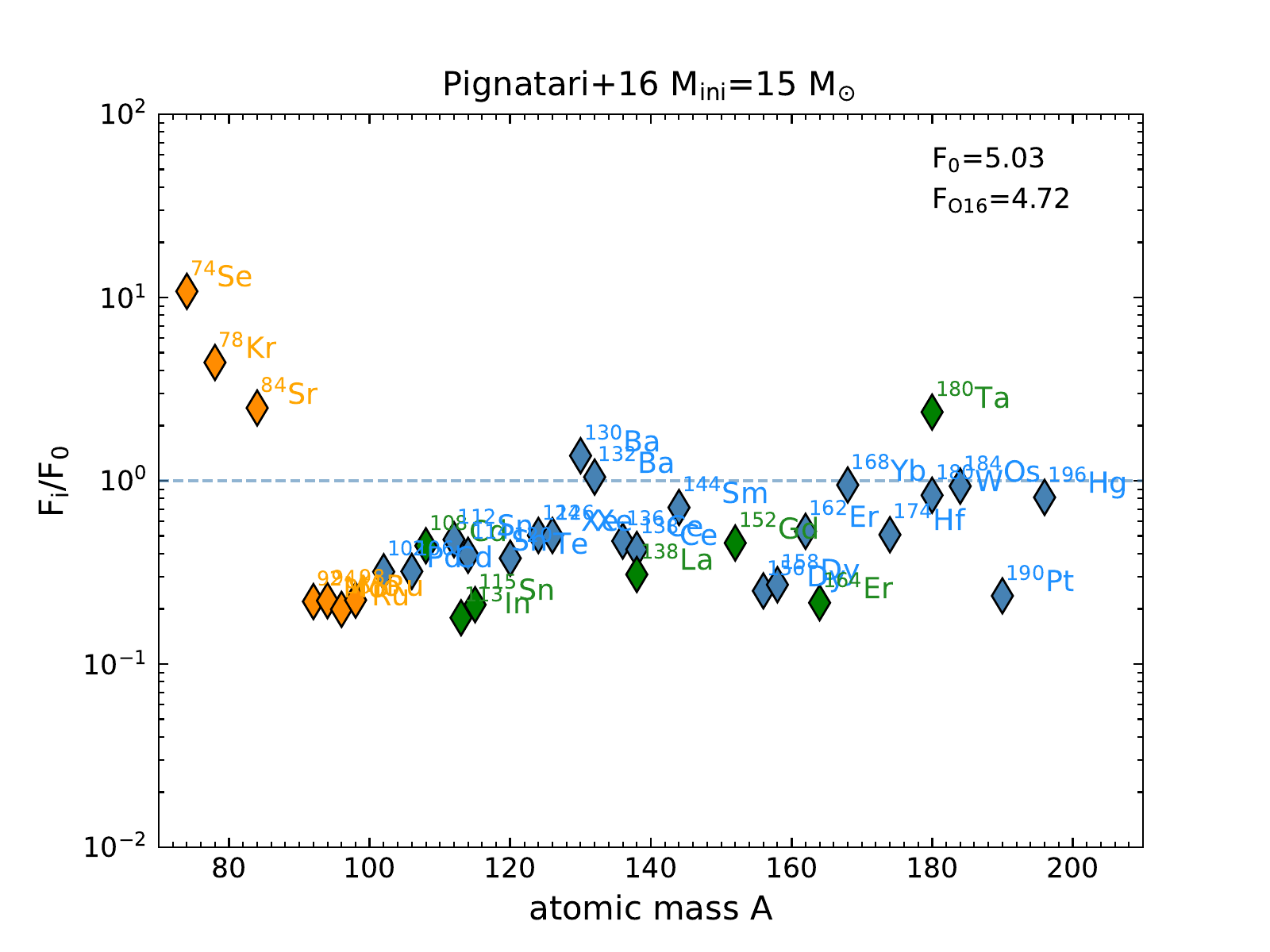}
            \includegraphics[scale=0.56]{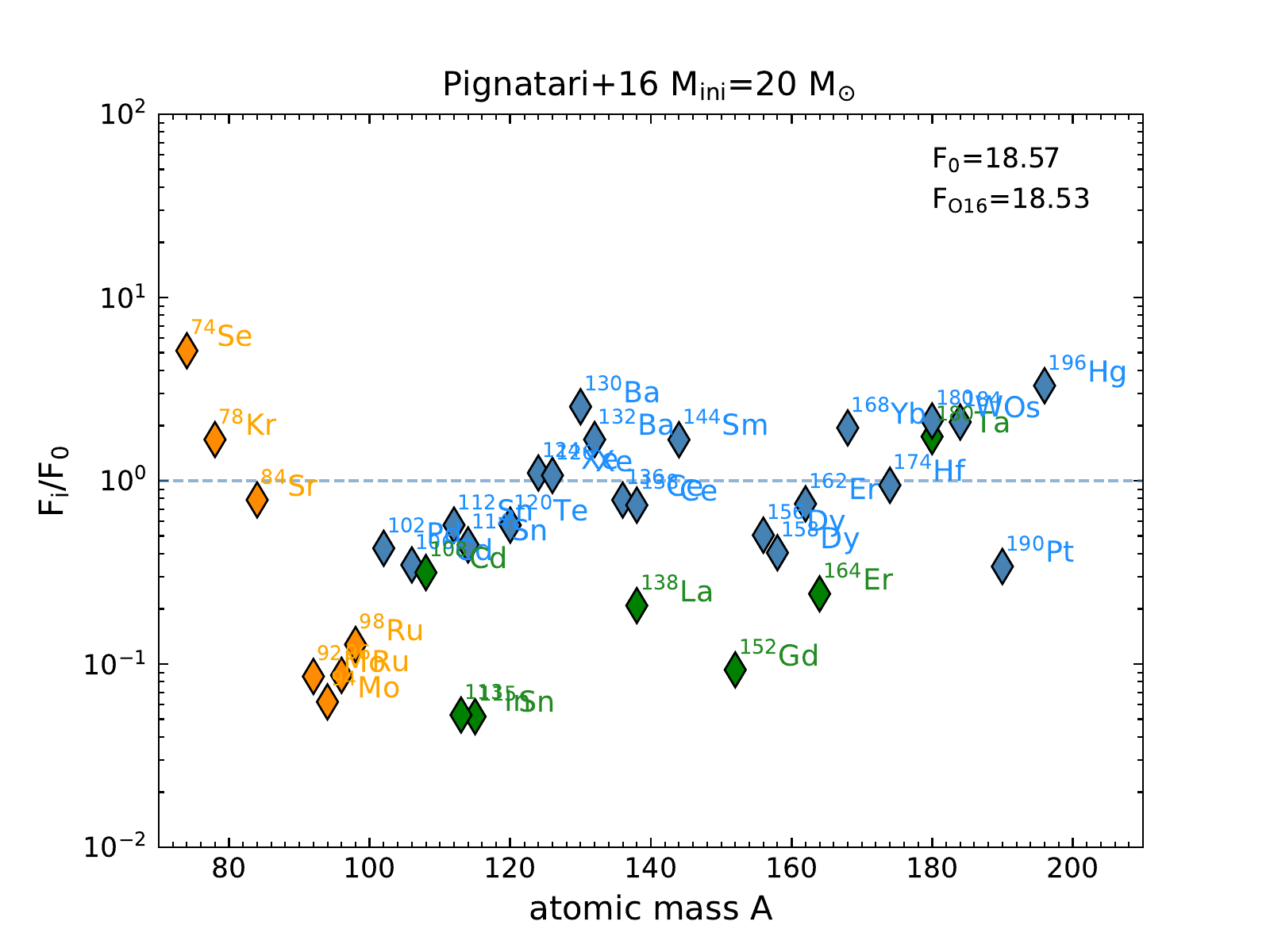}
            \includegraphics[scale=0.56]{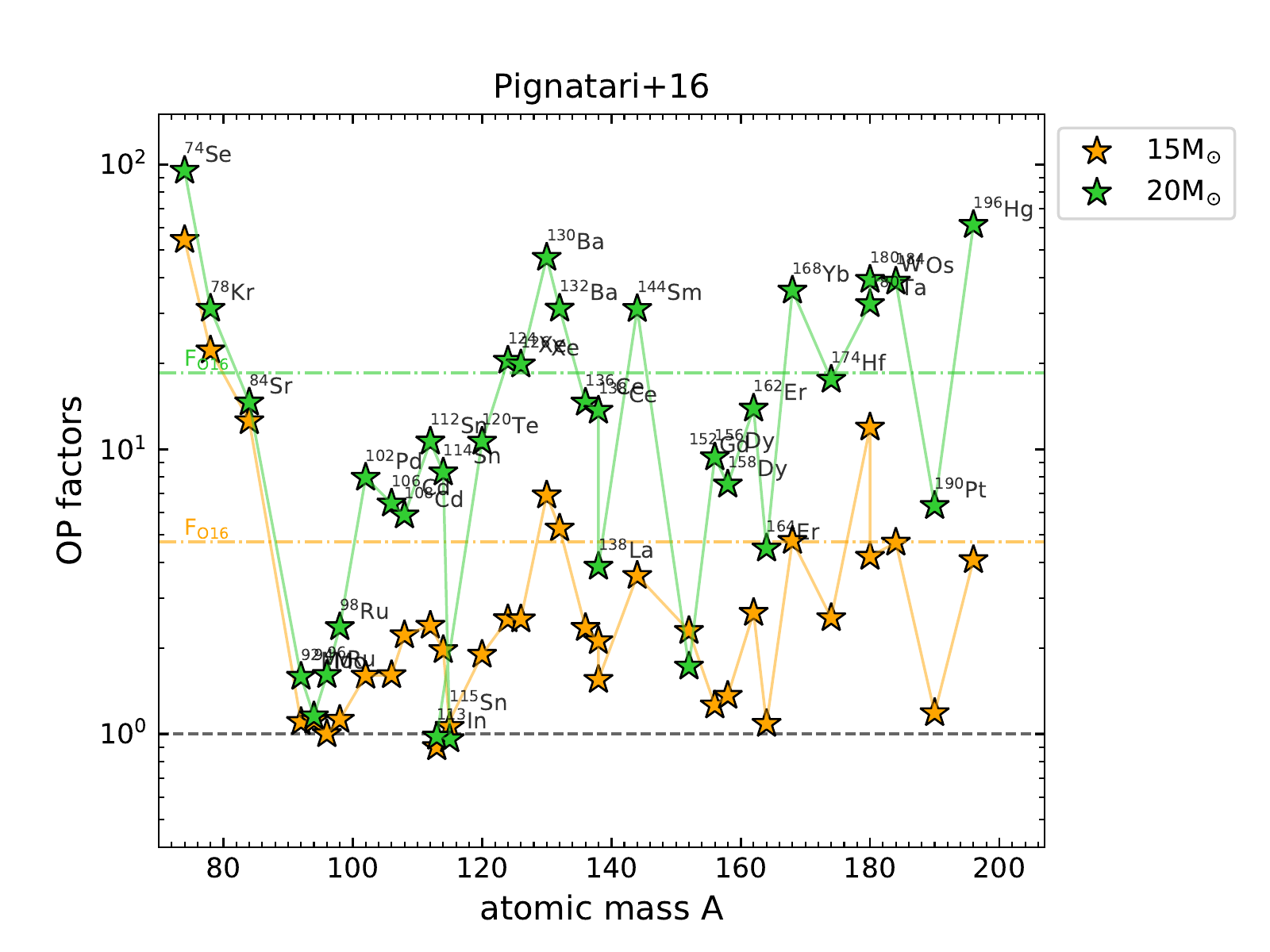}
            \caption{Same as \figurename~\ref{fig:app_op_rau} but for \cite{pignatari:16a} models.}
            \label{fig:app_op_pgn}
        \end{figure*}

\begin{figure*}
            \centering
            \includegraphics[scale=0.56]{f0/Rit_F0_15z0p02.pdf}
            \includegraphics[scale=0.56]{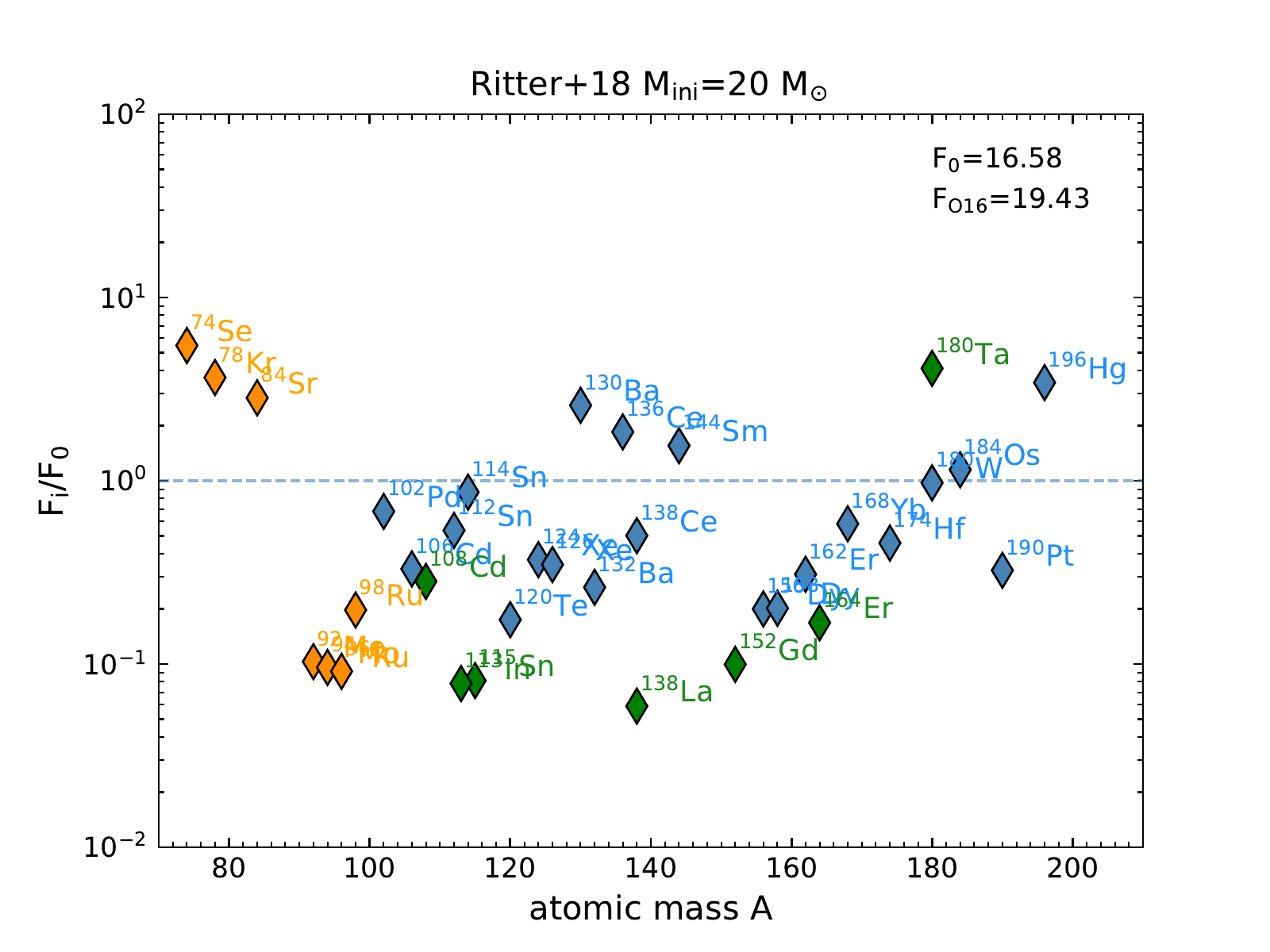}
            \includegraphics[scale=0.56]{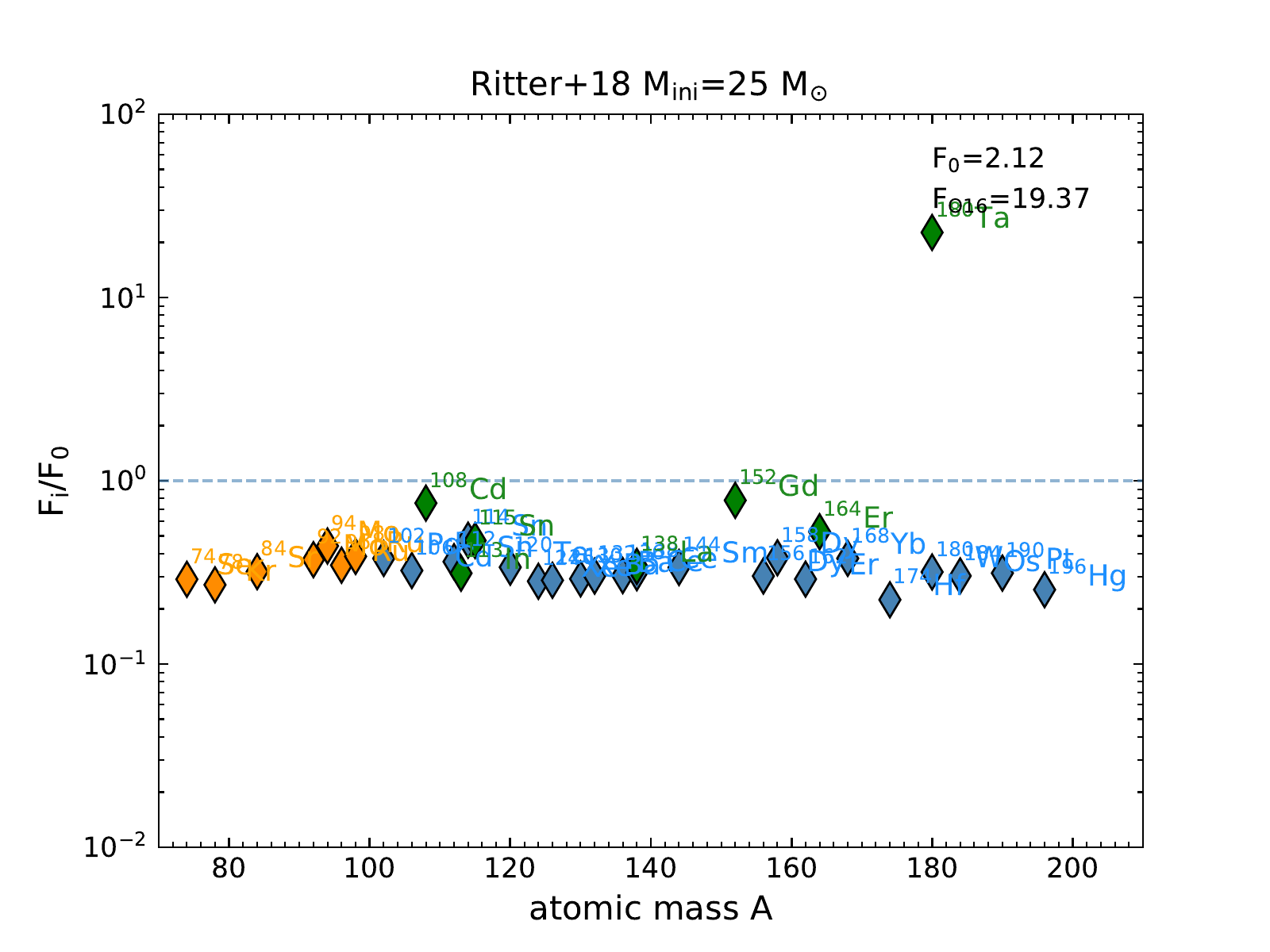}
            \includegraphics[scale=0.56]{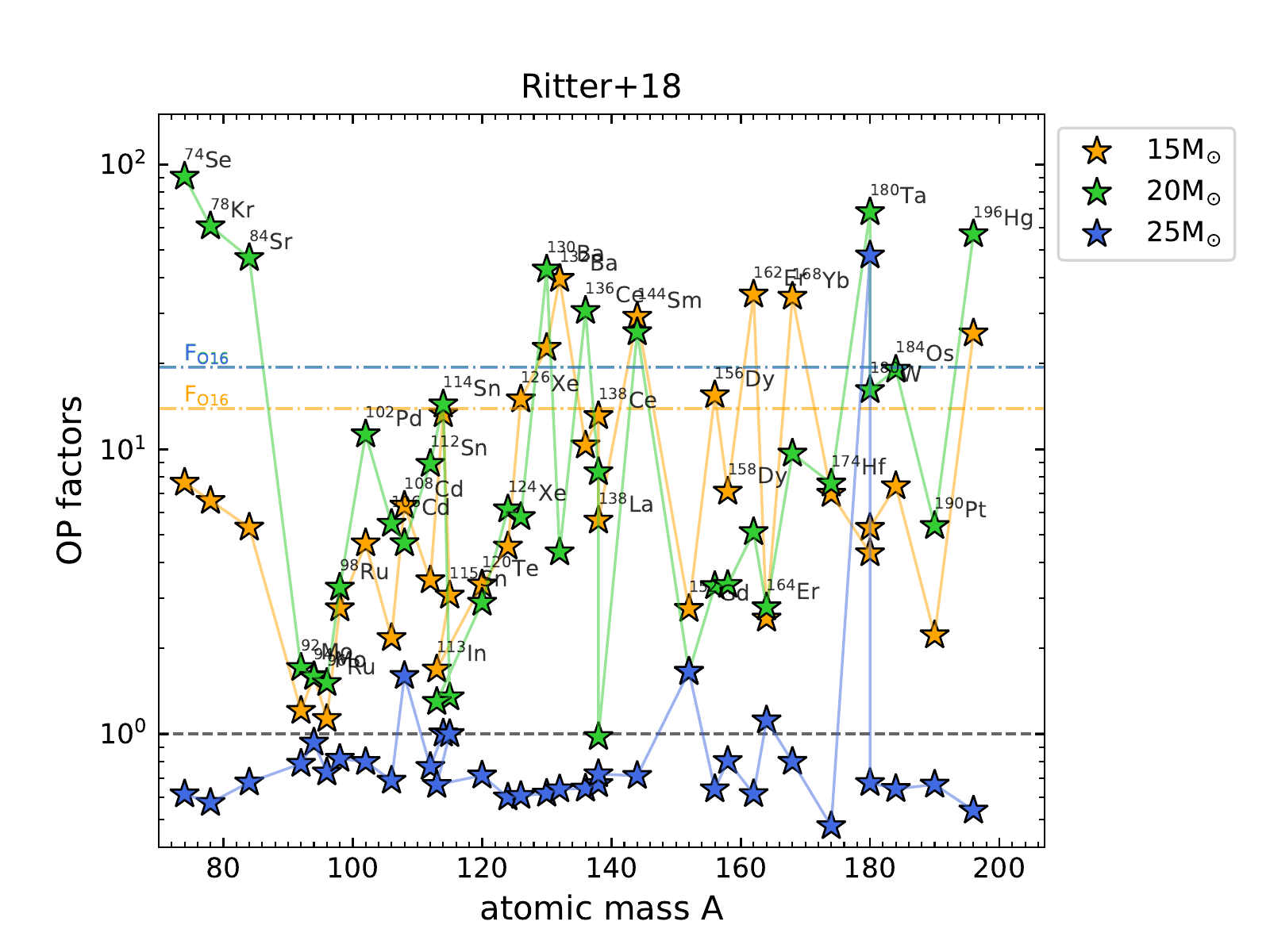}
            \caption{Same as \figurename~\ref{fig:app_op_rau} but for \cite{ritter:18} models.}
            \label{fig:app_op_rit}
        \end{figure*}

\begin{figure*}
            \centering            
            \includegraphics[scale=0.56]{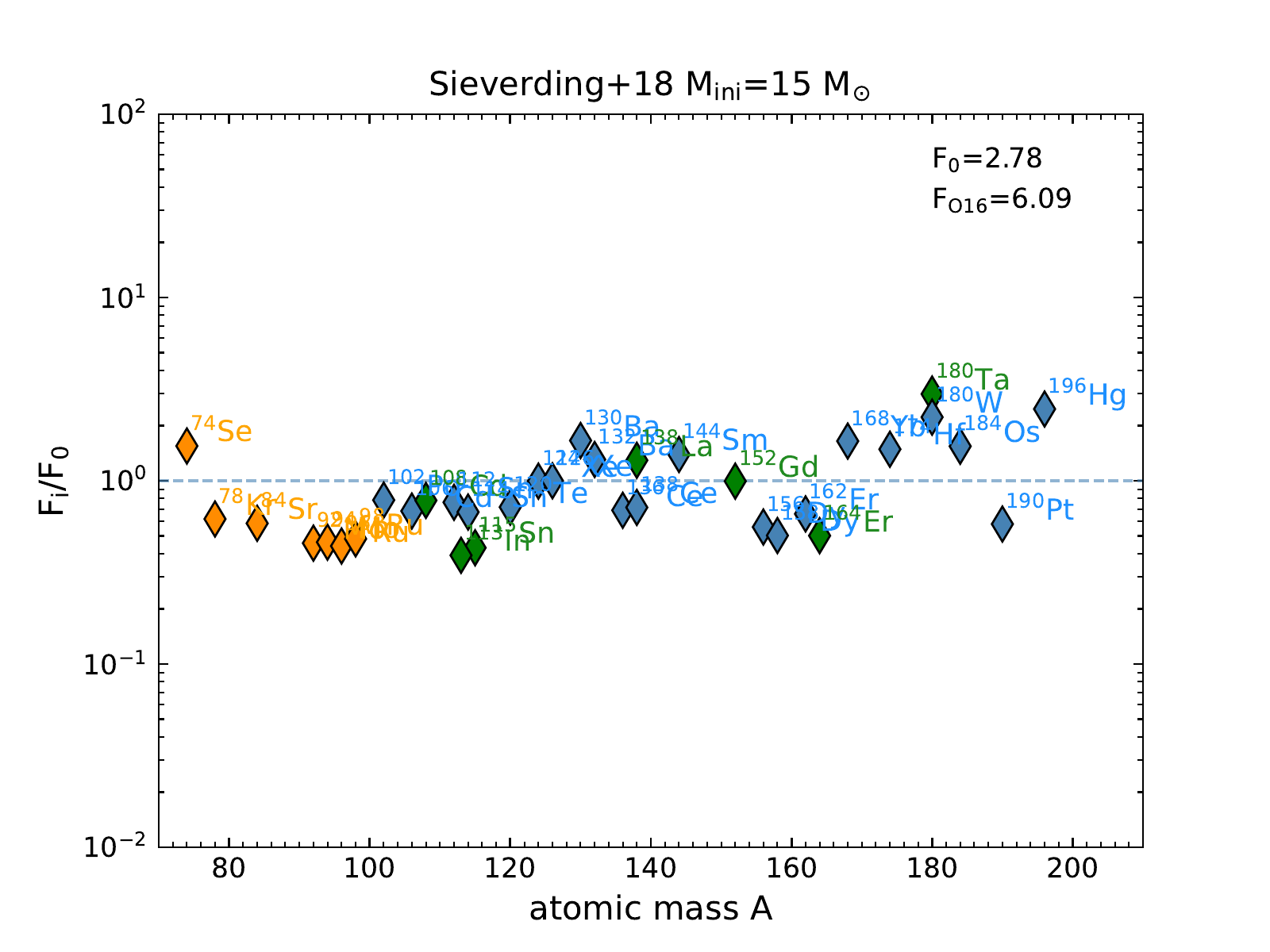}
            \includegraphics[scale=0.56]{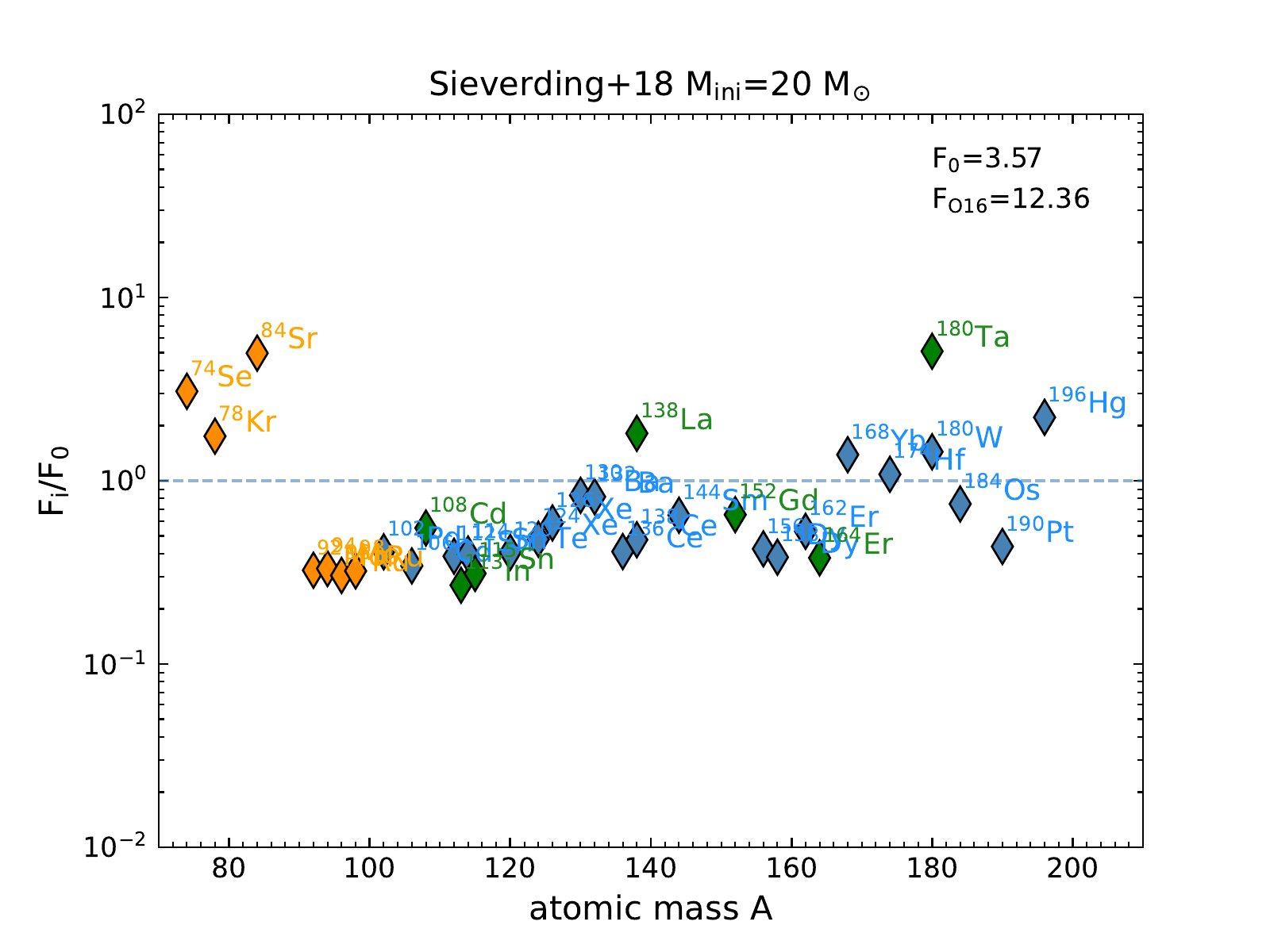}
            \includegraphics[scale=0.56]{f0/Sie_F0_25.pdf}
            \includegraphics[scale=0.56]{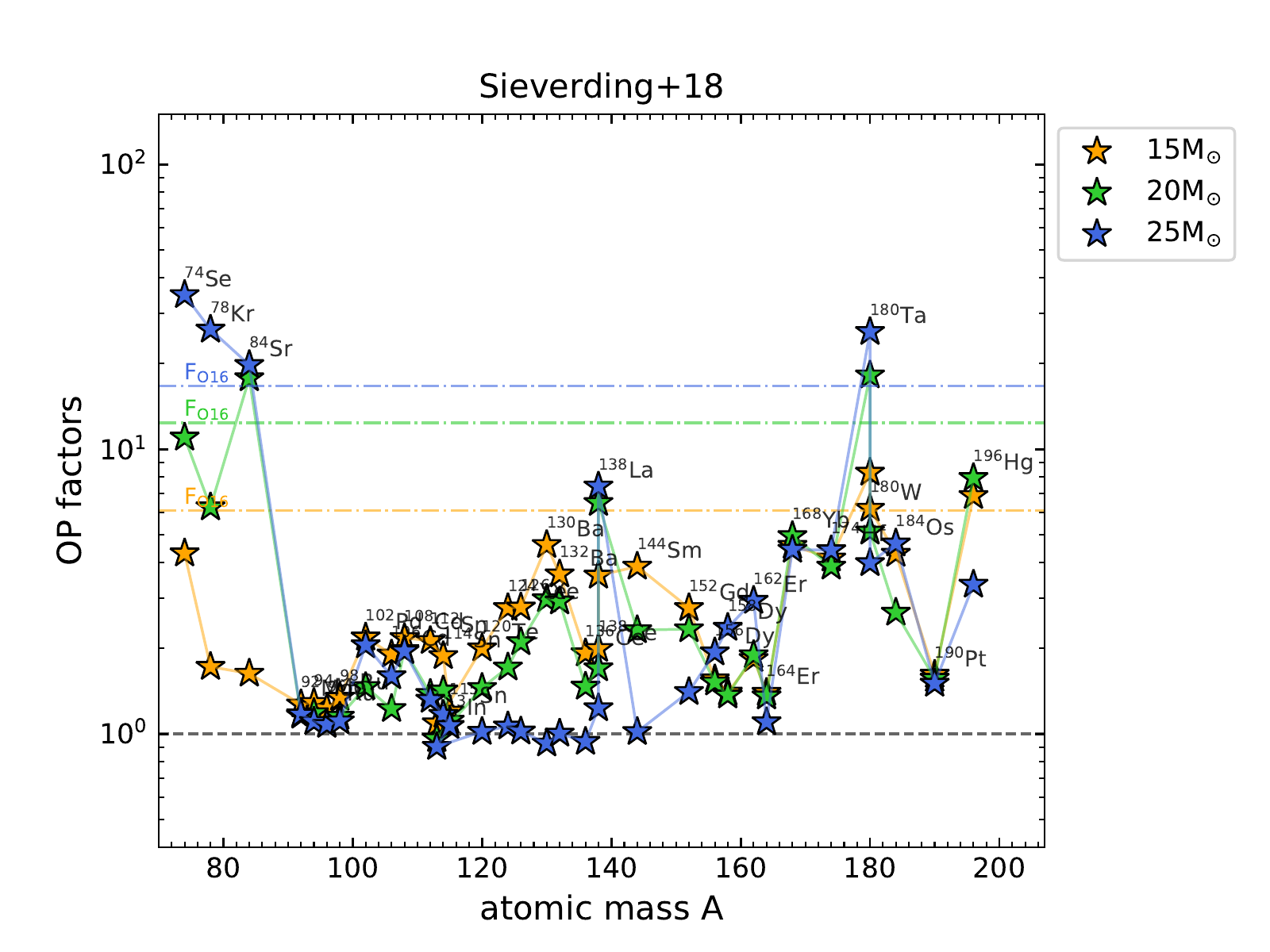}
            \caption{Same as \figurename~\ref{fig:app_op_rau} but for \cite{sieverding:18} models.}
            \label{fig:app_op_sie}
        \end{figure*}

\begin{figure*}
            \centering            
            \includegraphics[scale=0.56]{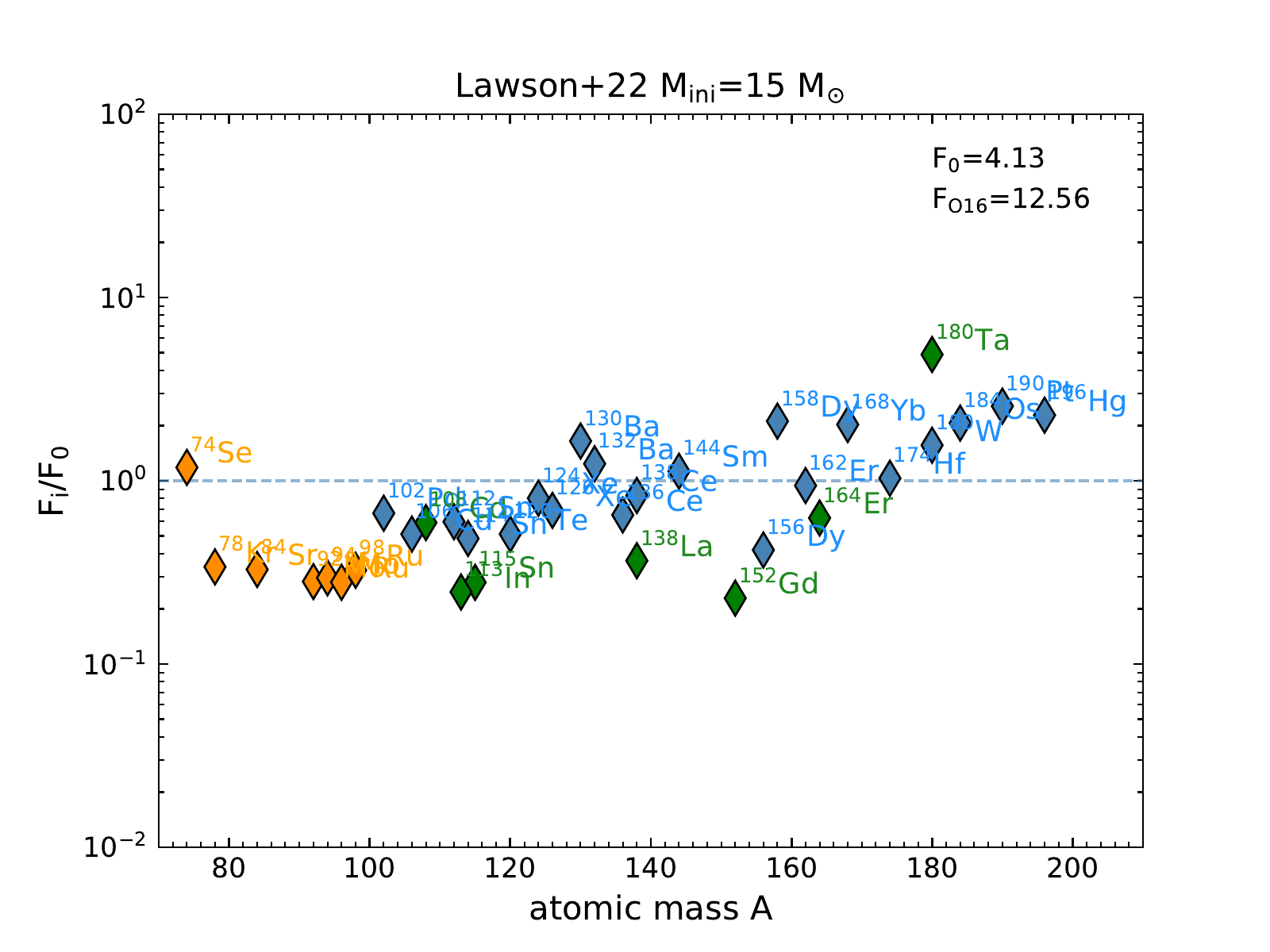}
            \includegraphics[scale=0.56]{f0/Law_F0_20.pdf}
            \includegraphics[scale=0.56]{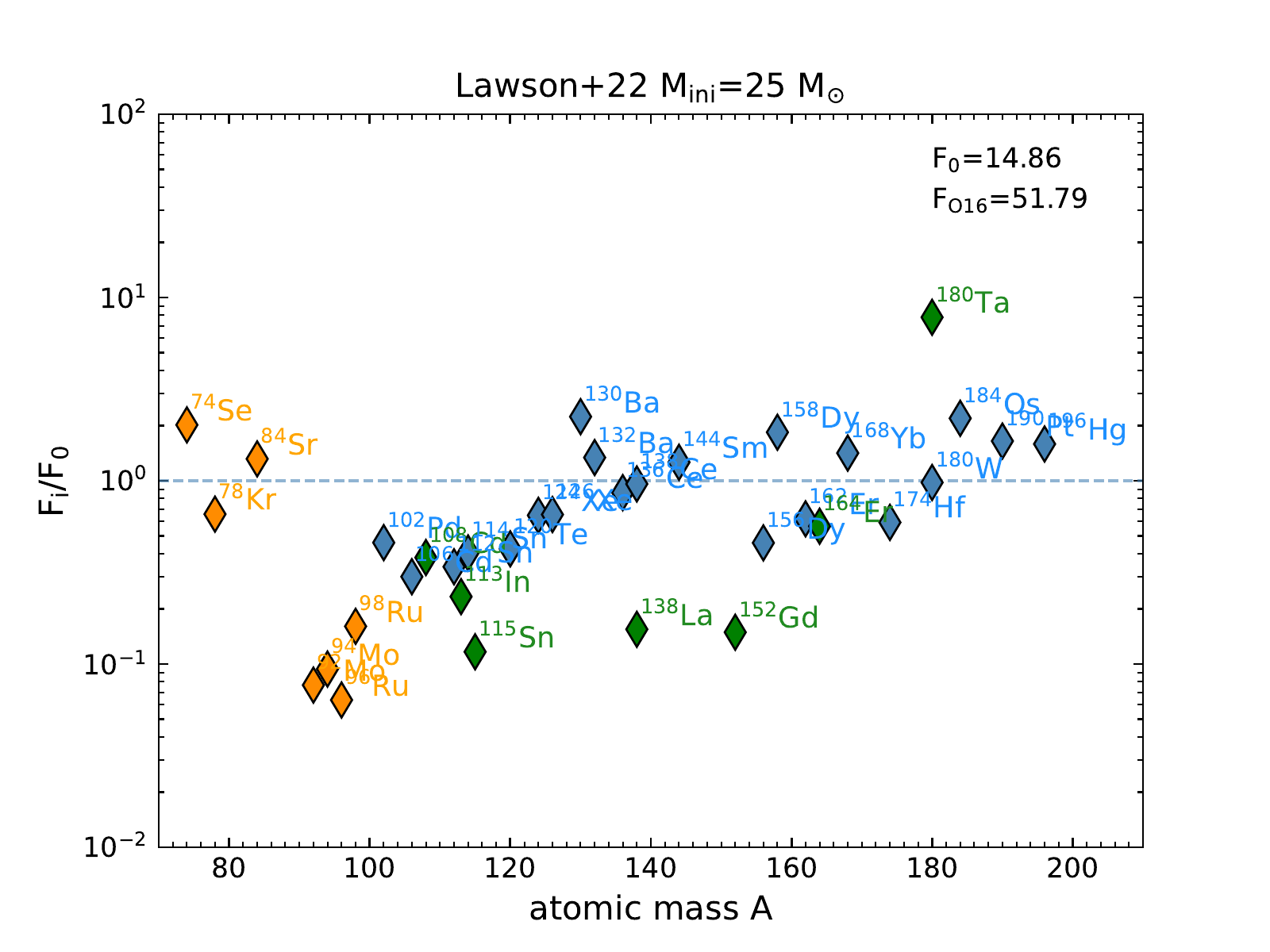}
            \includegraphics[scale=0.56]{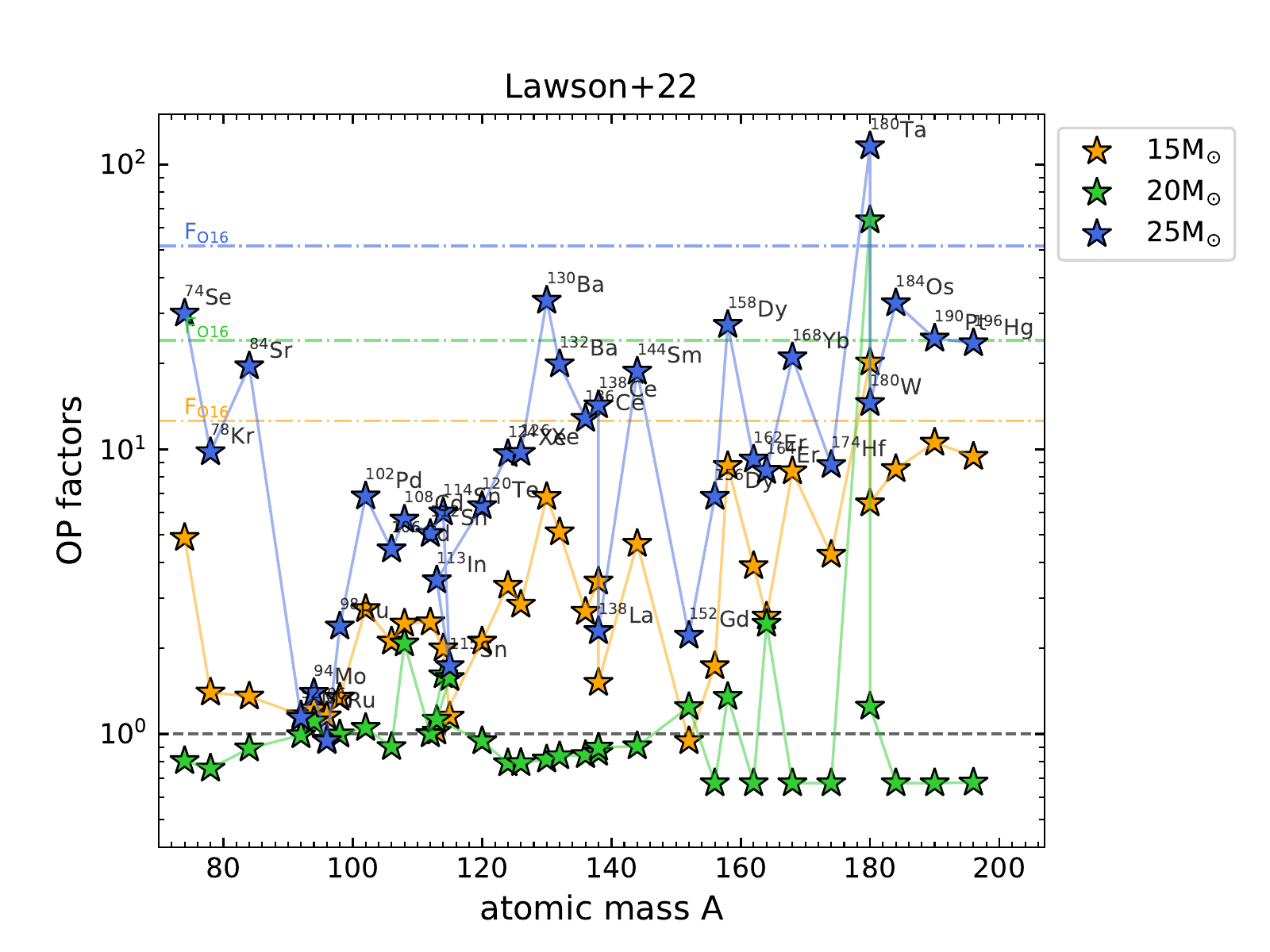}
            \caption{Same as \figurename~\ref{fig:app_op_rau} but for \cite{lawson:22} models.}
            \label{fig:app_op_law}
        \end{figure*}

        Here we present the overproduction (OP) factors and their \fz\ of the sets \cite{rauscher:02,pignatari:16a,ritter:18,sieverding:18,lawson:22} used in the analysis presented in this work. \figurename~\ref{fig:app_op_rau}~\ref{fig:app_op_pgn}~\ref{fig:app_op_rit}~\ref{fig:app_op_sie}~\ref{fig:app_op_law} show the F$_{\rm i}$/\fz\ and the isotopic OP factor distributions for the models from each set. As discussed in Section \ref{sec:f0}, we excluded from our analysis the 25 \msun\ from \cite{ritter:18} and the 20 \msun\ from \cite{lawson:22}, since the majority of the OP factors are lower than 1, due to the very large remnant mass obtained after the explosion, and the 25 \msun\ from \cite{pignatari:16a}, due to the low resolution in the \g--process peak abundances. The respective \fx[O16] is also reported as a reference.
        
        %RAU

        %PGN

        %RIT

        %SIE

        %LAW

    \section{Isotopic ratios} \label{app:isoratio} %Second appendix

        Here we present an extended and detailed discussion for 12 couples of isotopic ratios presented in \tablename~\ref{tab:isoratio}. The grey regions in the following plots represent regions of values a factor of two and three from the solar ratios. As already stated in Sec. \ref{sec:isoratio}, we consider models to be in relatively good agreement with solar when both following conditions are satisfied: (i) they fall into the grey areas and (ii) they have a overproduction at least twice as solar (OP $>$ 2).

\begin{figure}
                \centering
                \includegraphics[scale=0.5]{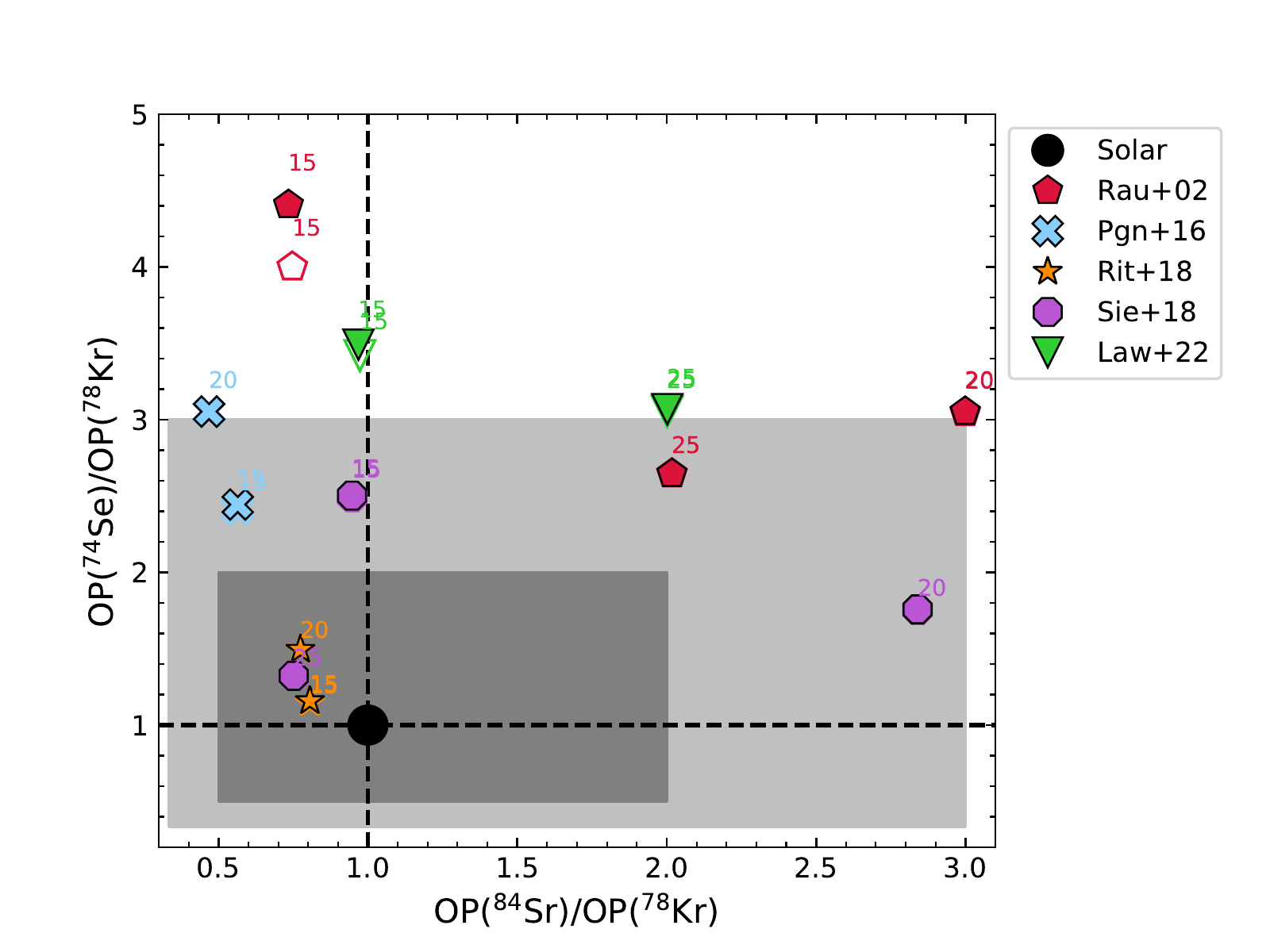}
                \caption{\isotope[74]{Se}/\isotope[78]{Kr} versus \isotope[84]{Sr}/\isotope[78]{Kr} from the model yields, with both ratios normalised to their respective solar values. The light and dark grey shaded areas represent, respectively, values a factor of two and three from the solar ratio (black filled circle). The different sets of models are represented by different types of symbols as indicated in the legend on the right-hand side. The mass of the star is indicated as a small number next to its corresponding symbol. Filled and empty symbols represent the yields including or not, respectively, the radiogenic contribution from all unstable isotopes. The only significant radiogenic contribution comes from \isotope[74]{Br} to \isotope[74]{Se}.}
                \label{fig:00}
            \end{figure}

        % === 00
        \subsection{\isotope[74]{Se}/\isotope[78]{Kr} vs \isotope[84]{Sr}/\isotope[78]{Kr} (\figurename~\ref{fig:00})} \label{sub:01}
    
            These two ratios include nuclei that may all have additional channels of production through explosive nucleosynthesis via the \A--rich freeze-out and the neutrino-driven winds (see Sec. \ref{sec:intro} and \ref{sec:f0}). In the models considered for this work, the bulk of the yields comes from the \g--process nucleosynthesis occurring in the explosive O/Ne burning regions. Only the PGN15 model ejects a significant fraction of material from \A--rich freeze-out, however, the total yield is still dominated by the O/Ne explosive burning region. Also in the two models with the C--O shell mergers (RIT15 and RAU20), the explosive \g--process has the most significant contribution to these three isotopes, in fact the explosion produces about 70--80$\%$ and 40--60$\%$, respectively, of the total yields (\tablename~\ref{tab:merper}).
            
            The \isotope[74]{Se}/\isotope[78]{Kr} ratio (y-axis) is always super-solar, i.e., OP(\isotope[74]{Se}) $>$ OP(\isotope[78]{Kr}). The \isotope[84]{Sr}/\isotope[78]{Kr} ratio (x-axis) of three out of the five 15 \msun\ models (RAU15, SIE15, and LAW15) mostly reflects the initial, solar abundances in the envelope of the stars because both OP(\isotope[84]{Sr}) and OP(\isotope[78]{Kr}) are lower than 2.
            In the other two 15 \msun\ models (PGN15 and RIT15) as well as the RIT20 and SIE25 models, instead, \isotope[78]{Kr} and \isotope[84]{Sr} are significantly produced during the explosive \g--process (see Figures in Appendix \ref{app:OP}) therefore the values plotted in the figure represent the \g--process composition, not the initial composition. 
            Among the remaining three 20 \msun\ models, PGN20 produces one of the highest OP(\isotope[78]{Kr}) of all the models, which it leads to the lowest OP(\isotope[84]{Sr})/OP(\isotope[78]{Kr}) ratio. The last two 20 \msun\ models (RAU20 and SIE20) have both a very similar yield of \isotope[78]{Kr} and \isotope[84]{Sr}. The solar value of \isotope[84]{Sr}/\isotope[78]{Kr} is instead $\sim 0.7$, which results in OP(\isotope[84]{Sr})/OP(\isotope[78]{Kr}) the factor of $\sim3$ over the solar ratio. In the 25 \msun\ case, RAU25 and LAW25 have different yields and different OP factors, but the ratio between \isotope[84]{Sr} and \isotope[78]{Kr} yields is the same, therefore, this leads to a similar behaviour of the ratio of the OP factors. 
    
\begin{figure}
                \centering
                \includegraphics[scale=0.5]{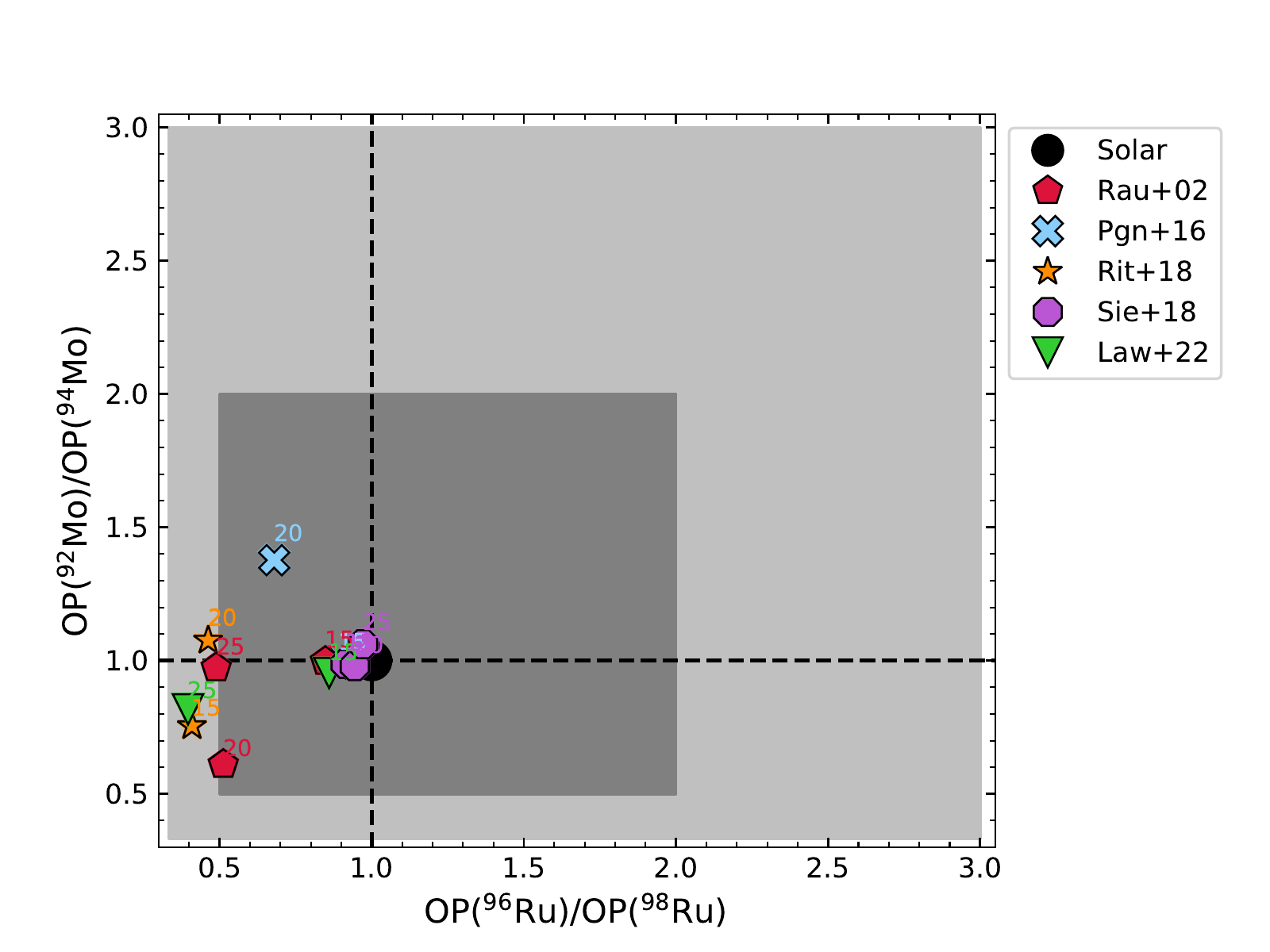}
                \caption{Same as \figurename~\ref{fig:00}, but for \isotope[92]{Mo}/\isotope[94]{Mo} versus \isotope[96]{Ru}/\isotope[98]{Ru}.}
                \label{fig:01}
            \end{figure}

        % === 01
        \subsection{\isotope[92]{Mo}/\isotope[94]{Mo} vs     \isotope[96]{Ru}/\isotope[98]{Ru} (\figurename~\ref{fig:01})} \label{sub:02}
        
            In almost all models, the OPs of these 4 isotopes are significantly lower than 2, which confirms that standard massive star models are unable to produce a significant amount of these isotopes and represents the reason why most of the models are in agreement with the solar abundance: they mostly reflect the envelope composition. All the model shows a sub-solar \isotope[96]{Ru}/\isotope[98]{Ru} ratio (x-axis), while the \isotope[92]{Mo}/\isotope[94]{Mo} ratio (y-axis) is more variable and there are 6 models roughly a factor of two different different from solar. In the case of the two models with C--O shell mergers (RIT15 and RAU20), the production of the neutron richer isotopes of each element is favoured, therefore both plotted ratios are sub-solar (\tablename~\ref{tab:merper}).         
            Out of the other 4 models different from solar, in RIT20 the \g--process takes place during explosive Ne burning, in an extended region of $\sim0.6$ \msun, with a wide \g--process peak, in particular of \isotope[92]{Mo} and \isotope[98]{Ru}, which results in higher yields of these two isotopes compared to the other models. Instead, in PGN20 there are two distinct \g--process regions, one in the typical O/Ne explosive burning zone and one just above the interface between the ONe and CO cores, where the shock wave accelerates again with a consequent increase of the peak temperature. This double contribution to the \g--process nucleosynthesis increases the OP factors of \isotope[92]{Mo}, \isotope[96]{Ru} and \isotope[98]{Ru}. However, %no one of those features gives a relevant production.
            those features do not affect significantly the overall distribution of the \p--nuclei.
            
\begin{figure}
                \centering
                \includegraphics[scale=0.5]{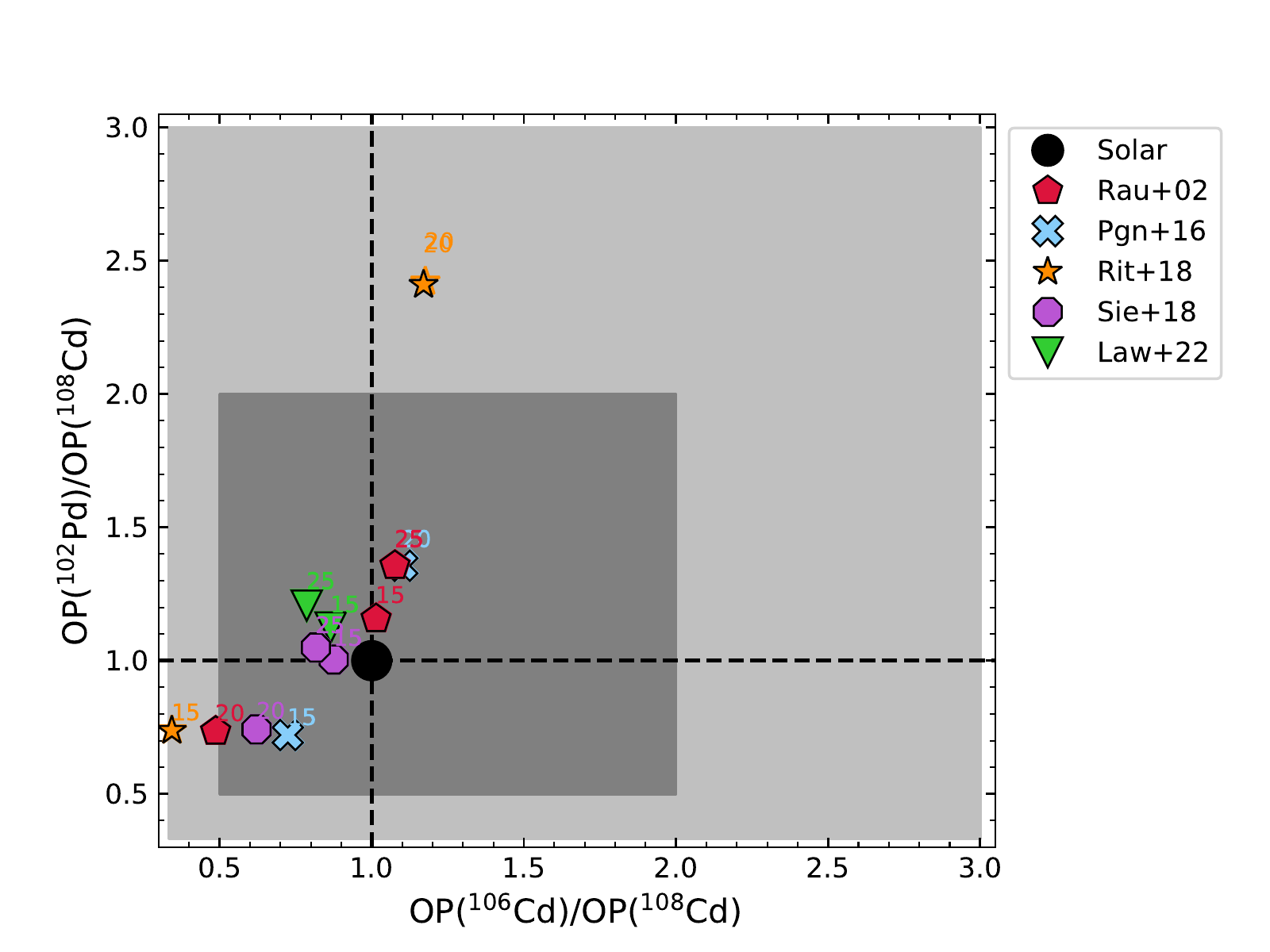}
                \caption{Same as \figurename~\ref{fig:00}, but for \isotope[102]{Pd}/\isotope[108]{Cd} versus \isotope[106]{Cd}/\isotope[108]{Cd}.}
                \label{fig:02}
            \end{figure}

        % === 02
        \subsection{\isotope[102]{Pd}/\isotope[108]{Cd} vs \isotope[106]{Cd}/\isotope[108]{Cd} (\figurename~\ref{fig:02})} \label{sub:03}
    
            Most of the models fall within a factor of 2 from the solar ratios. The 4 models RIT15, PGN15, SIE20, and RAU20 have all  
            sub-solar ratios and the same \isotope[102]{Pd}/\isotope[108]{Cd} (y-axis). In the case of PGN15 and SIE20 the sub-solar \isotope[102]{Pd}/\isotope[108]{Cd} ratio is due to OP(\isotope[102]{Pd}) $<$ 2, while the other two models (RIT15 and RAU20) are those that experience a C--O shell merger, where the production of \isotope[108]{Cd}, at denominator in both plotted ratios, is favoured over that of \isotope[106]{Cd} and \isotope[102]{Pd}. The production of \isotope[102]{Pd} is different between those two C--O shell merger models: in RIT15 it is dominated by the merger and in RAU20 by the explosion (see \tablename~\ref{tab:merper}). Instead, RAU15/25, PGN20 and RIT20 have all super-solar ratios because their OP(\isotope[102]{Pd}) and OP(\isotope[106]{Cd}) (i.e., of the isotopes at numerator) are significantly higher than the other models, while their OP(\isotope[108]{Cd}) is similar. All the other models have super-solar \isotope[102]{Pd}/\isotope[108]{Cd} and sub-solar \isotope[106]{Cd}/\isotope[108]{Cd}, mostly because OP(\isotope[108]{Cd}) and OP(\isotope[102]{Pd}) $\sim$ 2 while OP(\isotope[106]{Cd}) $<$ 2. No model falls into the quadrant identified by \isotope[106]{Cd}/\isotope[108]{Cd} $>$ solar and \isotope[102]{Pd}/\isotope[108]{Cd} $<$ solar, which means that \isotope[102]{Pd} and \isotope[106]{Cd} are always overproduced together.
    
\begin{figure}
                \centering
                \includegraphics[scale=0.5]{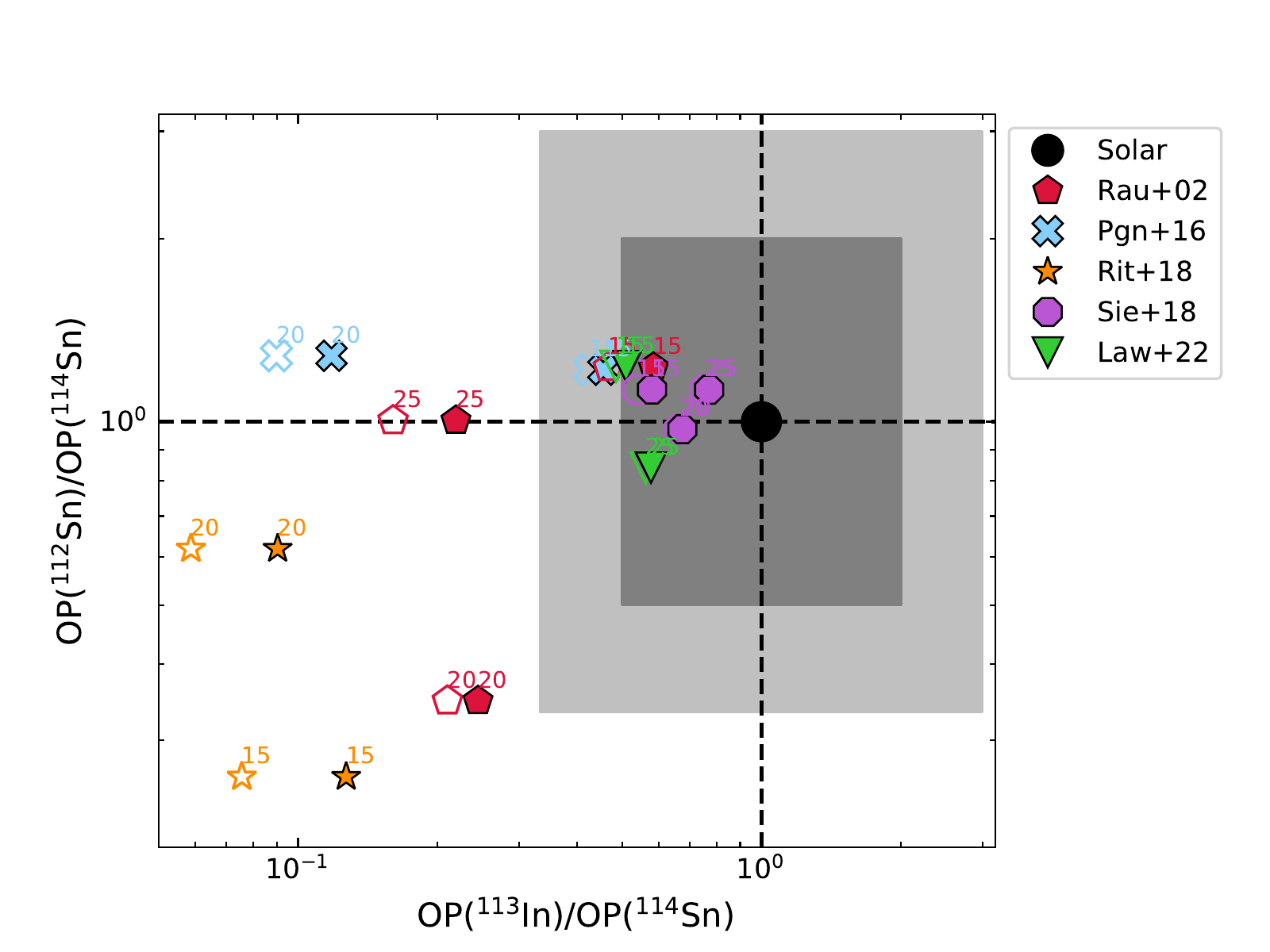}
                \includegraphics[scale=0.5]{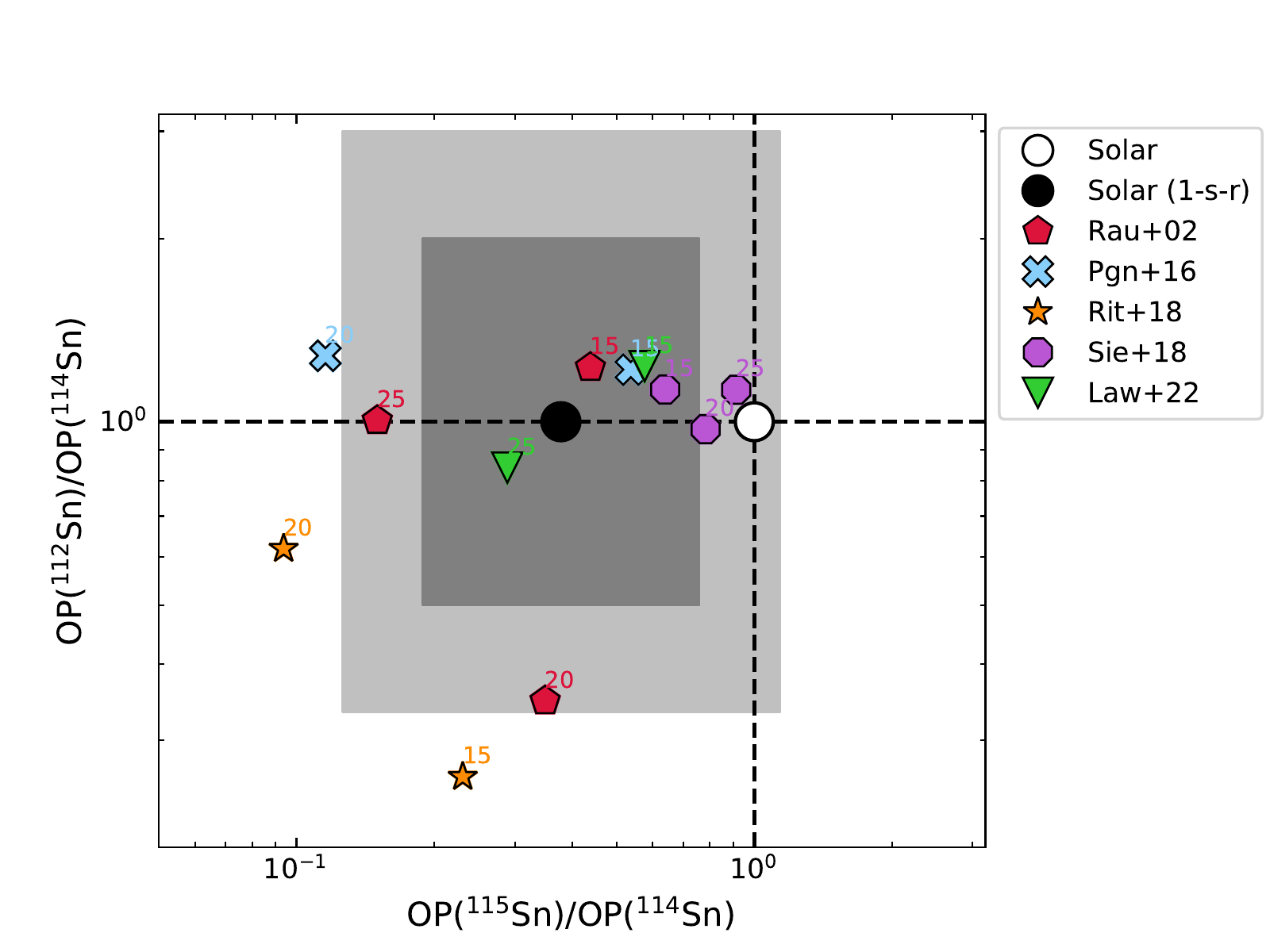}
                \caption{Same as \figurename~\ref{fig:00}, but for
                \isotope[112]{Sn}/\isotope[114]{Sn} versus \isotope[113]{In}/\isotope[114]{Sn} (upper panel) and \isotope[112]{Sn}/\isotope[114]{Sn} versus \isotope[115]{Sn}/\isotope[114]{Sn} (lower panel). \isotope[113]{In} has a radiogenic contribution from \isotope[113]{Sn}. In the lower panel, the empty circle represents the solar ratio, while the filled circle represents the solar ratio minus the \s-- and \rr--process contribution to \isotope[115]{Sn} (1-\s-\rr). In this plot, the gray shaded areas identify a factor of 2 and 3 from the ratio calculated using the (1-\s-\rr $\,$) value.}
                \label{fig:034}
            \end{figure}

        % === 03 & 04
        \subsection{\isotope[112]{Sn}/\isotope[114]{Sn} vs \isotope[113]{In}/\isotope[114]{Sn} and \isotope[112]{Sn}/\isotope[114]{Sn} vs \isotope[115]{Sn}/\isotope[114]{Sn} (\figurename~\ref{fig:034})} \label{sub:04}
    
            Of the four isotopes discusses here, the production of \isotope[113]{In}, \isotope[114]{Sn}, and \isotope[115]{Sn} can also occur via neutron captures because \isotope[113]{Cd} and \isotope[115]{In} have an unstable isomer (see Sec. \ref{sec:intro} and \ref{sec:f0}). The galactic chemical evolution (GCE) computation of \cite{bisterzo:14} found that the \s--process nucleosynthesis in low-mass AGB stars provides a negligible contribution to these nuclei (0, 0.1, and 4.2$\%$, to \isotope[113]{In}, \isotope[114]{Sn}, and \isotope[115]{Sn}, respectively).
            The estimated \rr--process contributions to \isotope[113]{In} and \isotope[115]{Sn}, instead, are between 2 and 16$\%$, and $60\pm10\%$, respectively \citep[see, e.g.,][and references therein]{dillmann:08,nemeth:94,theis:98}. Therefore, the \g--process nucleosynthesis should have a significant role in the production of all of these isotopes. In \figurename~\ref{fig:034} we accounted for the neutron-capture contribution to \isotope[115]{Sn} only, and subtracted its $s$- and $r$-process contributions, because the contributions to the other isotopes are negligible. In general, the values close to the solar ratios do not represent the result of \g--process nucleosynthesis because the yields are dominated by the initial abundances. The cases that do not behave like this are discussed in detail below.
            Specifically, OP(\isotope[113]{In}) is lower than 2 in all the models, therefore \isotope[113]{In}/\isotope[114]{Sn} (x-axis, top panel) is always lower than solar. The models that effectively activate the \g--process are characterised by a radiogenic component to \isotope[113]{In} dominated by \isotope[113]{Sn}. However, this contribution is never enough to produce OP(\isotope[113]{In}) $>$ 2. 
    
            \figurename~\ref{fig:034} shows that the models that have a \g--process component contributing to the isotopes of Sn are typically two to three times away from the solar value, except for the LAW25 model which has a significant \g--process component but the ratios are all close to solar because \isotope[113]{In} and \isotope[114]{Sn} are similarly overproduced by an \s--process component in the C shell, therefore \isotope[112]{Sn}/\isotope[114]{Sn} is slightly sub-solar. 
            
            Out of the 5 most extreme models, the C--O shell merger in the RIT15 and RAU20 models favors the production of the Sn neutron richer isotopes, which results in sub-solar \isotope[113]{In}/\isotope[114]{Sn} and \isotope[112]{Sn}/\isotope[114]{Sn} ratios. The other three most extreme models (PGN20, RIT20 and RAU25) have the highest OP(\isotope[112]{Sn}) and OP(\isotope[114]{Sn}) and therefore their \isotope[113]{In}/\isotope[114]{Sn} ratio (top panel, x-axis) is strongly sub-solar. 
            In PGN20, the OP(\isotope[112]{Sn}) dominates over OP(\isotope[114]{Sn}). This is due to the peculiar structure of this star (see Sec. \ref{sec:datasets}) with two distinct abundance peaks of \g--process nucleosynthesis, in both of which the abundance of \isotope[112]{Sn} is higher than the abundance of \isotope[114]{Sn}. In RIT20, an additional production of \isotope[114]{Sn} occurs in the C shell, therefore \isotope[112]{Sn}/\isotope[114]{Sn} decreases to a value below solar. In the RAU25 model OP(\isotope[112]{Sn}) and OP(\isotope[114]{Sn}) are similar, therefore their \isotope[112]{Sn}/\isotope[114]{Sn} ratio is close to one. 
            
            In the lower panel of \figurename~\ref{fig:034} the y-axis is the same as in the top panel, in the x-axis \isotope[115]{Sn} substitutes \isotope[113]{In}, and the discussion is similar. In fact, OP(\isotope[115]{Sn}) as OP(\isotope[113]{In}) is always lower than 2, except in the case of the C--O shell merger. Most of the models, including those that have a significant \g--process production, fall within a factor of 3 of the value derived by subtracting the neutron-capture contributions from the solar value. 
    
\begin{figure}
                \centering
                \includegraphics[scale=0.5]{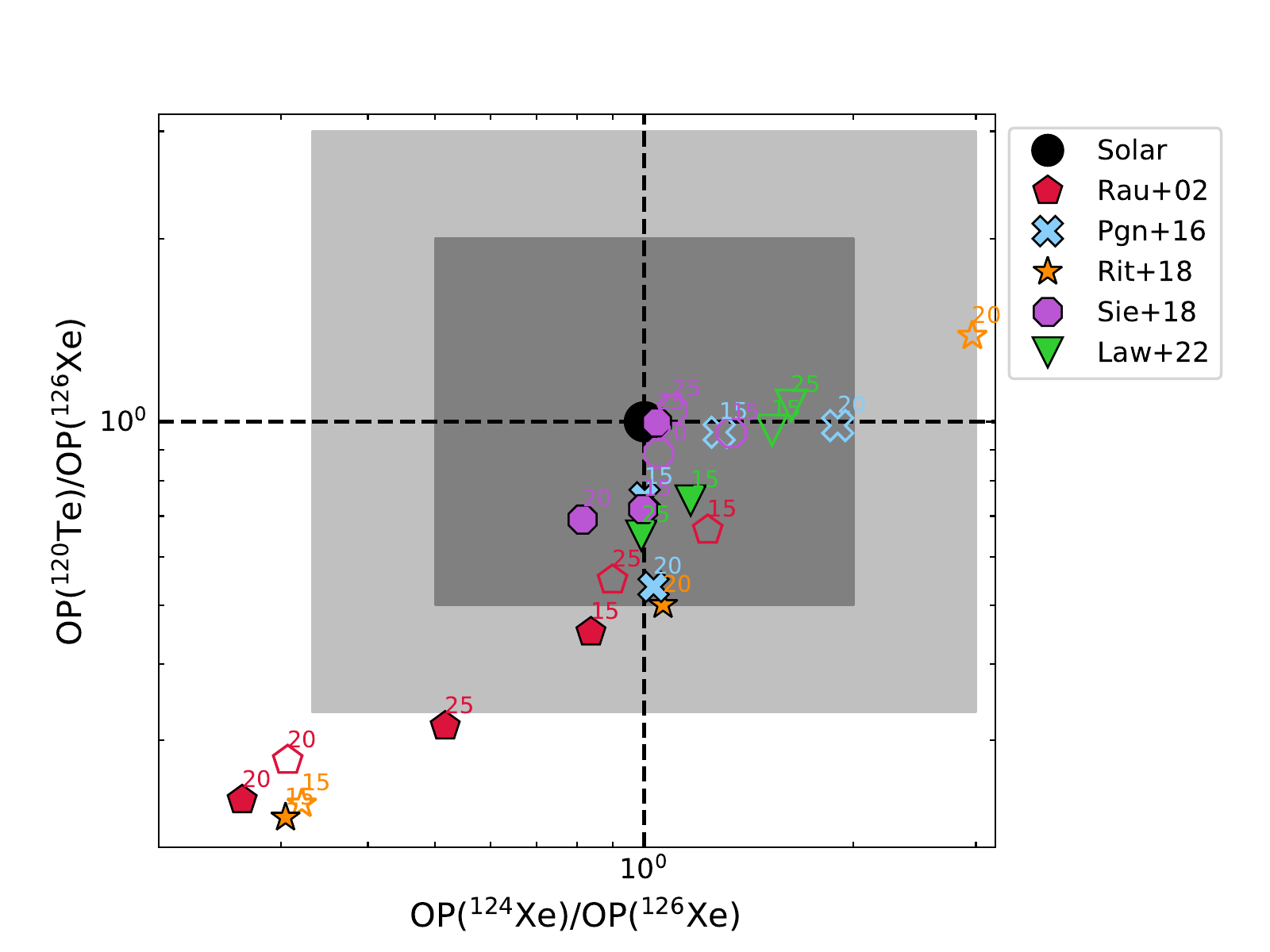}
                \caption{Same as \figurename~\ref{fig:00}, but for \isotope[120]{Te}/\isotope[126]{Xe} versus \isotope[124]{Xe}/\isotope[126]{Xe} ratio. \isotope[126]{Xe} has a radiogenic contribution from \isotope[126]{Ba} and \isotope[126]{Cs}.}
                \label{fig:05}
            \end{figure}

        % === 05
        \subsection{\isotope[120]{Te}/\isotope[126]{Xe} vs \isotope[124]{Xe}/\isotope[126]{Xe}(\figurename~\ref{fig:05})} \label{sub:06}
    
            Most of the models move to a sub-solar \isotope[120]{Te}/\isotope[126]{Xe} (y-axis) and solar \isotope[124]{Xe}/\isotope[126]{Xe} (x-axis) once the radiogenic contribution to \isotope[126]{Xe} is taken into account. Note that the SIE25 model is close to solar because all the OPs are close to one, therefore it is non relevant for the \g--process analysis (see \figurename~\ref{fig:f0}).
            Three models are instead different from solar: RIT15, RAU20, and RAU25. Two of these are the models with C--O shell merger (RIT15 and RAU20), where we found a similar overproduction of \isotope[120]{Te} and \isotope[124]{Xe} (i.e., the two isotopes at the numerator), and a relatively larger overproduction of \isotope[126]{Xe}. This results in both \isotope[120]{Te}/\isotope[126]{Xe} and \isotope[124]{Xe}/\isotope[126]{Xe}) more than three times lower than solar. Also in the RAU25 model, the overproduction of \isotope[126]{Xe} is larger than that of \isotope[124]{Xe} and \isotope[120]{Te}, however, in this case OP(\isotope[124]{Xe}) $\sim$ 2OP(\isotope[120]{Te}). 
    
\begin{figure}
                \centering
                \includegraphics[scale=0.5]{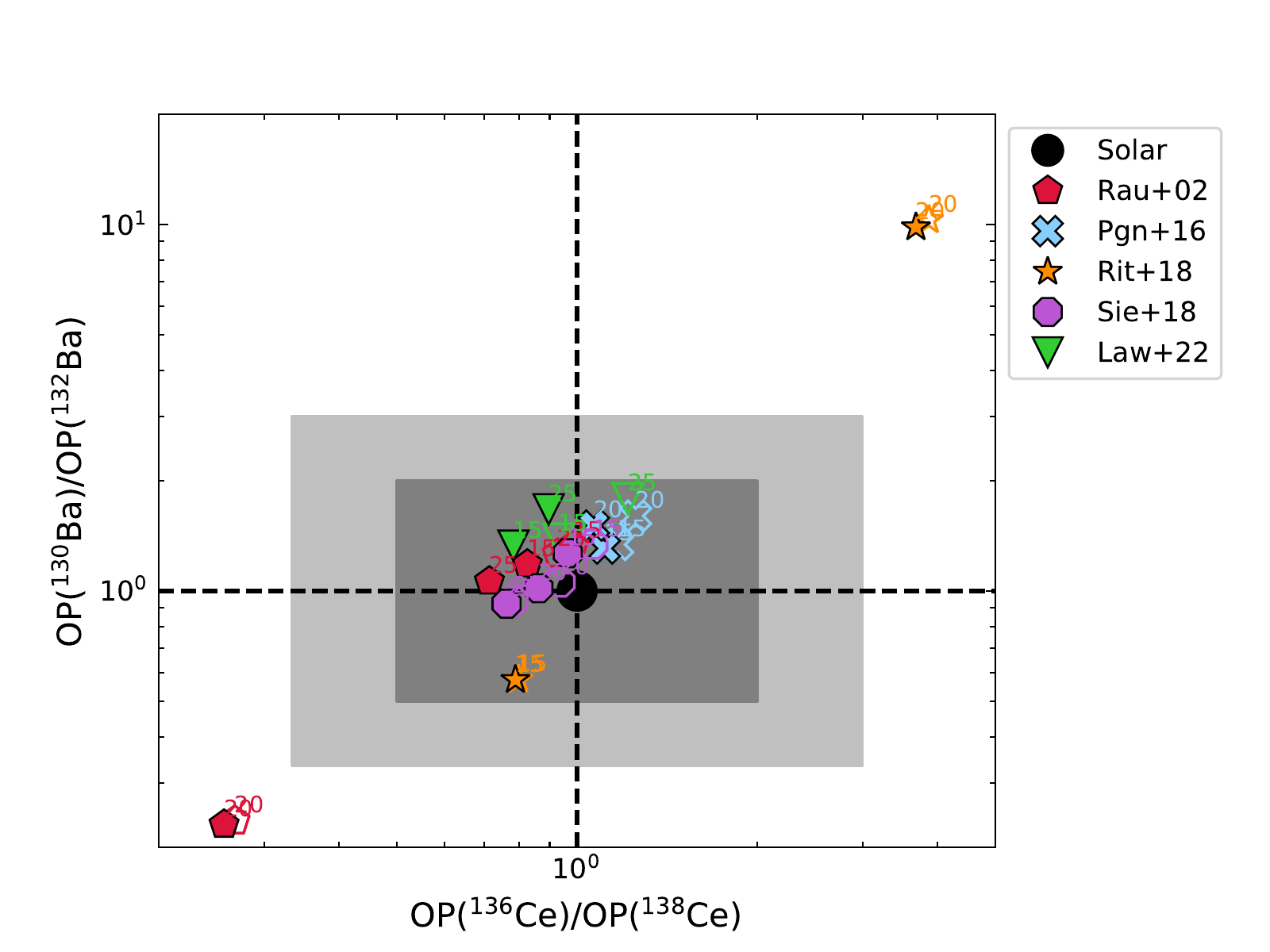}
                \caption{Same as \figurename~\ref{fig:00}, but for \isotope[130]{Ba}/\isotope[132]{Ba} versus the \isotope[136]{Ce}/\isotope[138]{Ce} ratio. \isotope[132]{Ba} and \isotope[138]{Ce} have radiogenic contributions from \isotope[132]{Ce}, \isotope[132]{La} and \isotope[138]{Nd}, respectively.}
                \label{fig:06}
            \end{figure}

        % === 06
        \subsection{\isotope[130]{Ba}/\isotope[132]{Ba} vs \isotope[136]{Ce}/\isotope[138]{Ce} (\figurename~\ref{fig:06})} \label{sub:07}
    
            The \g--process nucleosynthesis in the majority of the models reproduces very closely the solar ratios on both axis (although note that as in the previous Sec. \ref{sub:06}, the SIE25 model has all the OP factors close to 1), with two exceptions: RIT20 and RAU20.
            The RIT20 ratios are both super-solar because of a combination of the higher OPs ($\sim$30--40) of the more neutron-deficient isotopes (at numerator) relative to the lower OPs ($\sim$4--8) of the neutron-richer isotopes (at denominator). The RAU20 ratios are, instead, both sub-solar due to the effect of the C--O shell merger, which favours the neutron-richer isotopes. The same effect is present in the other model with C--O shell merger (RIT15), albeit to a lesser extent because OP(\isotope[132]{Ba}) and OP(\isotope[138]{Ce}) are similar in the two models, while OP(\isotope[130]{Ba}) and OP(\isotope[136]{Ce}) are higher in RIT15 than in RAU20.
            
\begin{figure}
                \centering
                \includegraphics[scale=0.5]{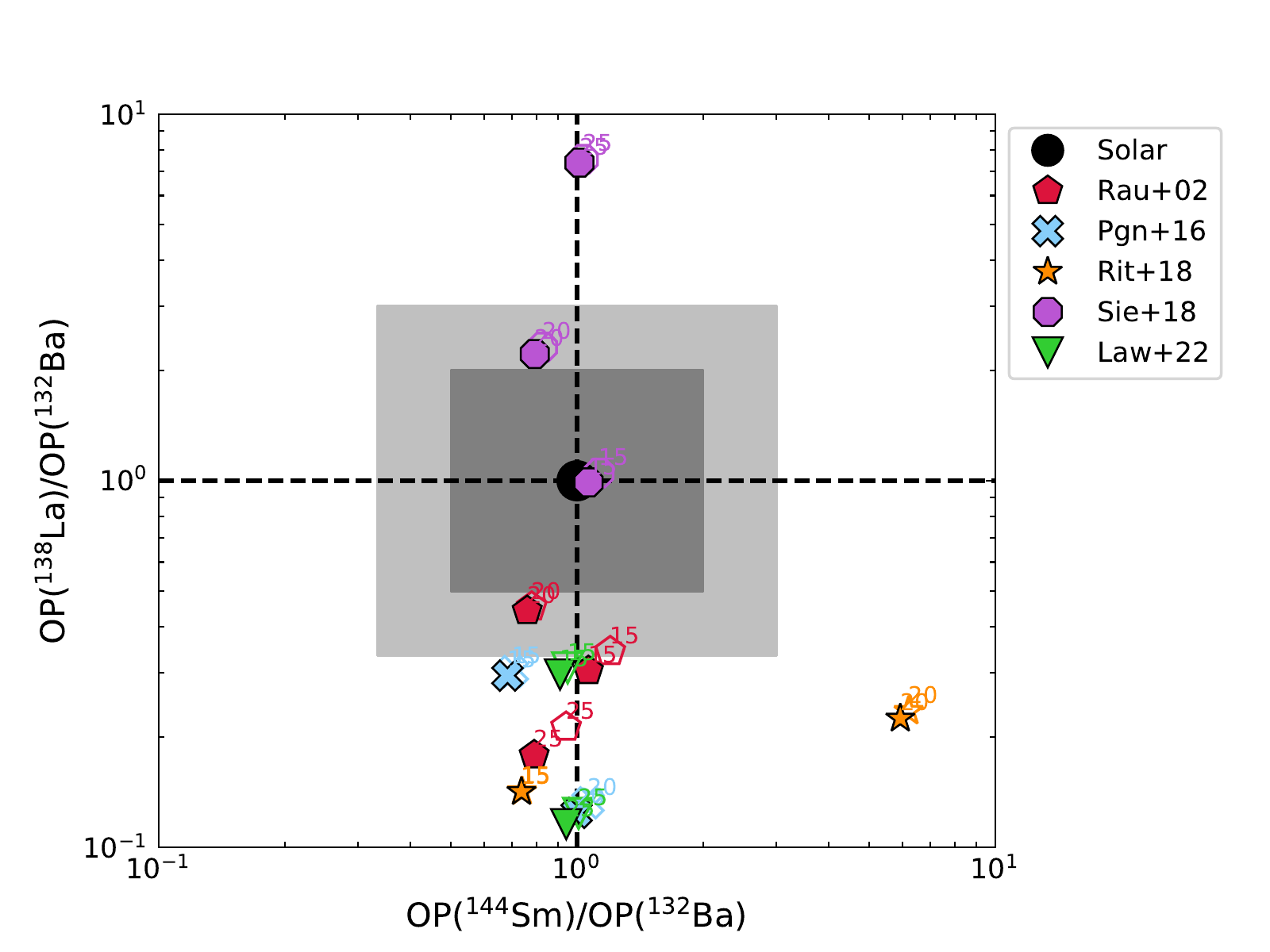}
                \caption{Same as \figurename~\ref{fig:00}, but for \isotope[138]{La}/\isotope[132]{Ba} versus the \isotope[144]{Sm}/\isotope[132]{Ba} ratio. \isotope[132]{Ba} has radiogenic contribution from isotope[132]{Ce} and \isotope[132]{La}.}
                \label{fig:07}
            \end{figure}

        % === 07
        \subsection{\isotope[138]{La}/\isotope[132]{Ba} vs \isotope[144]{Sm}/\isotope[132]{Ba} (\figurename~\ref{fig:07})} \label{sub:08}
    
            The \isotope[144]{Sm}/\isotope[132]{Ba} ratio (x-axis) of all the model falls within a factor of two of the solar value, except for RIT20, where the \isotope[132]{Ba} is underproduced. The \isotope[138]{La}/\isotope[132]{Ba} ratio (y-axis), instead, is more scattered and in most of the models it is more than three times lower than solar, except for the SIE models. This is because, as mentioned in Sec. \ref{sec:intro} and \ref{sec:f0}, \isotope[138]{La} may have an additional neutrino capture contribution, which is only included in the nuclear network of the SIE models. These models also have the lowest OP(\isotope[132]{Ba}), further contributing to higher \isotope[138]{La}/\isotope[132]{Ba} ratio. 
    
\begin{figure}
                \centering
                \includegraphics[scale=0.5]{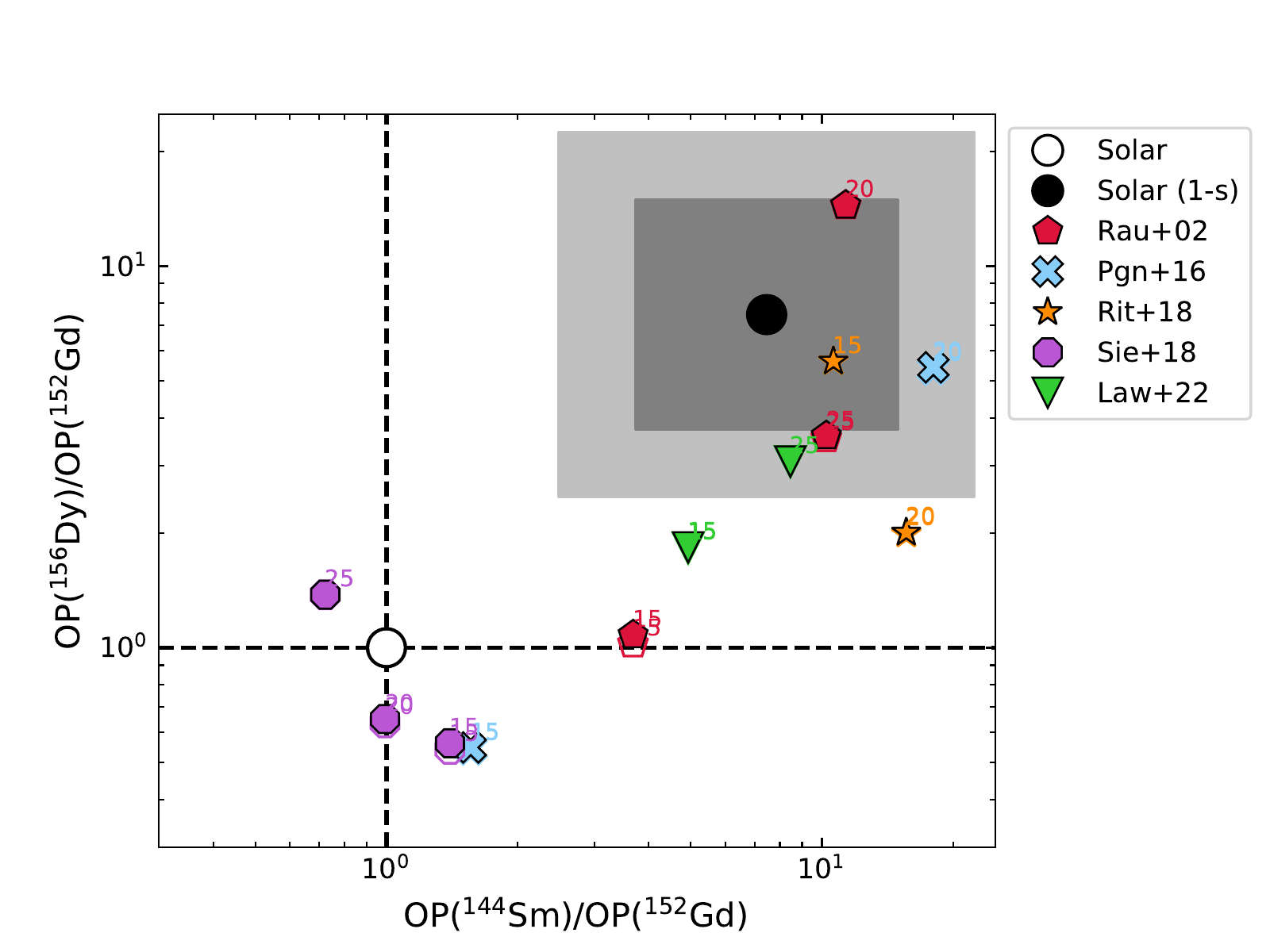}
                \caption{Same as \figurename~\ref{fig:00}, but for \isotope[156]{Dy}/\isotope[152]{Gd} versus the \isotope[144]{Sm}/\isotope[152]{Gd} ratio. The empty circle represents the solar ratio, while the filled symbol represents the solar ratio minus the \s--process contribution to the isotope \isotope[152]{Gd} (1-\s). Similarly to \figurename~\ref{fig:034}, the gray shaded areas identify a factor of 2 and 3 from the ratio calculated using the (1-\s) value.}
                \label{fig:08}
            \end{figure}

        % === 08
        \subsection{\isotope[156]{Dy}/\isotope[152]{Gd} vs \isotope[144]{Sm}/\isotope[152]{Gd} (\figurename~\ref{fig:08})} \label{sub:09}
    
             Of the three isotopes considered here, as discussed in Sec. \ref{sec:intro}  and \ref{sec:f0}, a fraction of the solar abundance of \isotope[152]{Gd} comes from the \s--process in AGB stars. The GCE computation of \cite{bisterzo:14} reported this fraction to be $86.6\%$. Therefore, in \figurename~\ref{fig:08} we represent the value to be matched as the solar (1-$s$) fraction only, i.e., $13.4\%$. Furthermore, in the models presented here \isotope[152]{Gd} has a dominant \s--process contribution from the pre-supernova He and C shells and the explosive \g--process component is only a small fraction of the total yield. \isotope[156]{Dy} and \isotope[144]{Sm}, instead, are \p--only nuclei that are produced by \g--process during the explosion, and their OPs are significantly higher than that of \isotope[152]{Gd}. 
    
            As in the cases of Sec. \ref{sub:06} and \ref{sub:07}, the SIE25 model is close to solar because all the OPs are close to one, therefore this is not discussed further. The \isotope[144]{Sm}/\isotope[152]{Gd} ratio (x-axis) of most of the models falls within a factor of three of the ratio calculated with the (1-$s$) value. The three SIE15, SIE20, and PGN15 models, instead, are close to the solar ratio because they have OP(\isotope[152]{Gd}) similar to the other models, but a relatively lower OP(\isotope[144]{Sm}). The \isotope[156]{Dy}/\isotope[152]{Gd} ratio (y-axis) of five models (RIT15, RAU20/25, PGN20 and LAW25) falls close to the ratio calculated with the (1-$s$) value. Among those, the models with C--O shell mergers (RIT15 and RAU20) have the best agreement. All the other models are closer to solar because they have OP(\isotope[156]{Dy}) $<2$, except for RIT20, where OP(\isotope[156]{Dy}) $\sim$ 4.
    
\begin{figure}
                \centering
                \includegraphics[scale=0.5]{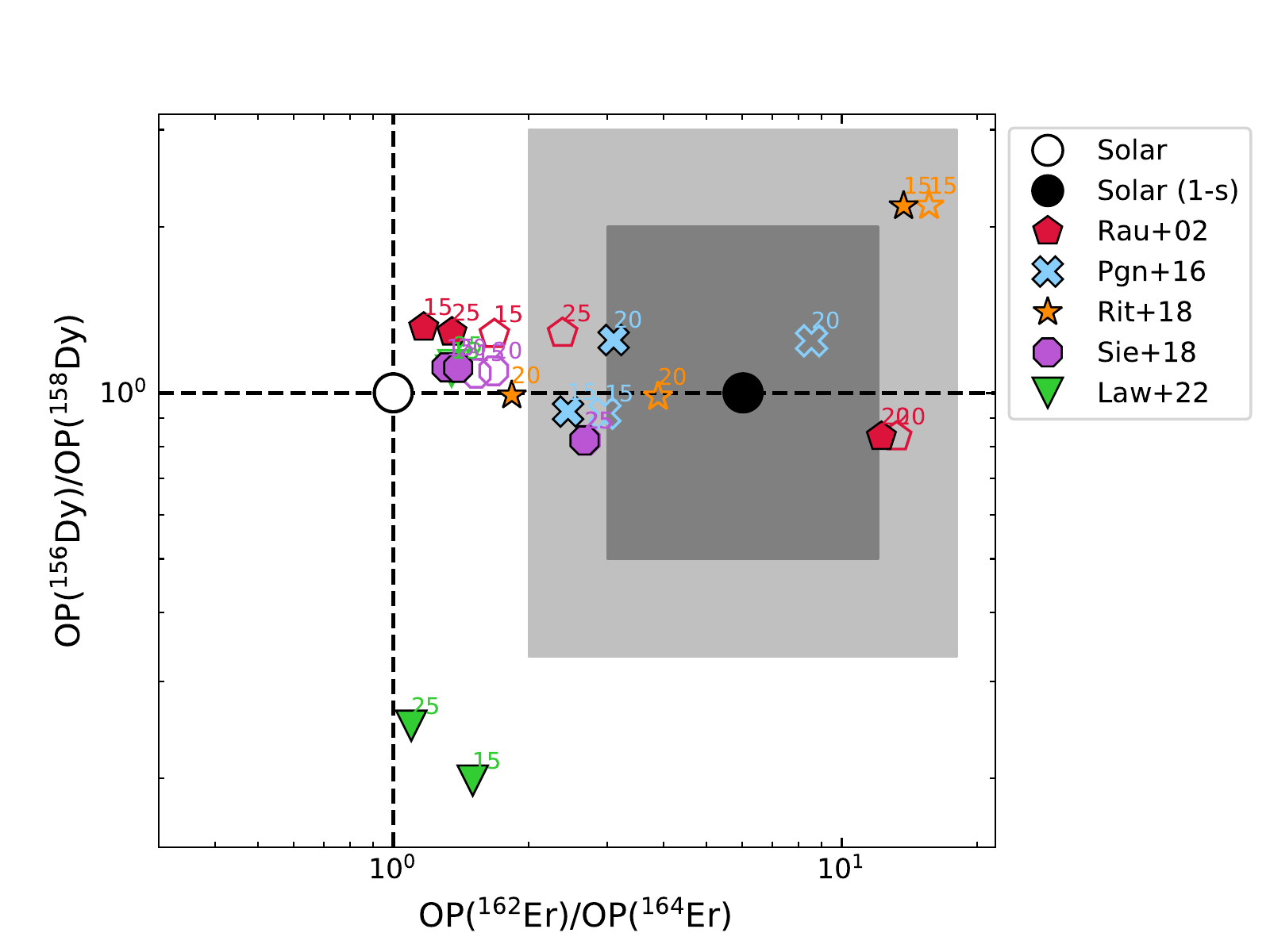}
                \caption{Same as \figurename~\ref{fig:00}, but for \isotope[156]{Dy}/\isotope[158]{Dy} versus the \isotope[162]{Er}/\isotope[164]{Er} ratio. As in \figurename~\ref{fig:08}, the empty circle represents the solar ratio, while the filled symbol represents the solar ratio minus the \s--process contribution to the isotope \isotope[164]{Er} (1-\s). The gray shaded areas identify a factor of 2 and 3 from the ratio calculated using the (1-\s) value.\isotope[158]{Dy} has strong radiogenic contribution from \isotope[158]{Er} in the LAW models only. \isotope[164]{Er} has a radiogenic contribution from \isotope[164]{Yb}, and a much less significant contribution from \isotope[164]{Tm}. \isotope[162]{Er} has radiogenic contribution from \isotope[162]{Yb}.}
                \label{fig:09}
            \end{figure}

        % === 09
        \subsection{\isotope[156]{Dy}/\isotope[158]{Dy} vs \isotope[162]{Er}/\isotope[164]{Er} (\figurename~\ref{fig:09})} \label{sub:10}
    
            As in the case of \isotope[152]{Gd} (Sec \ref{sub:09}), also the solar abundance of \isotope[164]{Er} (the denominator of the x-axis) has an \s--process component (see Sec.~\ref{sec:intro} and \ref{sec:f0}). Therefore, here we only focus on the possible \g--process origin of \isotope[164]{Er}, which corresponds to $16.6\%$ of its solar abundance \citep{bisterzo:14}. Five models (PGN15, PGN20, SIE25, plus the two models with C--O shell mergers RIT15 and RAU20) fall within a factor of 3 of the \isotope[162]{Er}/\isotope[164]{Er} ratio calculated with the (1-$s$) value.
            Unlike in the previous cases (Sec. \ref{sub:02}, \ref{sub:03}, \ref{sub:04}, \ref{sub:06}, and \ref{sub:07}) here the C--O shell merger favours the production of the neutron-deficient \isotope[162]{Er} relative to \isotope[164]{Er}. 
    
            The \isotope[156]{Dy}/\isotope[158]{Dy} ratio (y-axis) is within a factor of 2 from solar, except in the two LAW models, where the radiogenic contribution to \isotope[158]{Dy} is higher than in the other models and results in a shift of the ratio to a value more than 4 times lower than solar. Note that, as in the case of the Er isotopes discussed above, the C--O shell merger (in RIT15 and RAU20) favours the production of the neutron-deficient isotope. 
    
\begin{figure}
                \centering
                \includegraphics[scale=0.5]{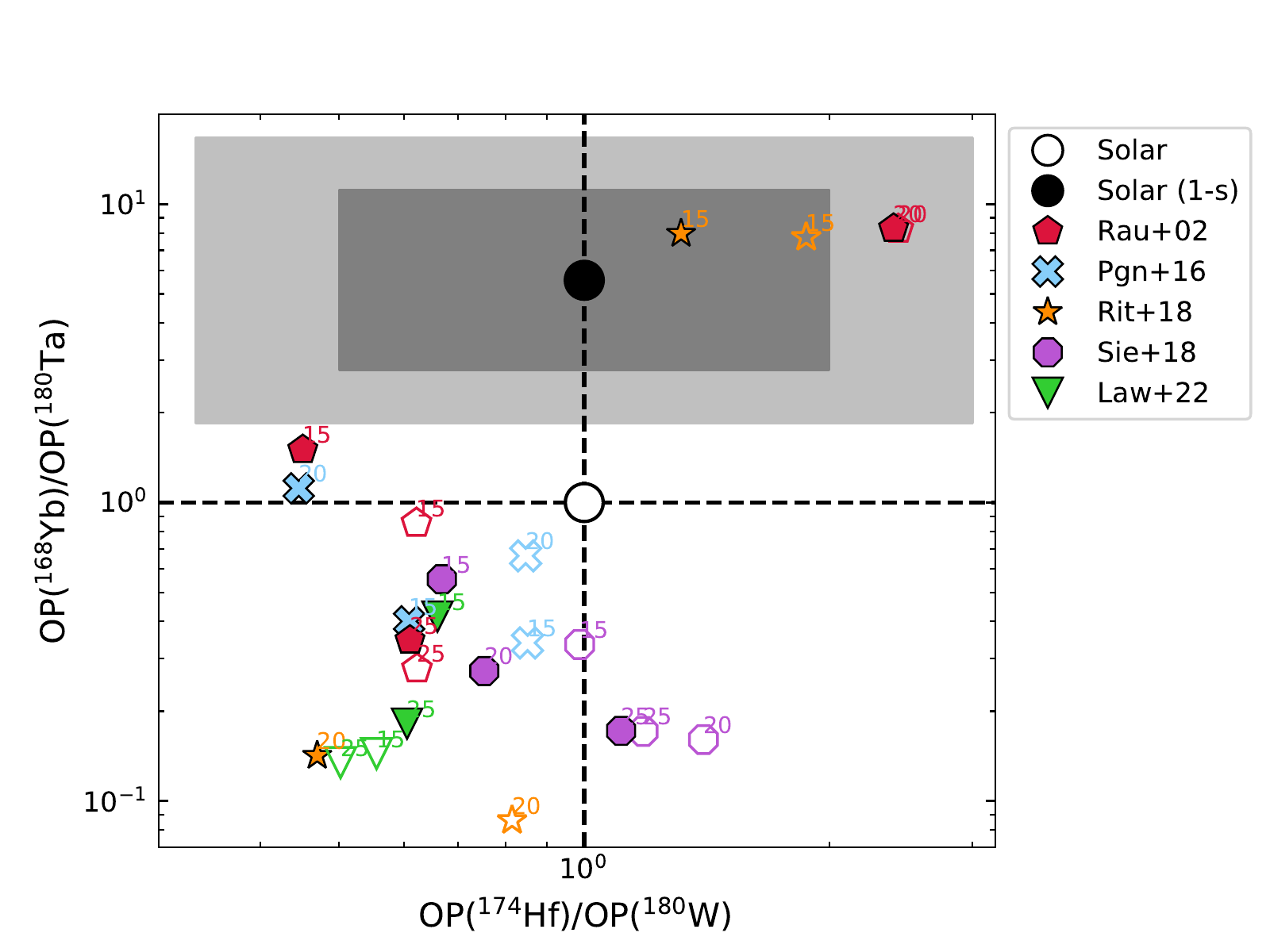}
                \caption{Same as \figurename~\ref{fig:00}, but for \isotope[168]{Yb}/\isotope[180]{Ta} versus the \isotope[174]{Hf}/\isotope[180]{W} ratio. As in \figurename~\ref{fig:08}, the empty circle represents the solar ratio, while the filled symbol represents the solar ratio minus the \s--process contribution to the isotope \isotope[180]{Ta} (1-\s). The gray shaded areas identify a factor of 2 and 3 from the ratio calculated using the (1-\s) value. \isotope[168]{Yb}, \isotope[174]{Hf}, and \isotope[180]{W} have a radiogenic contribution from \isotope[168]{Hf}, \isotope[174]{W}, and \isotope[180]{Os}, respectively. In the RAU models the same three isotopes have a further radiogenic contribution from \isotope[168]{Lu}, \isotope[174]{Ta}, and \isotope[180]{Re}, respectively.}
                \label{fig:10}
            \end{figure}

        % === 10
        \subsection{\isotope[168]{Yb}/\isotope[180]{Ta} vs \isotope[174]{Hf}/\isotope[180]{W} (\figurename~\ref{fig:10})} \label{sub:11}
    
            As in the case of \isotope[163]{Dy} (Sec \ref{sub:10}), \isotope[179]{Hf} becomes unstable at stellar temperatures, activating a \s--process branching point. Therefore, \isotope[180]{Ta} (and \isotope[180]{W} to a lesser extent) may show a neutron-capture contribution due to the chain \isotope[179]{Hf}($\beta^-$)\isotope[179]{Ta}(n,\g)\isotope[180]{Ta}($\beta^-$)\isotope[180]{W}. Therefore, we consider only the [(1-$s$) solar] abundance of \isotope[180]{Ta}, which is $18.0\%$ \citep{bisterzo:14}. Note that this residual solar abundance of \isotope[180]{Ta} is not only due to the \g--process, since this isotope may also receive a contribution from neutrino-capture on \isotope[180]{Hf} \citep{byelikov:07,sieverding:15}. The models mostly populate the region of the plot where the production of \isotope[180]{W} and \isotope[180]{Ta} dominate, respectively, over the production of \isotope[174]{Hf} and \isotope[168]{Yb} and they are more than a factor of three away from [(1-$s$) solar] ratios. The C--O shell mergers (RIT15 and RAU20), instead, result in the opposite behaviour. Furthermore, these models are the within a factor of 3 from the [(1-$s$) solar] ratios.
            
            In all the models (except the two LAW models) the radiogenic contributions to \isotope[174]{Hf} and \isotope[180]{W} shift the \isotope[174]{Hf}/\isotope[180]{W} ratio (x-axis) to lower values. In the two LAW models, instead, the \isotope[180]{W} yield is much larger than the \isotope[174]{Hf} yield, due to an efficient production of \isotope[180]{W} via the \s--process in the C-shell. 
    
\begin{figure}
                \centering
                \includegraphics[scale=0.5]{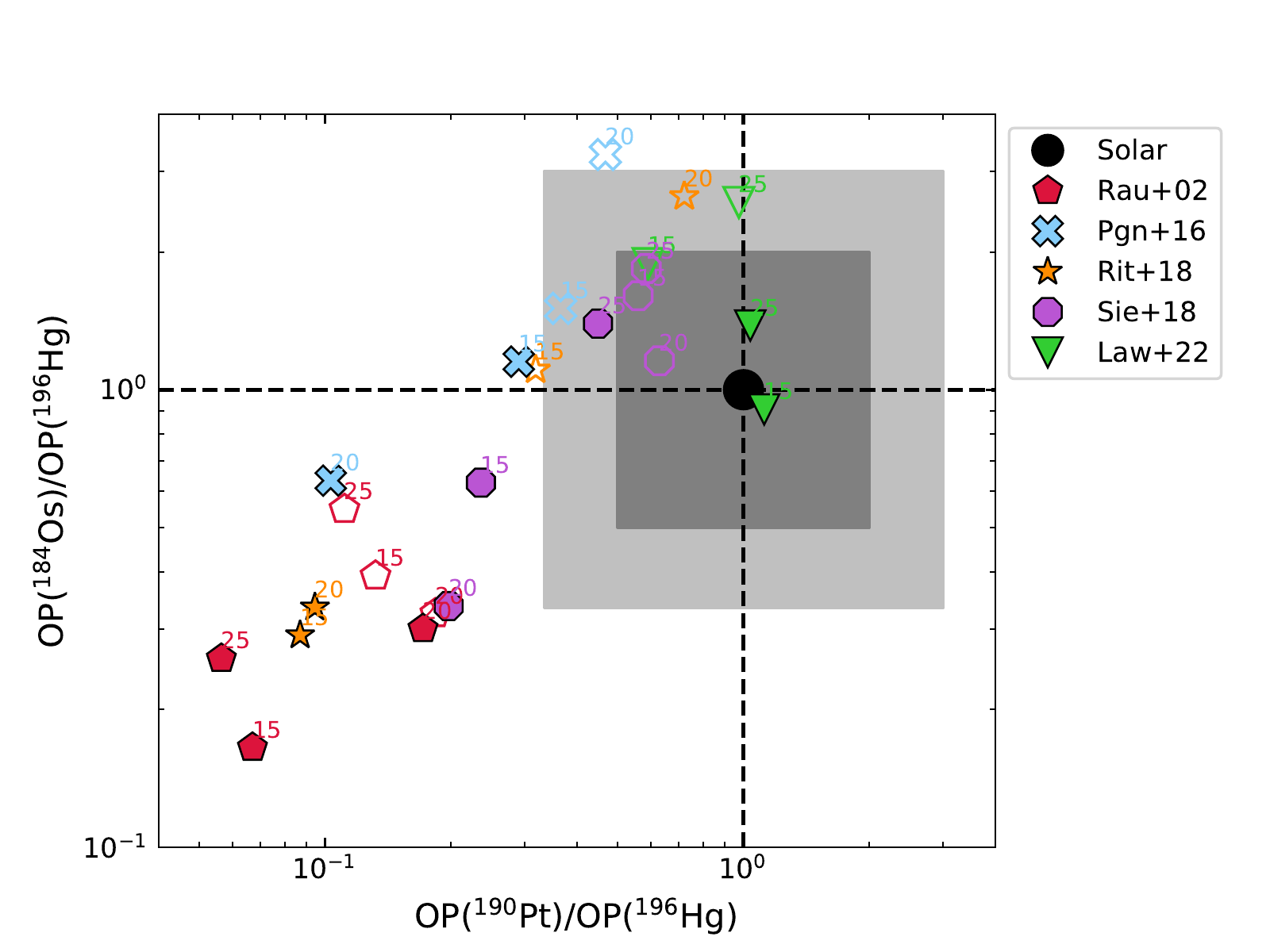}
                \caption{Same as \figurename~\ref{fig:00}, but for \isotope[184]{Os}/\isotope[196]{Hg} versus the \isotope[190]{Pt}/\isotope[196]{Hg} ratio. \isotope[196]{Hg} has a radiogenic contribution from \isotope[196]{Pb} and, only in RAU models, from \isotope[196]{Tl}. \isotope[190]{Pt} has a radiogenic contribution from \isotope[190]{Hg} and, only in LAW models, from\isotope[190]{Au}.}
                \label{fig:11}
            \end{figure}

        % === 11
        \subsection{\isotope[184]{Os}/\isotope[196]{Hg} vs \isotope[190]{Pt}/\isotope[196]{Hg} (\figurename~\ref{fig:11})} \label{sub:12}
    
            The majority of the models overproduce \isotope[196]{Hg} (the denominator in both ratios), relative to \isotope[184]{Os} and \isotope[190]{Pt} and predict ratios more than three times lower than solar. The main responsible for this result is the strong radiogenic contribution from \isotope[196]{Pb} to \isotope[196]{Hg}. In fact, without the inclusion of this contribution most of the models (except the RAU models) fall within a factor of three from solar ratio. 
            Specifically, these models have OP(\isotope[196]{Hg}) $>$ OP(\isotope[184]{Os}) $>$ OP(\isotope[190]{Pt}). It follows that 
            the discrepancy between the models and the solar ratios is more severe in the case of \isotope[190]{Pt}/\isotope[196]{Hg} (x-axis), where these models show ratios between a factor of 4 to 20 lower than solar. For the \isotope[184]{Os}/\isotope[196]{Hg} ratio (y-axis), instead, the most extreme model gives a value a factor of 6 lower than solar. 
            
            Of the remaining four models closer to the solar ratios (PGN15, SIE25, LAW15, and LAW25), PGN15 and SIE25 have OP(\isotope[190]{Pt}) $< 2$ and OP(\isotope[184]{Os}) $>$ OP(\isotope[196]{Hg})$>$ OP(\isotope[190]{Pt}), while the LAW models behave differently. This is because they have OP(\isotope[190]{Pt}) significantly higher than the other models, 
            also due to the radiogenic contribution of \isotope[190]{Hg} and \isotope[190]{Au}. Specifically, in LAW15 OP(\isotope[190]{Pt}) $>$ OP(\isotope[196]{Hg}) $>$ OP(\isotope[184]{Os}). In this model, the radiogenic contribution to \isotope[190]{Pt} is twice as large as that to \isotope[196]{Hg}, therefore, the inclusion of the radiogenic nuclei decreases the  ratio by a factor of two. In LAW25, OP(\isotope[184]{Os}) $>$ OP(\isotope[190]{Pt}) $>$ OP(\isotope[196]{Hg}). In this model, the radiogenic contribution to \isotope[190]{Pt} and \isotope[196]{Hg} is roughly equal, so the ratio does not change.
    
\end{appendix}

\end{document}